\pdfoutput=1
\documentclass[11pt]{article}
\usepackage{amssymb}
\usepackage{graphicx}
\parskip \baselineskip
\newcommand{\ds}{\displaystyle}
\newcommand{\R}{\mathbb{R}}

\newcommand{\Z}{\mathbb{Z}}

\newcommand{\ol}{\overline}

\newcommand{\ra}{\rightarrow}
\newcommand{\Ra}{\Rightarrow}

\newcommand{\bmax}{\hbox{\boldmath$\max$}}
\newcommand{\bempt}{\hbox{\boldmath$\emptyset$}}

\begin{document}
\title{The joy of implications, aka pure Horn formulas: mainly a survey}

\author{Marcel Wild}

\date{}
\maketitle

\begin{quote}
A{\scriptsize BSTRACT}: Pure Horn clauses have also been called (among others) functional dependencies, strong association rules, or simply implications. We survey the mathematical theory of implications with an emphasis on the progress made in the last 30 years.
\end{quote}


\section{Extended introduction}

This article is devoted to the mathematics and (to lesser extent) algorithmics of implications; it is mainly a survey of results obtained in the past thirty years but features a few novelties as well. 
The theory of implications mainly developed, often under mutual ignorance, in these five fields:

 Boolean Function Theory, Formal Concept Analysis, Lattice Theory, Relational Database Theory, Learning Theory.

As standard text-books in these fields we recommend [CH], [GW], [Bi] $+$  [G], [MR2] $+$ [M], and [RN, ch.VI] $+$ [FD] respectively. Broadly speaking we collect from each field only those major results that concern (or can be rephrased in terms of) ``abstract implications'', and {\it not} the substance matter of the field itself. There are three minor exceptions to this rule. First, there will be two detours (Subsections 4.1, 4.2) into lattice theory; among the five fields mentioned this is the one the author is most acquainted with. Second, in Subsection 1.1 just below, in order to motivate the theory to come, we glance at three ``real life'' occurencies of implications in these areas:
Relational Databases,
Formal Concept Analysis, and Learning Spaces. The third exception concerns 3.6; more on that later.
The second part (1.2) of our extended introduction gives the detailed section break up of the article.

{\bf 1.1} We shall only give very rudimentary outlines of three areas mentioned above; more detailed accounts of 1.1.1 to 1.1.3 are found in [MR2], [GW], [FD].  The sole purpose here is to convey a feeling for the many meanings that a statement ``$A$ implies $B$'' can have. This will contrast with the uniform mathematical treatment that all ``abstract'' implications $A \ra B$ obey.

\begin{center}
\includegraphics[scale=0.5]{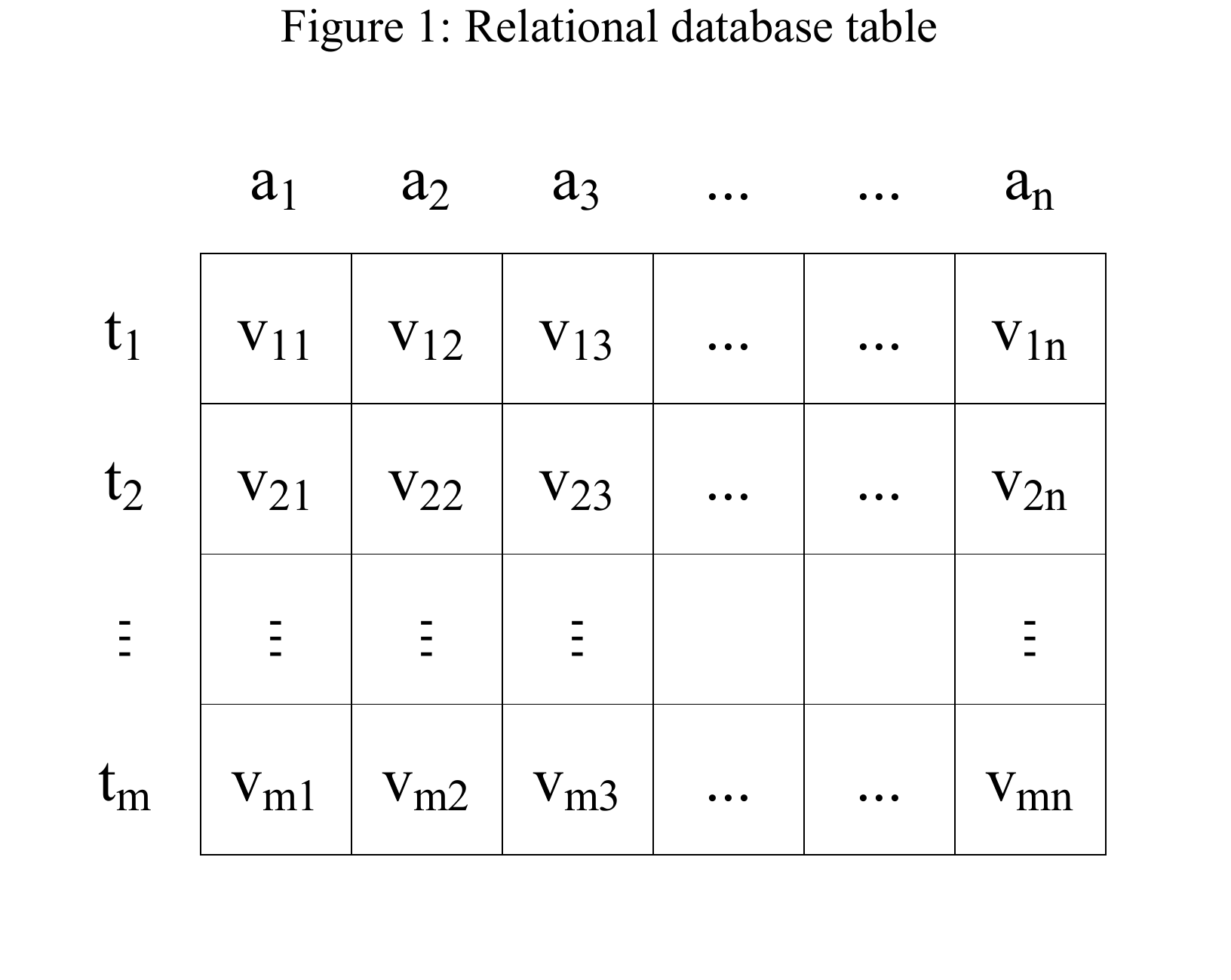}
\end{center}

{\bf 1.1.1} As to relational database (RDB), imagine this as a large array in which every row (called record) corresponds to a particular object $t_i$, and in which the columns correspond to the various attributes  $a_j$
that apply. See Figure 1. Each attribute has a domain which is the set of values that it may assume. Following an example of J. Ullman, take a relational database whose records match the ``teaching events'' occuring at a university in a given semester. The attributes are $C=$ course, $T=$ teacher, $H=$ hour, $R =$ room, $S=$ student. The domain of $C$ may be $\{$algebra, analysis, lattice theory, $\cdots \}$, the domain of $T$ could be $\{$Breuer, Howell, Janelidze, $ \cdots \}$, and so forth. If $A, B$ are sets of attributes then the validity of $A \ra B$  means that {\it any two objects which have identical values for all attributes in $A$, also have identical values for all attributes in $B$.} Examples of implications $A \ra B$ (also called {\it functional dependencies}) that likely hold in a well designed database include the following: $\{C\} \ra \{T\}$ (each course has one teacher), $\{H,R\} \ra \{C\}$ (only one course meets in a room at one time), $\{H, S\} \ra \{R\}$ (a student can be in only one room at a given time).

{\bf 1.1.2} Let now $G$ and $M$ be any sets and $I \subseteq G \times M$ be a binary relation. In Formal Concept Analysis (FCA) one calls the triple $(G, M, I)$ a {\it context}, and $gIm$ is interpreted as the {\it object} $g \in G$ having the {\it attribute} $m \in M$. If $A, B \subseteq M$ then the validity of $A \ra B$  has a different\footnote{One may view a context as a RBD all of whose attribute domains are Boolean, thus $\{\mbox{True}, \mbox{False}\}$ or $\{1,0\}$. But depending on viewing it as RBD or context, different implications hold.} ring from before:  {\it Any object that has all attributes in $A$, also has all attributes in $B$} (see also 2.1.2 and 2.2.3). 

Let us focus on particular contexts of type $(G, M, \ni)$. Thus the objects $g \in G$ become {\it subsets} $X$ of some set $M$ of {\it items}. Saying that $g \in G$ ``has attribute'' $m \in M$ now just means $X \ni m$. Often the sets $X$ are called {\it transactions}, and the elements $m \in M$ are called {\it items}. 
If $A, B \subseteq M$ then  $A \ra B$ is a valid implication iff {\it every transaction $X$ that contains the itemset $A$, also contains the itemset $B$}. For instance, each transaction can contain the items a customer  bought at a supermarket on a particular day. In this scenario a plausible implication e.g. is $\{$butter, bread$\} \ra \{$milk$\}$. Notice that $A \ra B$ may be a valid implication simply because many transactions do not contain $A$ at all. To exclude this possibility one often strengthens the previous definition of ``valid implication'' by additionally demanding that say 70\% of all transactions must contain the itemset $A$. The terminology ``transaction'' and ``itemset'' is borrowed from Frequent Set Mining (FSM), a paradigm that developed in parallel to FCA for a long time, despite of close ties.  See also 3.6.3.4.

{\bf 1.1.3} As to Learning Spaces [FD], these are mathematical structures applied in mathematical modeling of education. In this framework (closer in spirit to [GW] than to [RN] type learning theory) the validity of an implication $A \ra B$ means the following: {\it Every student mastering the (types of) problems in set $A$ also masters the problems in set $B$.} See also Expansion 16.

{\bf 1.2} Some readers may have guessed that this zoo of implications fits the common hat of pure Horn functions, i.e. Boolean functions like $(x_1 \wedge x_2 \wedge x_3) \ra x_4$ and conjunctions thereof. While this is true the author, like others, has opted for a more {\it stripped down formalism}, using elements and sets rather than literals and truth value assignments, etc. Nevertheless, discarding pure Horn function terminology altogether would be short-sighted; certain aspects can only be treated, in any sensible way, in a framework that provides immediate access to the empire of general Boolean function theory that e.g. houses prime implicates and the consensus algorithm.

Without further mention, all structures considered in this article will be {\bf finite}. Thus we won't point out which concepts extend or can be adapted to the infinite case. A word on [CH, chapter 6, 56 pages] is in order. It is a survey on Horn functions to which the present article (PA) compares as follows. Briefly put, the intersection $CH \cap PA$ is sizeable (though not notation-wise), and so are $CH\backslash PA$ (e.g. applications, dualization, special classes), as well as $PA \backslash CH$ (e.g. 3.6 and 4.1 to 4.4). We note that 4.1 also features special classes but {\it others}.

Here comes the section break up. Section 2 recalls the basic connections between closure operators $c$ and closure systems ${\cal F}$ (2.1), and then turns to implications ``lite'' in 2.2. Crucially, each family $\Sigma$ of implications $A \ra B$ gives rise to a closure operator $c(\Sigma, -)$ and whence to a closure system ${\cal F} = {\cal F}(\Sigma)$. Furthermore, {\it each} closure operator $c$ is of type $c= c(\Sigma, -)$ for suitable $\Sigma$. Section 3 is devoted to the finer theory of implications. Centerpieces are the Duquenne-Guigues implicational base (3.2) and the canonical direct base in 3.3. Subsection 3.4 is about mentioned pure Horn functions, 3.5 is about acyclic and related closure operators, and 3.6 surveys the connections between two devices to grasp closure systems ${\cal F}$. One device is any implicational base, the other is the subset $M({\cal F}) \subseteq {\cal F}$ of meet-irreducible closed sets. 

Section 4 has the title ``Selected topics''.
In 4.1 the attention turns from meet to join-irreducibles, i.e. we show that {\it every} lattice ${\cal L}$ gives rise to a closure system ${\cal F}_J$ on its set $J = J({\cal L})$ of join irreducibles. Consequently it makes sense to ask about optimum implicational bases $\Sigma$ for various types of lattices. We have a closer look at modular, geometric and meet-distributive lattices. The other topics in brief are: an excursion into universal algebra (4.2), {\it ordered} direct implicational bases (4.3), an algorithm for generating ${\cal F}(\Sigma)$ in compact form (4.4), and general (impure) Horn functions in 4.5. According to Theorem 6 implications ``almost'' suffice to capture even impure Horn functions.

In order to have full proofs of some results without interrupting the story line, we store these proofs in little ``boxes'' (called Expansion 1 to Expansion 20) in Section 5. Most of these results are standard; nevertheless we found it worthwile to give proofs fitting our framework. Some Expansions simply contain further material. Due to space limitations the full versions of some Expansions are only available in the preliminary draft [W7].

Recall that this article attempts to survey the {\it mathematical theory} of pure Horn functions ($=$ implications), and apart from mentioned exceptions {\it not} their applications.  Our survey also includes a couple of new results, mainly in 2.2.5, 3.3.2,  3.4.3, 4.1.6, in Expansion 8 and in (33). Further Theorem 3 and 6 are new. In order to stimulate research four {\it Open Problems} are dispersed throughout the text (in 3.6.2, \ Expansion 5, \ Expansion 15).

\section{The bare essentials of closure systems and implications}

Everything in Section 2 apart from 2.2.5 is standard material.
Because of the sporadic appearance of contexts (1.1.2) a good reference among many is [GW].

\subsection{Closure systems and closure operators}

A {\it closure system} ${\cal F}$ with universe $E$ is a subset of the powerset ${\cal P}(E)$ with the property that

(1) \quad $\bigcap {\cal G} \in {\cal F}$ for all ${\cal G} \subseteq {\cal F}$.

Here $\bigcap {\cal G}$ denotes the intersection of all sets contained in ${\cal G}$. Its smallest element is $\bigcap {\cal F}$ and, crucially, it has a largest element as well. Namely, as a matter of taste, one may either postulate that $E$ belongs to ${\cal F}$, or one may argue that $\emptyset \subseteq {\cal F}$ implies $\bigcap \emptyset \in {\cal F}$, and that $\bigcap \emptyset = E$. Thus ${\cal F}: = {\cal P}(E)$ is the largest closure system with universe $E$, and ${\cal F}: = \{E\}$ is the smallest. The members $X\in {\cal F}$ are called {\it closed} sets, and $X \in {\cal F} \backslash \{E\}$ is 
{\it meet-irreducible} if there are no strict closed supersets $A$ and $B$ of $X$ with $A \cap B = X$. We write $M({\cal F})$ for the set of meet irreducibles of ${\cal F}$. It is clear that

(2)  \quad $(\forall X \in {\cal  F}) \ \ ({\cal F} \backslash \{X\}$ is closure system $\Leftrightarrow X \in M({\cal F})$)


{\bf 2.1.1} Closure systems are linked to closure operators\footnote{We recommend [BM, sec.6] for a historic account of the origins of these two concepts.}. (The link to lattices is postponed to 4.1.) Namely, {\it closure operators} are maps $c: {\cal P}(E) \ra {\cal P}(E)$ which are extensive ($U \subseteq c(U)$), idempotent ($c(c(U)) = c(U))$ and monotone ($U \subseteq U'\Ra c(U) \subseteq c(U'))$. In this situation (see Expansion 1)

(3) \quad ${\cal F}_c : = \{X \in {\cal P}(E) : c(X) = X \}$ is a closure system.

As to the reverse direction, 
if ${\cal F} \subseteq {\cal P}(E)$ is a closure system then
$c_{\cal F}(U) : = \bigcap \{S \in {\cal F}: \ S \supseteq U \}$
yields a closure operator $c_{\cal F} : {\cal P}(E) \ra {\cal P}(E)$. 
One can show [GW, Theorem 1] that ${\cal F}_{(c_{\cal F})} = {\cal F}$ and $c_{({\cal F}_c)} = c$. One calls $U$ a {\it generating set} of $X \in {\cal F}$ if $c_{\cal F} (U) = X$. On a higher level ${\cal H} \subseteq {\cal P}(E)$ is a {\it generating set} of ${\cal F}$ if ${\cal F} ({\cal H}) : = \{\bigcap {\cal G}: {\cal G} \subseteq {\cal H}\}$ equals ${\cal F}$. It is easy to see that ${\cal H}$ is a generating set of ${\cal F}$ iff ${\cal H} \supseteq M({\cal F})$. In this case
$c_{\cal F}(U)$ can also be calculated as 

(4) \quad $c_{\cal H}(U) = \bigcap \{S \in{\cal H}: S \supseteq U\}$.

The first idea that springs to mind to calculate ${\cal F}({\cal H})$ from ${\cal H}_1 : = {\cal H}$ is to keep on calculating ${\cal H}_{k+1} = {\cal H}_k \ast {\cal H}_1 : = \{X \cap Y : X \in {\cal H}_k, Y \in {\cal H}_1\} \ (k  = 1, 2, \ldots)$ until ${\cal H}_{k+1} = {\cal H}_k = {\cal F}({\cal H})$. Unfortunately the approach is doomed by the frequent recalculation of closed sets, and the need to keep large chunks of ${\cal F}({\cal H})$ in central memory. A clever idea of C.E. Dowling [FD, p.50] avoids the recalculations, but not the space problem; see also Expansion 4.

{\bf 2.1.2.} Here comes a frequent source of closure operators. Let $E_1, E_2$ be sets and let $R \subseteq E_1 \times E_2$ be a binary relation. 
For all $X \subseteq E_1$ and $Y \subseteq E_2$ put 
$$\begin{array}{lllll}
X^\dagger & : = & f(X) & : = & \{y \in E_2: (\forall x \in X) (x,y) \in R\}\\
\\
Y^\ast & : = & g(Y) & : =& \{x \in E_1: (\forall y \in Y) (x,y) \in R\} \end{array}$$
Then the pair $(f, g)$ yields a {\it Galois connection}. It is easy to see that $X \subseteq Y^\ast$ iff $X^\dagger \supseteq Y$. Furthermore, it holds [GW, Section 0.4] that $c_1 : = g \circ f$ is a closure operator ${\cal P}(E_1) \ra {\cal P}(E_1)$, and $c_2 : =f \circ g$ is a closure operator ${\cal P}(E_2) \ra {\cal P}(E_2)$.  
For instance, let $(G, M, I)$ be a context in Formal Concept Analysis as glimpsed in 1.1.2. If $A \subseteq M$ is any set of attributes then $c_2(A) = A{^{\ast \dagger}}$ is the set of attributes $m$ enjoyed by every object $g \in A^\ast$, i.e. by every object $g$ that has all attributes of $A$.  Put another way, $A \ra c_2(A)$ is a ``valid'' implication in the sense that whenever $g$ has all attributes in $A$, then $g$ has all attributes in $c_2(A)$. This matches our discussion of ``implications'' $A \ra B$ in 1.1.2. See [PKID1] for a survey of 1072 papers dedicated to applications of FCA.

\subsection{Implications ``lite''}

A pair of subsets $(A, B) \in {\cal P}(E) \times {\cal P}(E)$ will be called an {\it implication}. Both $A = \emptyset$ or $B = \emptyset$ are allowed. (See 3.4.2 for the full picture). We shall henceforth write $A \ra B$ instead of $(A,B)$ and call $A$ the {\it premise} and $B$ the {\it conclusion} of the implication.  Any family

(5) \quad $\Sigma : = \{A_1 \ra B_1, A_2 \ra B_2, \cdots, A_n \ra B_n\}$

of implications gives rise to a closure operator as follows. Putting $[n] : = \{1, 2, \ldots, n\}$ for any set $S \subseteq E$ we define

(6) \quad $S': = S \cup \bigcup \{B_i: \ i \in [n], A_i \subseteq S \}$.

By finiteness the chain $S \subseteq S'\subseteq (S')'\subseteq \cdots$ stabilizes at some set $c(\Sigma, S)$. This algorithm matches {\it forward chaining} in [CH, 6.2.4]. We call $c(\Sigma, S)$ the $\Sigma$-{\it closure} of $S$. It is clear that the function $c(\Sigma, -)$ is a closure operator on ${\cal P}(E)$. As to speeding up the calculation of $c(\Sigma, X)$ see Expansion 2.
 It is evident that $\Sigma \subseteq \Sigma'$ implies $c(\Sigma, U) \subseteq c(\Sigma', U)$ for all $U \subseteq E$, but say $\Sigma = \Sigma_1 \cup \Sigma_2$ does not entail $c(\Sigma, U) = c(\Sigma_2, c(\Sigma_1, U))$.
 By (3) the closure operator $c(\Sigma, -)$ induces a closure system ${\cal F}(\Sigma)$. Hence for all $X \subseteq E$ it holds that
 
 (7) \quad $X \in {\cal F}(\Sigma) \ \Leftrightarrow \ X = c(\Sigma, X) \ \Leftrightarrow \ \forall (A \ra B) \in \Sigma: \ A \not\subseteq X$ \ or \ $B \subseteq X$

Skipping $c(\Sigma, -)$, it is easy to show directly that for any given family $\Sigma$ of implications the sets $X \subseteq E$ with $(A \subseteq X \Ra B \subseteq X$, for all $(A \ra B) \in \Sigma)$ constitute a closure system.

{\bf 2.2.1}  We say that $\Sigma$ is {\it equivalent} to $\Sigma'$ (written $\Sigma \equiv \Sigma'$) if the closure operators $c(\Sigma, -)$ and $c(\Sigma', -)$ coincide. There are three obvious (and others in 3.4) notions of ``smallness'' for families $\Sigma$ of implications as in (5):
\begin{itemize}
	\item $\Sigma$ is {\it nonredundant} if $\Sigma \backslash \{A_i \ra B_i\}$ is not equivalent to $\Sigma$ for all $1 \leq i \leq n$.
	\item $\Sigma$ is {\it minimum} if $ca (\Sigma):=|\Sigma|$ equals $\min \{  |\Sigma'|: \Sigma' \equiv \Sigma  \}$.
	\item $\Sigma$ is {\it optimum} if  $s(\Sigma) : = |A_1 | + \cdots +|A_n| + |B_1| + \cdots + |B_n|$ equals\\
	 $\min \{s(\Sigma') : \Sigma' \equiv \Sigma \}$.
\end{itemize}
For instance, $\Sigma_1 : = \{\{1\} \ra \{2\}, \ \{1 \} \ra \{3\}, \ \{1\} \ra \{2,3\} \}$ is {\it redundant} ($=$ not nonredundant) because say $\{1\} \ra \{2,3\}$ can be dropped. 
Both $\Sigma_2 : = \{\{1\} \ra \{2\}, \{1\} \ra \{3\} \}$ and $\Sigma_3: = \{\{1\} \ra \{2, 3\} \}$ are equivalent to $\Sigma_1$, and are clearly nonredundant. The latter is minimum, in fact optimum. Generally each minimum family is nonredundant. Less obvious, each optimum family is minimum as proven in Theorem 1.

{\bf 2.2.2} From $\{1\} \ra \{2\}$ and $\{2\} \ra \{3\}$ ``somehow follows'' $\{1\} \ra \{3\}$, but this notion needs to be formalized. We thus say that $A \ra B$ {\it follows} from (or: is a {\it consequence} of) a family $\Sigma$ of implications, and write $\Sigma \vDash (A \ra B)$, if $\Sigma \cup \{A \ra B\}$ is equivalent to $\Sigma$. The following fact is often useful:

(8) \quad $\Sigma \vDash (A \ra B)$ if and only if $B \subseteq c(\Sigma, A)$

{\it Proof of (8)}. As to $\Ra$, by assumption the two closure operators $c(\Sigma, -)$ and $c(\Sigma \cup \{A \ra B\}, -)$ coincide. Thus in particular $B \subseteq c(\Sigma \cup \{A \ra B\}, A) = c(\Sigma, A)$. As to $\Leftarrow$, it suffices to show that $c(\Sigma \cup \{A \ra B\}, U)$ which clearly coincides with $c(\Sigma \cup \{A \ra B\}, c(\Sigma, U))$, is contained in $c(\Sigma, U)$ for $U \subseteq E$. {\it Case 1:} $A \not\subseteq c(\Sigma, U)$. Then $c(\Sigma \cup \{A \ra B\}, c(\Sigma, U)) = c(\Sigma, U)$ by the very definition of the closure operator $c(\Sigma \cup \{A \ra B\}, -)$, {\it Case 2:} $A \subseteq c(\Sigma, U)$. Then by assumption $B \subseteq c(\Sigma, A) \subseteq c(\Sigma, U)$, and so again $c(\Sigma \cup \{A \ra B\}, c(\Sigma, U)) = c(\Sigma, U)$. \quad $\square$

In Expansion 3 we introduce among other things a ``syntactic'' notion $\vdash$ of {\it derivability} and show that $\Sigma \vdash (A \ra B)$ is equivalent to $\Sigma \vDash (A \ra  B)$.

{\bf 2.2.3}  Conversely, let  us {\it start out} with any closure operator $c: {\cal P}(E) \ra {\cal P}(E)$. Then a family $\Sigma$ of implications is called an {\it implicational base} or simply {\it base} of $c$ if $c(S) = c(\Sigma, S)$ for all $S \subseteq E$. Each closure operator $c$ {\it has} an implicational base, in fact $\Sigma_c : = \{X \ra c(X) : X \subseteq E\}$ does the job\footnote{This is slightly less trivial than it first appears.  Clearly $c(Y) \subseteqq c(\Sigma_c, Y)$, but why not $\subsetneqq$?}. Unfortunately, $\Sigma_c$ is too large to be useful. How to find smaller ones is the theme of Section 3.


{\bf 2.2.4} Putting $B = c(\Sigma, A)$ in (8) we see that $A \ra c(\Sigma, A)$ is a consequence of $\Sigma$. 
Thus for any closure operator $c$ the implication $A \ra c(A)$ is a consequence of any $\Sigma$ that happens to be an implicational base of $c$. But implications $A \ra c(A)$ often carry a natural meaning ``on their own'', such as $A \ra c_2(A)$ in 2.1.2.

{\bf 2.2.5} Streamlining the proof of [KN, Theorem 20] here comes an example of a visually appealing closure operator $c$, all of whose optimum bases can be determined ``ad hoc'', i.e. without the theory to be developed in Section 3.2. Namely, $c$ arises from an affine point configuration $E \subseteq \R^2$ by setting $c(A) : = E\cap ch(A)$ where $ch(A)$ is the ordinary (infinite) convex hull of $A$. For instance, if $E = [8]$ is as in Figure 2, then $c(\{1,2,4\}) = \{1, 2, 4, 5, 8\}$.

\begin{center}
\includegraphics[scale=0.8]{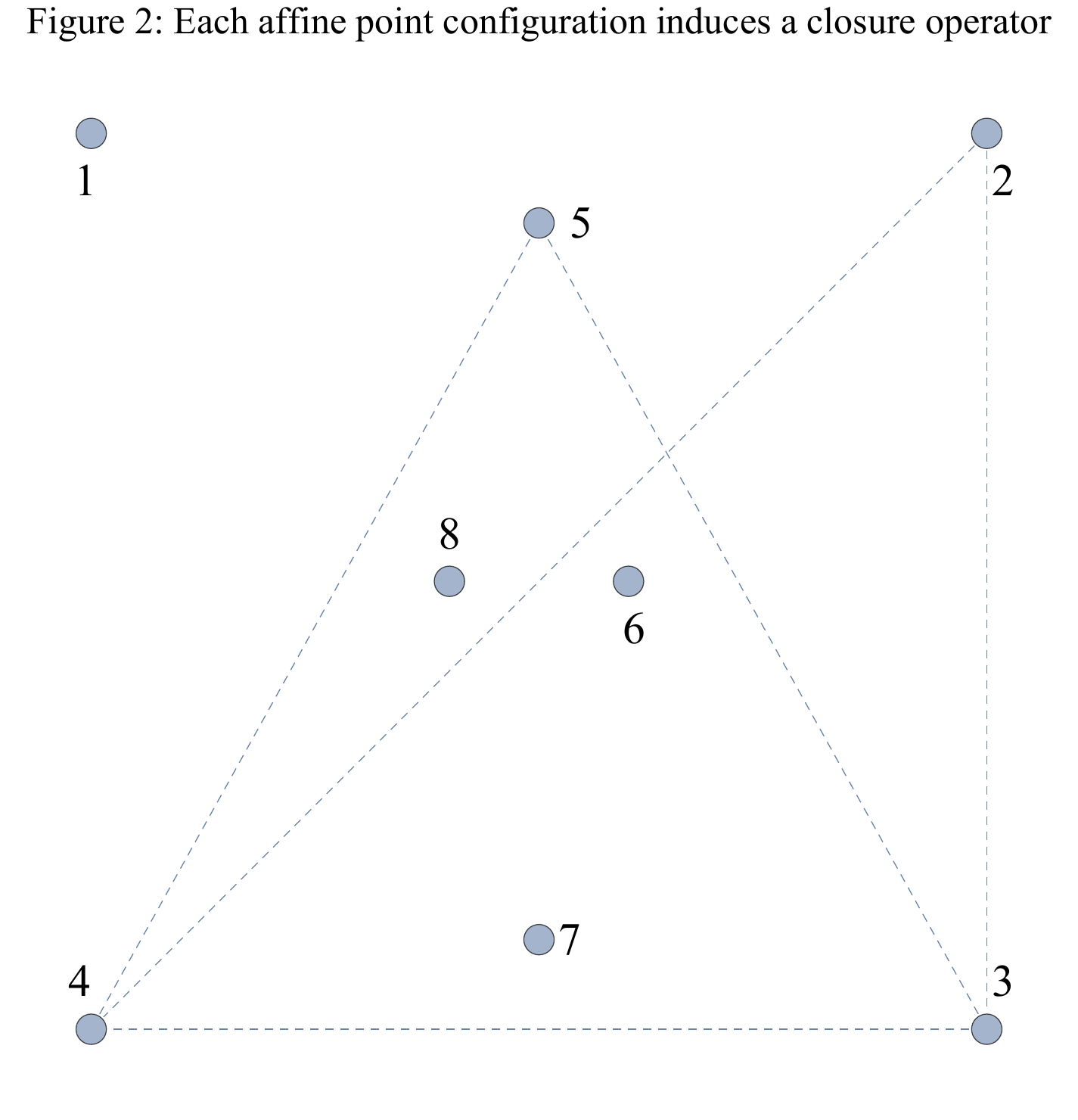}
\end{center}

From the deliberations below (which generalize to point sets in $\R^n$ without $n+1$ points in a hyperplane) it will readily follow that $c$ has exactly $144$ optimum bases. 
Let ${\cal T}$ be the set of all 3-element subsets $T \subseteq E$ with $ch(T) \cap (E \backslash T)\neq \emptyset$.
Let $\Sigma$ be any base of $c$ and let $T \in {\cal T}$ be arbitrary. From $c(\Sigma, T) = c(T) \varsupsetneqq T$, and the fact that all proper subsets of $T$ are closed, follows that $\Sigma$ must contain an implication with premise $T$. Now consider a set $\Sigma_{op}$ of implications $T \ra \{e_T\}$ where $T$ scans ${\cal T}$ and where $e_T \in ch(T) \cap (E \backslash T)$ is arbitrary. Obviously, $c(\Sigma_{op}, S) \subseteq c(S)$ for all $S \subseteq E$. If we can show that $\Sigma_{op}$ is a base at all, then it must be optimum by the above. By way of contradication assume that $\Sigma_{op}$ is no base, and fix a set $S \subseteq E$ with $c(\Sigma_{op}, S) \varsubsetneqq c(S)$ for which $ch(S)$ is minimal. From $S \varsubsetneqq c(S)$ follows\footnote{This follows from the well-known fact that convex hulls like $ch(S)$ can be obtained by repeatedly taking closures of $3$-element sets.} that $T \subseteq S$ for at least one $T \in {\cal T}$, and thus $e_T \in c(\Sigma_{op}, S)$. Consider the unique triangulation of $ch(S)$ into triangles $ch(T_i)(i \in I)$ all of whose (3-element) vertex sets $T_i$ contain $e_T$. Then $T_i \subseteq c(\Sigma_{op}, S)$, and so $c(\Sigma_{op}, T_i) \subseteq c(\Sigma_{op}, S)$. Furthermore from $ch(T_i)\varsubsetneqq  ch(S)$ follows $c(\Sigma_{op}, T_i) = c(T_i)$, and so

$$c(\Sigma_{op}, S) \supseteq \ds\bigcup_{i\in I} c(\Sigma_{op},T_i)  \ = \ \ds\bigcup_{i \in I} c(T_i) \ \stackrel{4}{=} \ c(S),$$
which contradicts $c(\Sigma_{op}, S) \varsubsetneqq c(S)$.  The mentioned number $144$ arises as $2^4 \cdot 3^2$ in view of the fact that exactly four $T \in {\cal T}$ have $|c(T) \backslash T| =2$ (namely $T = 123, 124, 134, 234)$, and exactly two $T \in {\cal T}$ have $|c(T) \backslash T|=3$  (namely $T = 127, 345$). Here we e.g. wrote 124 instead of $\{1, 2, 4\}$. This kind of shorthand will be used frequently.

\section{The finer theory of implications}

In 3.1 we couple to each closure operator $c$ some quasiclosure operator $S \mapsto S^\bullet$  which will be crucial in the sequel. In [W3] it is shown that certain minimization results independently obtained by Guigues-Duquenne [GD] and Maier [M] are equivalent. By now the formalisation of Guigues-Duquenne has prevailed (mainly due to the beneficial use of closure operators), and also is adopted in Section 3.2. Section 3.3 introduces the canonical direct implication base. Section 3.4 finally introduces pure Horn functions, and 3.5 addresses the acyclic case. 
It seems that the link between implications and the meet-irreducibles of the induced closure system (Section 3.6) must be credited to Mannila and R\"{a}ih\"{a} [MR1]. As indicated in the introduction, in 3.6 we also shed some light on {\it why} it is important to go from $\Sigma$ to $M({\cal F})$ and vice versa.

\subsection{Quasiclosed and pseudoclosed sets}

Given any closure operator $c: {\cal P}(E) \ra {\cal P}(E)$ and $S \subseteq E$ we put 

(9) \quad $S^\circ : = S \cup \bigcup \{c(U): \ U \subseteq S,\quad c(U) \neq c(S) \}$.

Because $E$ is finite the chain $S \subseteq S^\circ \subseteq (S^\circ)^\circ \subseteq \cdots$ will stabilize at some set $S^\bullet$. It is clear that $S \mapsto S^\bullet$ is  a closure operator and that $S^\bullet \subseteq c(S)$ for all $S \subseteq E$. We call $S \mapsto S^\bullet$ the $c${\it -quasiclosure}, or simply {\it quasiclosure} operator when $c$ is clear from the context.

\includegraphics[scale=0.6]{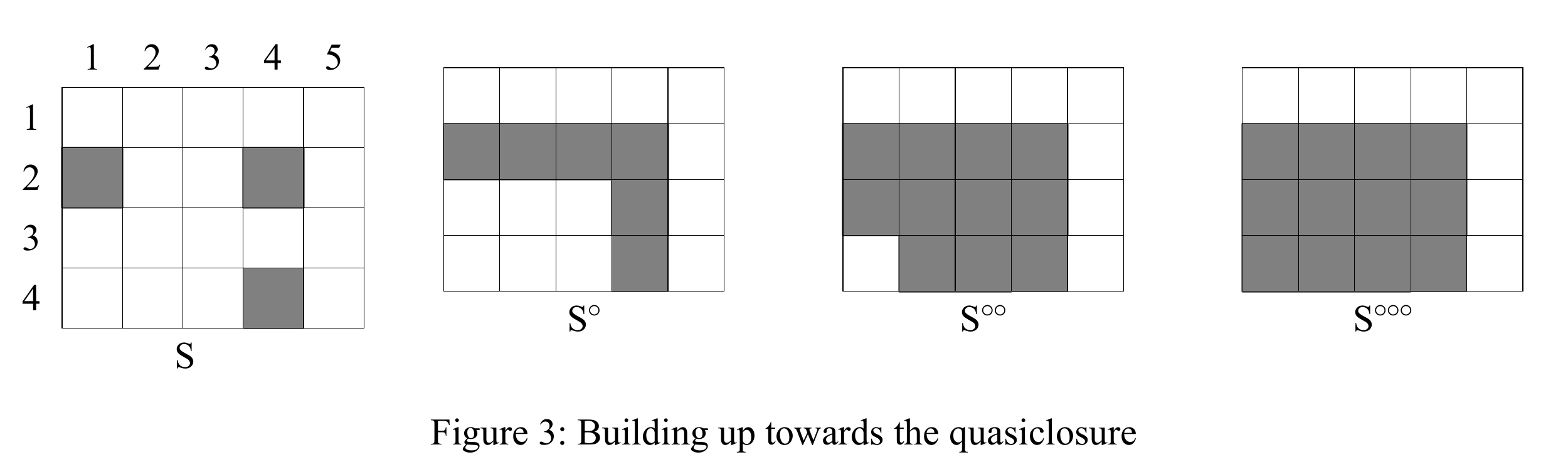}

As an example, consider the $4 \times 5$ grid $E$ in Figure 3 and the closure system ${\cal F} \subseteq {\cal P}(E)$ of all contiguous rectangles $I \times J$ (thus $I \subseteq [4]$ and $J \subseteq [5]$ are {\it intervals}). Let $c: = c_{\cal F}$ be the coupled closure operator. For $S: = \{(2,1), (2,4), (4,4)\}$ (matching the three gray squares on the left in Figure 3) all singleton subsets are closed, and for the 2-element subsets we have
$$\begin{array}{lll}
c(\{(2,1), (2,4))\}) & =& \{(2,1), (2,2),(2,3), (2,4)\} = : S_1 \neq c(S),\\
\\
c(\{(2,4), (4,4)\}) &= & \{(2,4), (3,4), (4,4) \} = : S_2 \neq c(S),\\
\\
c(\{(2,1), (4,4)\}) & =& \{2,3,4\} \times \{1,2,3,4\} = c(S). \end{array}$$
Hence $S^\circ = S_1 \cup S_2$. If $T \subseteq S^\circ$ is any set with $(4,1) \in c(T)$ then necessarily $(2,1), (4,4) \in T$ (why?), whence $c(T) = c(S)$. Hence $S^{\circ \circ} \subseteq c(S) \backslash \{(4,1)\}$. Jointly with 
$$c(\{(2,2), (4,4)\}) \cup c(\{(2,1), (3,4)\}) = c(S) \backslash \{(4,1)\}$$
follows that $S^{\circ \circ} =c(S) \backslash \{(4,1)\}$. Finally $S^{\circ \circ \circ} = S^\bullet = c(S)$ because e.g. $(4,1) \in c(\{(3,1), (4,2)\}) \neq c(S)$.
We call\footnote{Unfortunately no standard terminology exists. It holds that $Y \subseteq X^\bullet$ iff $X$ {\it directly determines} $Y$ (modulo some ``cover of functional dependencies'') in the sense of [M, Def.5.9]. Do not confuse this notion of ``direct'' with the one in Section 3.3.} a subset {\it properly quasiclosed} if we like to emphasize that it is quasiclosed but {\it not} closed.
For instance the set $S = \{(2,1), (2,4)\}$ in Fig.3 is properly quasiclosed.

\includegraphics[scale=0.5]{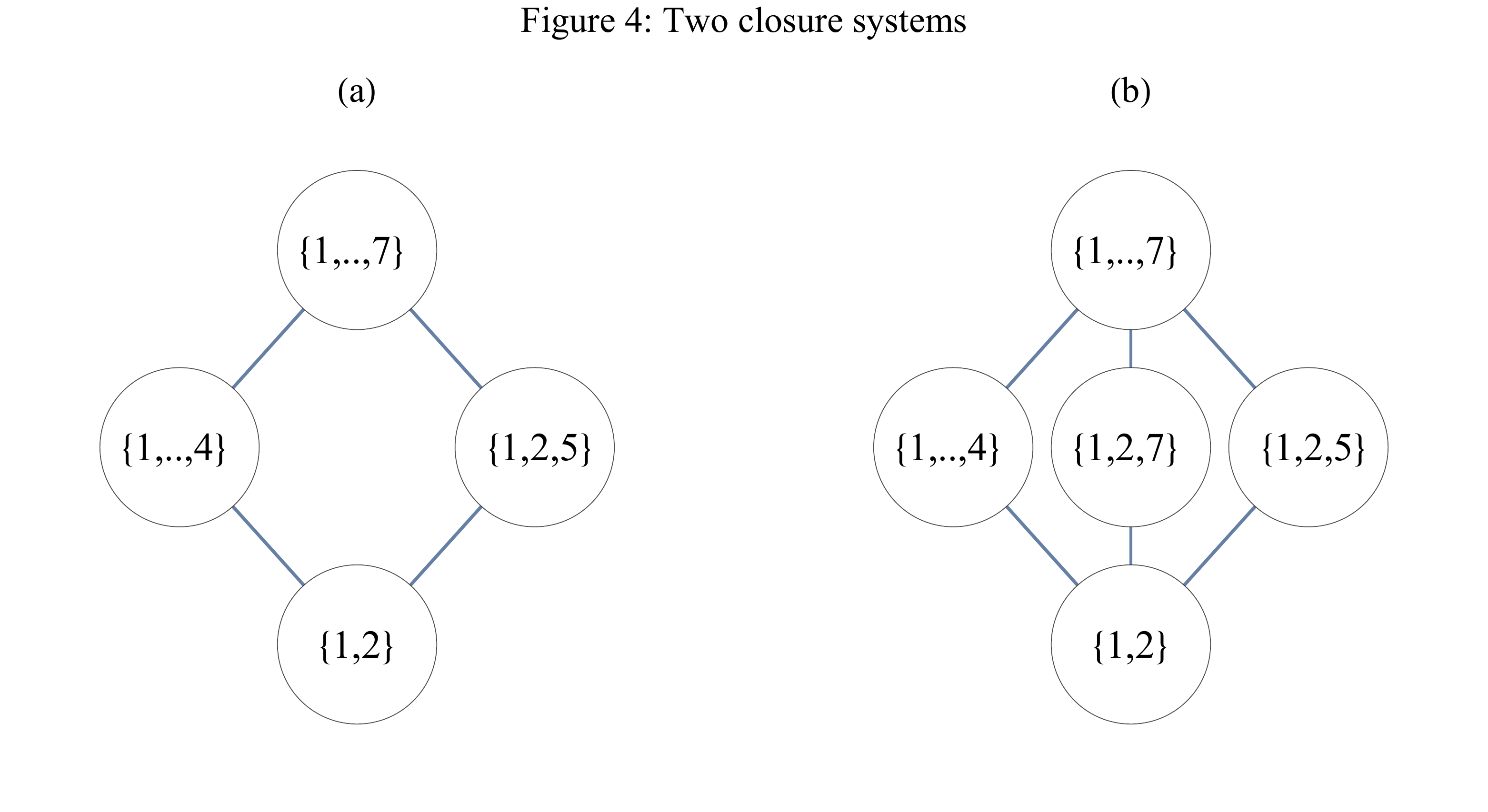}

{\bf 3.1.1} As another example take $E = [7]$ and let ${\cal F} \subseteq {\cal P}(E)$ be the closure system of Figure 4(a) with associated closure operator $c: = c_{\cal F}$. For our $c$ at hand the properly quasiclosed generating sets for each closed set are these:

$\begin{array}{cll}
12 & : & \bempt, 1,2\\
1234 & : & {\bf 123}, {\bf 124} \ (\mbox{why not 13?})\\
125 & : & \mbox{none}\\
1234567 & : & {\bf 126}, {\bf 127}, 1256, 1257, 1267, {\bf 12345}, 12346, 12347,
12567, 123456, 123457, 123467
\end{array}$

Let ${\cal F} \subseteq {\cal P}(E)$ be a closure system. As opposed to (2) one can show that

(10) \quad $(\forall Q \subseteq E) \ \ {\cal F} \cup \{Q\}$ is a closure system \ $\Leftrightarrow \ Q$ is quasiclosed

See Figure 4(b) where $Q : = \{1, 2, 7\}$ was added to ${\cal F}$. One checks that indeed $Q \cap X \in {\cal F}$ for all $X \in {\cal F}$.

\subsection{The canonical Guigues-Duquenne base}

For closure operators $c: \ {\cal P}(E) \ra {\cal P}(E)$ we define the equivalence relation $\theta \subseteq {\cal P}(E) \times {\cal P}(E)$ by

(11) \quad $(U, U') \in \theta \ : \ \Leftrightarrow \ c(U) = c(U')$.

For any implicational base $\Sigma$ of $c$ and for any $X \subseteq E$ let $\Sigma (X)$ be the set of those implications $A \ra B$  in $\Sigma$ for which $c(A) = c(X)$. It holds that

(12) \quad $Y^\bullet = c(\Sigma \backslash \Sigma (Y), Y)$ \quad for all \quad $Y \subseteq E$,

where $Y \mapsto Y^\bullet$ is the $c$-quasiclosure operator. Being a key ingredient for establishing Theorem 1 below let us repeat and slightly amend the proof of (12) given in [W5, Lemma 4]. For starters we replace  $\Sigma$ by the equivalent family  $\ol{\Sigma}$ of  implications which has each $U \ra V$ from $\Sigma (Y)$ replaced by the {\it full} implication $U \ra c(\Sigma, U)$. Because $\Sigma \setminus \Sigma (Y)$ equals $\ol{\Sigma} \setminus \ol{\Sigma}(Y)$ it suffices to prove that

$(\ol{12})$ \quad $Y^\bullet = c(\ol{\Sigma} \setminus \ol{\Sigma} (Y), Y)$ for all $Y \subseteq E$.

The inclusion $\supseteq$ being obvious  it suffices to show that $B \subseteq c(\ol{\Sigma} \setminus \ol{\Sigma}(Y), Y)$ implies $B^\circ \subseteq c(\ol{\Sigma} \setminus \ol{\Sigma}(Y), Y)$. Since $B^\circ= B \cup \bigcup \{c(\ol{\Sigma}, X): X \subseteq B$  and $c (\ol{\Sigma}, X) \subsetneqq c(\ol{\Sigma}, B)\}$ this further reduces to show that $c(\ol{\Sigma}, X) \subsetneqq c(\ol{\Sigma}, Y)$ implies that $c(\ol{\Sigma}, X) = c(\ol{\Sigma} \setminus \ol{\Sigma}(Y), X)$. But this holds since by construction all implications from $\ol{\Sigma} (Y)$ are of type $(U \ra c(\ol{\Sigma}, U)) = (U \ra c(\ol{\Sigma}, Y))$, and thus cannot be used in the generating process of $c(\ol{\Sigma}, X)$. This proves $(\ol{12})$ and hence (12).

A properly quasiclosed set $P$ is {\it pseudoclosed}\footnote{From an algorithmic point of view this equivalent defintion is more appropriate: $P$ is pseudoclosed iff $P \neq c(P)$ and $c(P_0) \subseteq P$ for all pseudoclosed sets $P_0$ strictly contained in $P$. Another name for pseudoclosed is {\it critical} (not to be confused with ``critical'' in 4.1.5).} if it is minimal among the properly quasiclosed sets in its $\theta$-class. (In the set listing of 3.1.1 these are the boldface sets.) Consider now the family of implications

(13) \quad $\Sigma_{GD} : = \{P \ra c(P): \ P \subseteq E$ is pseudoclosed$\}$,

where $GD$ stands for Guigues-Duquenne. Clearly $c(\Sigma_{GD}, Y) \subseteq c(Y)$ for all $Y \subseteq E$, and so $\Sigma_{GD}$ will be an implicational base of $c$ if we can show that $c(\Sigma_{GD}, Y) \supseteq c(Y)$ for all $Y \subseteq E$. Indeed, by (12) applying the implications from $\Sigma_{GD}\backslash \Sigma_{GD}(Y)$ blows up $Y$ to $Y^\bullet$. If $Y^\bullet \neq c(Y)$ then by definition there is a pseudoclosed set $P\subseteq Y^\bullet$ with $c(P) = c(Y^\bullet) = c(Y)$. Applying the implication $(P \ra c(P))\in \Sigma_{GD}$ to $Y^\bullet$ shows that $c(\Sigma_{GD}, Y) \supseteq c(Y)$.

This establishes part (a) of Theorem 1 below. For the remainder see [W3, Thm.5] which draws on [GD] and again uses (12). Two more concepts are in order.
One calls $X \in {\cal F}_c$ {\it essential} if $X$ contains a properly quasiclosed generating set. Thus the essential sets coincide with the closures of the pseudoclosed sets.
The {\it core}  [D] of a closure operator $c: {\cal P}(E) \ra {\cal P}(E)$ is

(14) \quad $\mbox{core}(c) =\, \mbox{core}({\cal F}_c): = \{X \in {\cal F}_c : \ X \ \mbox{is essential} \}$.

\begin{tabular}{|l|} \hline \\
{\bf Theorem 1:} Let $c: \ {\cal P}(E) \ra {\cal P}(E)$ be a closure operator.\\
\\
(a) The family of implications $\Sigma_{GD}$ is an implicational base of $c$.\\
\\
(b) If $\Sigma$ is any implicational base then $|\Sigma| \geq |\Sigma_{GD}|$. More specifically, for each pseudoclosed\\
\hspace*{.5cm} $P \subseteq E$ there is some $(A_P \ra B_P) \in \Sigma$ with $A_P \subseteq P$ and $A^\bullet_P = P$.\\
\\
(c) If $\Sigma$ is a nonredundant implicational base then $\{c(A): (A \ra B) \in \Sigma\}$ equals $\mbox{core} ({\cal F}_c)$.\\
\\
(d) If $\Sigma$ is a nonredundant implicational base which moreover consists of {\it full} implications\\
\hspace*{.5cm} $A \ra c(A)$ then $\Sigma$ is minimum.\\
\\
(e) If $\Sigma$ is optimum then $\Sigma$ is minimum. Furthermore for each of the implications $A_P \ra B_P$\\
\hspace*{.5cm} defined in (b) the cardinality of $A_P$ is uniquely determined by $P$ as\\
\hspace*{.5cm} $\min \{|X|: X \subseteq P, \quad c(X) = c(P)\}.$\\ \\ \hline 
\end{tabular}

Because of (b) the Guigues-Duquenne base is often called {\it canonical}\footnote{Some authors as [GW] speak of the {\it stem base} but for us ``stem'' has another meaning (see 3.3).}. 
Those families $\Sigma$ of implications that are of type $\Sigma = \Sigma_{GD}$ for some closure operator $c$ were {\it inherently} characterized by Caspard [C]. The whole of Theorem 1 can be raised to the level of semilattice congruencies\footnote{For a glimpse on semilattice congruences in another but related context see 4.2.1.} [D2] but this further abstraction hasn't flourished yet. For practical purposes any minimum base $\Sigma$ is as good as $\Sigma_{GD}$.  For instance, a trivial way to shorten $\Sigma_{GD}$ to $\Sigma'_{GD}$ is to replace each $P \ra c(P)$ in $\Sigma_{GD}$ by $P \ra (c(P) \backslash P)$. The extra benefit of $\Sigma_{GD}$ is its beauty on a theoretical level as testified by Theorem 1.

{\bf 3.2.1} To illustrate Theorem 1 we consider $c: = c_{\cal F}$ where ${\cal F}$ is the closure system from 3.1.1. Hence the canonical base of $c$ is

$\Sigma_{GD} = \{\emptyset \ra 12, \ \ 123 \ra 1234, \ \ 124 \ra 1234, \ \ 126 \ra [7], \ \ 127 \ra [7], \ \ 12345 \ra [7]\}$.

It happens that all premises (apart from $12345$ which has $35$ and $45$) contain {\it unique} minimal generating sets of the conclusions, and so by Theorem 1(e) each optimum base of $c$ must be of type

$\Sigma_{op} = \{\emptyset \ra B_1, \ \ 3 \ra B_2, \ \ 4 \ra B_3, \ \ 6 \ra B_4, \ \ 7 \ra B_5, \ \ 35 \ra B_6  \ \ (\mbox{or} \ 45 \ra B_6)\}$.

It turns out that e.g.

$\Sigma_1 = \{\emptyset \ra 12,\ \ 3 \ra 4,  \ \ 4 \ra 3, \ \ 6 \ra 357, \ \ 7 \ra 6, \ 35 \ra 6\}$

is optimum. To prove it one must (a) show that $\Sigma_1$ is a base at all, and (b) show that the sum $2+1+1+3+1+1=9$ of the sizes of the conclusions is minimum. We omit the argument. See also Problem 4 in Expansion 15.

{\bf 3.2.2} In this section and (only here) $[n]$ denotes the strong component of $n$, i.e. not $\{1, 2, \cdots, n\}$. As a less random application of Theorem 1 consider the case where $c$ admits a base $\Sigma$ of {\it singleton} premise implications\footnote{We disallow $\emptyset$ as premise in order to avoid distracting trivial cases. Further we point to 4.1.2 for the connection to lattice distributivity.}.  Such a situation can be captured by a directed graph. For instance

(15) \quad $\Sigma : = \{ 1 \ra 6, \ 2 \ra 56, \ 3 \ra 2, \ 4 \ra 3689, \ 5 \ra 347, \ 6 \ra 9, \ 7 \ra 8, \ 8 \ra 7 \}$

matches the arcs in the directed graph $G(\Sigma)$ in Figure 5(a). What, then, do $\Sigma_{GD}$ and the optimal bases $\Sigma_0$ look like? Being singletons, and because of $c(\emptyset)  =\emptyset$, all premises of implications in $\Sigma$ are pseudoclosed (note $\{9\}$ is closed), and so Theorem 1(b) implies that these are {\it all} pseudoclosed sets of $c$. From this and Figure 5(a) it follows that
$$\begin{array}{lll} \Sigma_{GD} & = & \{1 \ra 169,  \ 2 \ra 23456789, \ 3 \ra 23456789, \ 4 \ra 23456789, \\
\\
& & \ \ \ 5 \ra 23456789, \ 6 \ra 69, \ 7 \ra 78, \ 8 \ra 78 \} \end{array}$$

\begin{center}
\includegraphics[scale=0.6]{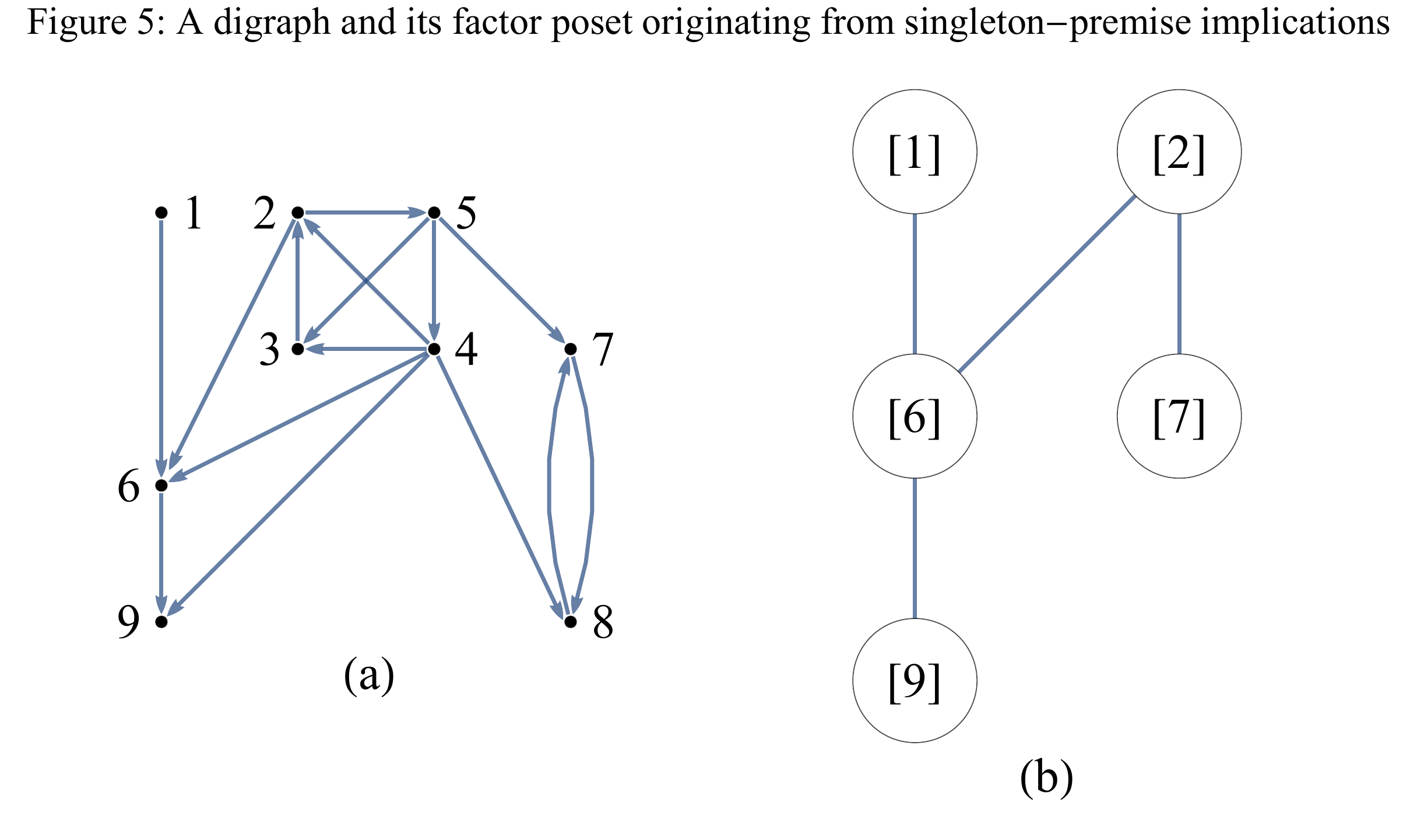}
\end{center}

The strong components of $G(\Sigma)$ are $\{1\}, \{2,3,4, 5\}, \{6\}$, $\{7,8\}, \{9\}$ and the resulting factor poset $(P, \leq)$ is depicted in Figure 5(b). We claim that the optimal bases $\Sigma_0$ look like this: The elements in each strong component $K$ are set up, in arbitrary circle formation such as $2 \ra 5 \ra 4 \ra 3 \ra 2$ for $K = [2]$. (For $|K| =1$ the circle formation reduces to a point.) Furthermore, for any non-minimal $K$ choose any minimal transversal $T$ of the lower covers of $K$ in $(P, \leq)$ and distribute $T$ to the circle formation of $K$ in arbitrary fashion. Thus $K = [2]$ admits $T_1 = \{6,7\}$ and $T_2 = \{6,8\}$. Choosing $T_1$ one can e.g. pad up $\{2 \ra 5, 4 \ra 3\}$ to $\{2 \ra 56, 4 \ra 37\}$ or alternatively $\{5 \ra 4\}$ to $\{5 \ra 467\}$. Choosing $T_2$ one can e.g. pad up $\{2 \ra 5, 3 \ra 2\}$ to $\{2 \ra 56, 3 \ra 28\}$. The latter choice yields an optimum base
$$\Sigma_0 = \{2 \ra 56, \ 5 \ra 4, \ 4 \ra 3, \ 3 \ra 28; \ 1 \ra 6; \ 6 \ra 9; \ 7 \ra 8, \ 8 \ra 7\}.$$
To prove the claim, first note that families of type $\Sigma_0$ obviously {\it are} implicational bases. We next show that {\it each} family $\Sigma'$ equivalent to $\Sigma$ in (15) must contain implications that link [2] to both lower covers [6] and [7]. Indeed, suppose each $\{\alpha \} \ra B$ in $\Sigma'$ with $\alpha \in [2]$ has $B \cap [6] = \emptyset$. Then we get the contradiction that $[2] \cup [7] \cup [9]$ is $\Sigma'$-closed but not $\Sigma$-closed. From this it readily follows that the bases of type $\Sigma_0$ have minimum size $s(\Sigma_0)$. This kind of argument carries over to the optimization of all families $\Sigma$ with merely singleton premises.

Calculating $\Sigma_{GD}$ depends in which way $c$ is given. The two most prominent cases are $c = c_{\cal H}$ and $c = c(\Sigma, -)$.
The first is hard (3.6.3), the second easy (Expansion 11).

\subsection{The canonical direct implicational base}

An implicational base $\Sigma$ of $c$ is {\it direct} if $c(\Sigma, X) = X'$ for all $X \subseteq E$ (see (6)).
Analogous to Theorem 1 each closure operator again admits a {\it canonical} direct implicational base $\Sigma_{cd}$. In order to state this in Theorem 2 we need a few definitions.
Let $U \cup \{e\} \subseteq E$ with $e \not\in U$. Following [KN] we call $U$ a {\it stem for} $e$, and $e$ a {\it root for} $U$, if $U$ is minimal with the property that $e \in c(U)$. (Other names have been used by other authors.) Further $U \subseteq E$ is a {\it stem} if it is a stem for some $e$, and $e \in E$ is a {\it root} if it is a root for some $U$. If $U$ is a stem, we put

(16) \quad roots$(U) : = \{ e \in E: \ e \ \mbox{is a root for} \ U\}$,

For instance, if $c(\emptyset) \neq \emptyset$ then roots$(\emptyset) = c(\emptyset)$. Dually, if $e$ is root, we put

(17) \quad stems$(e): = \{U \subseteq E: \ U \ \mbox{is a stem for} \ e \}$.

Note that $e \in E$ is {\it not} a root iff $E \backslash \{e\}$ is closed. Vice versa, a subset $S$ does {\it not} contain a stem iff all subsets of $S$ (including $S$ itself) are closed. Such sets $S$ are called\footnote{An equivalent definition occurs in 3.3.1. Note that in [W3] the meaning of ``free''  is ``independent''.} {\it free}.

\begin{tabular}{|l|} \hline \\
{\bf Theorem 2:} Let $c : {\cal P}(E) \ra {\cal P}(E)$ be a closure operator. Then\\
\\
\hspace*{3cm} $\Sigma_{cd}: = \{X \ra \ \mbox{roots}(X) : \ X \subseteq E \ \mbox{is a stem} \}$\\
\\
is a direct implicational base of $c$ of minimum cardinality.\\ \\ \hline \end{tabular}

{\it Proof.} Let $Y \subseteq E$. We first show that $Y' = c(Y)$. We may assume that $c(Y) \neq Y$ and pick any $e \in c(Y) \backslash Y$. Obviously there is $X \in \, \mbox{stems}(e)$ with $X \subseteq Y$. From $(X \ra \, \mbox{roots}(X)) \in \Sigma_{cd}$ it follows that $e \in Y'$. Thus $\Sigma_{cd}$ is a direct implicational base of $c$.

To show that $|\Sigma| \geq |\Sigma_{cd}|$  for any direct base $\Sigma$ of $c$ we fix any stem $X$ (say with root $e$).  It suffices to show that at least one implication in $\Sigma$ has the premise $X$. Consider the $\Sigma$-closure
$$c(X) = c(X, \Sigma)=X' = X \cup \{B_i : \ (A_i \ra B_i) \in \Sigma, \ A_i \subseteq X\}.$$
Suppose we had $A_i \neq X$ for all premises $A_i$ occuring in $\Sigma$. Then each $A_i$ contained in $X$ is a {\it proper} subset of $X$, and so the minimality of $X$ forces $e \not\in c(A_i)$, whence $e \not\in B_i \subseteq c(A_i)$, whence $e \not\in X'$. The contradiction $e \not\in c(X)$ shows that at least one $A_i$ equals $X$.  \quad $\square$

We stress that ``minimum'' in Theorem 2 concerns only the {\it directness} of $\Sigma_{cd}$; as will be seen, small subsets of $\Sigma_{cd}$ can remain (non-direct but otherwise appealing) bases of $c$.  The base $\Sigma_{cd}$, has been rediscovered in various guises by various authors; see [BM] for a survey. We may add that in the context of FCA and the terminology of ``proper premises'' $\Sigma_{cd}$ seemingly was first introduced in [DHO]. In the relational database world $\Sigma_{cd}$ is called a ``canonical cover'' [M, 5.4] and (according to D. Maier) first appeared in Paredens [P].
We shall relate $\Sigma_{cd}$ to prime implicates of pure Horn functions in 3.4, and to $M({\cal F})$ in 3.6, and we consider {\it ordered} direct bases in 4.3. Other aspects related to $\Sigma_{cd}$ are discussed in Expansions 5 and 6. Furthermore, the following concept will be more closely investigated in the framework of 4.1.5. We define it here because it is of wider interest. Namely, a stem $X$ is {\it closure-minimal} with respect to its root $e$ if $c(X)$ is a minimal member of $\{c(U): U \in stems(e)\}$.

{\bf 3.3.1} If $c: {\cal P}(E) \ra {\cal P}(E)$ is a closure operator then $X \subseteq E$ is called {\it independent} if  $x \not\in c(X \backslash \{x\})$ for all $x \in X$. A closed independent set is {\it free}. Further, a minimal generating set $X$ of $S \in {\cal F}_c$ is a {\it minimal key for} $S$, or simply a {\it minimal key} (if $S$ is irrelevant). Recall that a {\it set ideal} is a set system ${\cal S} \subseteq {\cal P}(E)$ such that $Y \in S$ and $X \subseteq Y$ jointly imply $X \in {\cal S}$. The maximal members of ${\cal S}$ are its {\it facets}. The following facts  are easy to prove:
\begin{enumerate}
	\item [(a)] A subset is independent iff it is a minimal key.
	\item[(b)] The family Indep$(c)$ of all independent (e.g. free) sets is a set ideal.
	\item[(c)] Each stem is independent  but not conversely.
\end{enumerate}
Since each $S \in {\cal F}_c$ contains at least one minimal key for $S$, it follows that $|{\cal F}_c| \leq |\mbox{Indep}(c)|$.
 Instead of ``minimal key'' other names such as ``minimal generator'' are often used, and ``minimal key'' sometimes means ``minimal key of $E$''. Generating all minimal keys has many applications and many algorithms have been proposed for the task. See [PKID1, Section 5.1.1] for a survey focusing on FCA applications.

{\bf 3.3.2} Let us indicate an apparently new method to get all minimal keys; details will appear elsewhere. The facets $S_1, S_2, \cdots S_t$ of Indep$(c)$ can be calculated with the {\it Dualize and Advance} algorithm (google that). It is then clear that the minimal keys of any closed set $X \in {\cal F}_c$ are {\it among} the (often few) maximal members of $\{S_1 \cap X, \cdots, S_t \cap X\}$. For special types of closure operators more can be said (see 4.1.4 and 4.1.5).

\subsection{Pure Horn functions, prime implicates, and various concepts of minimization}

We recall some facts about Boolean functions with which we assume a basic familiarity; e.g. consult [CH] as reference. Having dealt with the consensus method and prime implicates on a general level in 3.4.1, we zoom in to pure Horn functions in 3.4.2 and link them to implications. (Impure Horn functions appear in 4.5.) In 3.4.3 we show that the canonical direct base $\Sigma_{cd}$ in effect is the same as the set of all prime implicates. Subsection 3.4.4 is devoted to various ways of measuring the ``size'' of an implicational base, respectively pure Horn function.

{\bf 3.4.1} Recall that a function $f: \{0,1\}^n \ra \{0,1\}$ is called a {\it Boolean function}. A {\it bitstring} $a \in \{0,1\}^n$ is called a {\it model} of $f$ if $f(a) =1$. We write Mod$(f)$ for the set of all models of $f$. For instance, $f$ is a {\it negative} (or {\it antimonotone}) Boolean function if $x \leq y$ implies $f(x) \geq f(y)$.  
Thus, if we identify $\{0,1\}^n$ with the powerset ${\cal P}[n]: = {\cal P}([n])$ as we henceforth silently do, then Mod$(f)$ is a set ideal in ${\cal P}[n]$ iff $f$ is a negative Boolean function. Using {\it Boolean variables} $x_1, \cdots, x_n$ one can represent each Boolean function $f$ (in many ways) by a {\it Boolean formula} $F(x) = F(x_1, \cdots, x_n)$. We then say that $F$ {\it induces} $f$. A {\it literal} is either a Boolean variable or its negation; thus $x_2$ and $\ol{x}_5$ are literals. A {\it clause} is a disjunction of literals, such as $x_1 \vee \ol{x}_3 \vee \ol{x}_4 \vee x_7$. A {\it conjunctive normal form} (CNF) is a conjunction of clauses.  The CNF is {\it irredundant} if dropping any clause changes the represented Boolean function. Let $f$ be a Boolean function and let $C$ be a clause. Then $C$ is an {\it implicate} of $f$ if every model of $f$ is a model of $C$. We emphasize that ``implicate'' should not be confused with ``implication'' $A \ra B$, but there are connections as we shall see. One calls $C$ a {\it prime implicate} if dropping any literal from $C$ results in a clause which is no longer an implicate of $f$. In Expansion 7 we show how {\it all} prime implicates of $f$ can be generated from an arbitrary CNF of $f$. A {\it prime} CNF is a CNF all of whose clauses are prime implicates.

{\bf 3.4.2} A Boolean function $f: {\cal P}[n] \ra \{0,1\}$ is a {\it pure Horn function} if Mod$(f) \subseteq {\cal P}[n]$ is a closure system\footnote{Some authors, e.g. [CH, chapter 6], use a different but dual definition, i.e. that $\{a \in \{0,1\}^n: f(a) =0 \}$ must be a closure system. Each theorem in one framework immediately translates to the dual one. Do not confuse this kind of duality with the kind of duality in [CH, 6.8].}. 
The induced closure operator ${\cal P}[n] \ra {\cal P}[n]$ we shall denote by $c_f$. Conversely, each closure operator $c: {\cal P}[n] \ra {\cal P}[n]$ induces the pure Horn function $f_c: {\cal P}[n] \ra \{0,1\}$ defined by $f_c^{-1} (1) = {\cal F}_c$. Similar to 2.1.1 one has $f_{(c_f)} =f$ and $c_{(f_c)} = c$. As mentioned in 3.4.1 many distinct formulas $F$ induce any given\footnote{For instance, using concatenation instead of $\wedge$, {\it one} formula $F$ for the Horn function $f$ induced by the closure system in Figure 4(a) is  $F(x_1, \cdots, x_7) = x_1 x_2 x_3 x_4 x_5 x_6 x_6 x_7 \vee x_1 x_2 x_3 x_4 \ol{x}_5 \ol{x}_6 \ol{x}_7 \vee x_1 x_2 \ol{x}_3 \ol{x}_4 x_5 \ol{x}_6 \ol{x}_7 \vee x_1 x_2 \ol{x}_3 \ol{x}_4 \ol{x}_5 \ol{x}_6 \ol{x}_7$.} pure Horn function $f$. 
As is common, we shall focus on the most ``handy'' kind of formula $F$, for which the letter $H$ will be reserved. 

In order to define $H$ we first define a {\it pure} (or {\it definite}) {\it Horn clause} as a clause with exactly one positive literal.
 Thus $\ol{x}_1 \vee \ol{x}_2 \vee \ol{x}_3 \vee x_4$ is a pure Horn clause $C$. Accordingly consider the implication $\{1,2,3\} \ra \{4\}$. One checks that the Boolean function induced by formula $C$ is a Horn function $f: {\cal P}[n] \ra \{0,1\}$ (for any fixed $n \geq 4$). In fact $\mbox{Mod}(f)  = {\cal F} (\{123 \ra 4\})$. However, this doesn't extrapolate to the implication $12 \ra 34$ which doesn't match $\ol{x}_1 \vee \ol{x}_2 \vee x_3 \vee x_4$! Rather  $\{12 \ra 34\}$ is equivalent to $\{12 \ra 3, \ 12 \ra 4\}$ and whence\footnote{This is a good place to address a source of confusion. The formula $x_1 \wedge x_2$ {\it also} is the conjunction of two pure Horn clauses; it matches the implication $\emptyset \ra \{1,2\}$. The formula $x_1 \wedge x_2 \ra {\tt True}$ is a {\it tautology} which matches the implication $\{1,2\} \ra \emptyset$. But $x_1 \wedge x_2 \ra {\tt False}$ matches {\it no implication}. Rather it amounts to the {\it impure} Horn clause $\ol{x}_1 \vee \ol{x}_2$, the topic of Section 4.5.} matches the conjunction $(\ol{x}_1 \vee \ol{x}_2 \vee x_3) \wedge (\ol{x}_1 \vee \ol{x}_2 \vee x_4)$ of {\it two} pure Horn clauses. Generally, a {\it pure Horn CNF} $H$ is defined as a conjunction of  pure Horn clauses. Thus $H$ matches a family $\Sigma_H$ of unit implications. In particular, this shows that the Boolean function $f$ induced by $H$ really {\it is} a pure Horn function: $\mbox{Mod}(f)$ equals ${\cal F}(\Sigma_H)$, which we know to be  closure system (2.2). Conversely, starting with any family $\Sigma$ of implications, the {\it unit expansion} $\Sigma^u$ is obtained by replacing each $(A \ra B) \in \Sigma$ by the unit implications $A \ra \{b\} \ (b \in B)$. By definition $H_\Sigma$ is the pure Horn CNF whose clauses match the members of $\Sigma^u$. Notice that special features of $\Sigma$ need not be mirrored in $H_\Sigma$, and vice versa for $H$ and $\Sigma_H$.
 For instance, if $\Sigma$ is optimum then the pure Horn clauses in $H_\Sigma$ need not  be prime. See also 3.4.4.1.

{\bf 3.4.3} It is evident from the definitions of stem, root and prime implicate, and from Theorem 2, that each implication in $(\Sigma_{cd})^u$ yields a prime implicate of the pure Horn function $f : {\cal P}[n] \ra \{0,1\}$ determined by $\Sigma_{cd}$. Do we get {\it all} prime implicates (Horn or not) of $f$ in this way? Yes. The traditional proof is e.g. in [CH, p.271], and a fresh one goes like this.
Suppose  $f$ had a prime implicate $C$ which is not a Horn clause, say without loss of generality $C$ is $\ol{x}_1 \vee \ol{x}_2 \vee  x_3 \vee x_4$. Then both $\ol{x}_1 \vee \ol{x}_2 \vee x_3$ and $\ol{x}_1 \vee \ol{x}_2 \vee x_4$ are no implicates of $f$. Hence there are $S, T \in \, \mbox{Mod}(f)$ such that $\{1, 2\} \subseteq S$ but $3 \not\in S$, and such that $\{1,2\} \subseteq T$ but $4 \not\in T$. Thus $\{1,2\} \subseteq S \cap T \in \, \mbox{Mod}(f)$ but both $3,4 \not\in S \cap T$. Hence $S \cap T$ is a model of $f$ but not of $C$, contradicting the assumption that $C$ is an implicate of $f$. \quad $\square$

Thus the members of $\Sigma_{cd}^u$ are in bijection with the prime implicates of $f$. Any (usually non-direct) base of implications $\Sigma \subseteq \Sigma_{cd}^u$ will henceforth be called a {\it base of prime implicates}. In other words, bases of prime implicates match prime pure Horn CNF's.

{\bf 3.4.4} We now drop pure Horn functions until 3.4.4.1. Apart from $ca(\Sigma )$ and $s(\Sigma)$ introduced in 2.2 there are other ways to measure families of implications. If say

(18) \qquad $\Sigma = \{\, \{a,b\} \ra \{c,d\},\quad \{a, c, e\} \ra \{b\},\quad \{d\} \ra \{b,f\}\,\}$

then $ca(\Sigma) = 3$ and $s(\Sigma) = 11$. Further the {\it left hand size} is defined as the sum of the cardinalities of the premises, thus $lhs (\Sigma) : = 2+3+1 = 6$. Similarly the {\it right hand size} is $rhs(\Sigma) : = 2+1+2=5$. 
What are the relations between ``usual'' optimality ($op$ as defined in 2.2) and the new kinds of optimality lhs-op and rhs-op? Suppose first $\Sigma_0$ is simultaneously lhs-op and rhs-op. If $\Sigma$ is any other base of ${\cal F}(\Sigma_0)$ then
$$s(\Sigma_0) = lhs(\Sigma_0 ) + rhs(\Sigma_0) \leq lhs (\Sigma) + rhs (\Sigma) = s(\Sigma),$$
and so $\Sigma_0$ is optimal. This was observed in [AN1] and likely elsewhere before. Conversely, it follows at once from Theorem 1(e) that op $\Ra$ lhs-op. In [ADS] it is shown (see Figure 6) that also op $\Ra$ rhs-op.  For instance, it is impossible that a closure operator has two optimum bases with implications $\ast \ast \ra \ast \ast \ast, \ \ \ast \ast \ra \ast \ast$  and 
$\ast \ast \ast \ra \ast, \ \ast \ast \ast \ra \ast \ast$ respectively. To summarize:

(19) \qquad op \ $\Leftrightarrow$ \ lhs-op and rhs-op

A slightly less natural parameter is (ca$+$rhs)$(\Sigma) : = |\Sigma| + rhs(\Sigma)$. According to [ADS] these implications (and their consequences, but no others) take place:

\begin{center}
\includegraphics{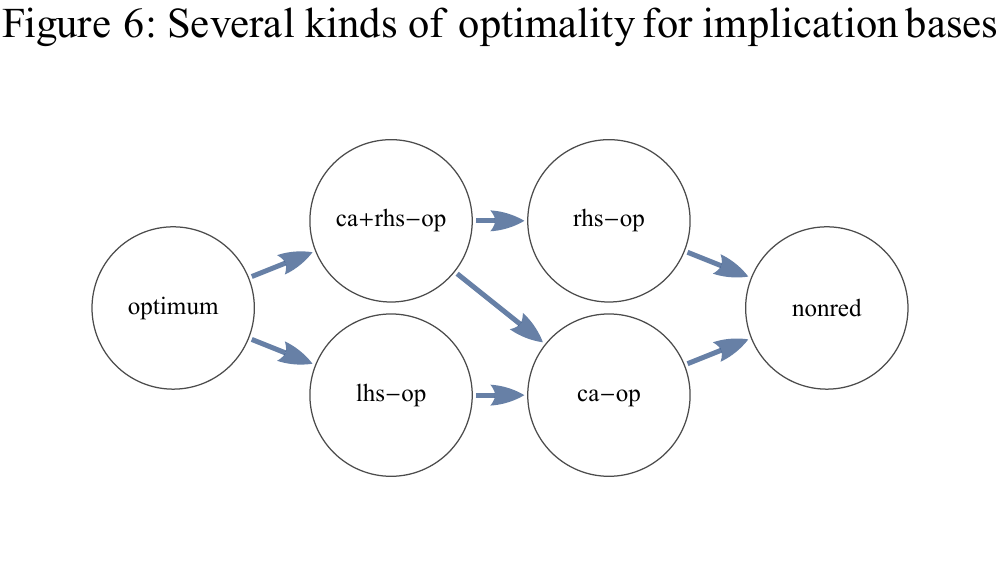}
\end{center}

{\bf 3.4.4.1} Let us stick with the measures above and re-enter pure Horn functions to the picture. For starters, when $\Sigma$ in (18) is translated in a pure Horn CNF we get

$(18')$ \qquad $H_\Sigma \ \ = \ \  (\ol{a} \vee \ol{b} \vee c) \quad \wedge \quad (\ol{a} \vee \ol{b} \vee d) \quad \wedge \quad (\ol{a} \vee \ol{c} \vee \ol{e} \vee b) \quad \wedge \quad (\ol{d} \vee b) \quad \wedge \quad (\ol{d} \wedge f)$

Notice that $rhs (\Sigma) = 5$ and $5$ is the number of clauses of $H_\Sigma$. Generally, for a fixed pure Horn function $f: {\cal P}[n] \ra \{0,1\}$ put
$$rhs(f) : = \min \{rhs(\Sigma) : \ \Sigma \ \mbox{is a base of} \ \mbox{Mod}(f)\}.$$
Thus $rhs(f)$ is the minimum number\footnote{Many other acronyms for this measure are dispersed throughout the literature. For instance, [CH, p.297] uses $\tau (f)$ for $rhs(f)$. On the side of uniformity, our notation $\lambda$ above matches the one in [CH, p.297].}
of pure Horn clauses needed to represent $f$. Rephrasing the [ADS] result above (which is reproven in [AN1, Thm.10]) one can say: If $\Sigma$ is any optimum base of $c$ then $H_\Sigma$ has rhs$(f_c)$ many clauses. The ``inverse'' operation of unit expansion is {\it aggregation}. Thus if $\Sigma = \{12, \ra 3, 12 \ra 4, 35 \ra 4, 35 \ra 1, 45 \ra 2\}$ then $\Sigma^{ag}: = \{12 \ra 34, 35 \ra 14, 45 \ra 2\}$. 

If similarly to $rhs(f)$ we define
$$ca(f) : = \min \{ca (\Sigma ) : \Sigma \ \mbox{is a base of Mod}(f)\},$$
then $ca(f)$ is not so succinctly expressed in terms of Horn clauses (but is e.g. useful in 4.5.2). Similarly the likewise defined parameters $lhs(f)$ and $s(f)$ are clumsier than their counterparts $lhs(\Sigma)$ and $s(\Sigma)$. Apart from   $rhs(f)$, the most natural measure for pure Horn functions is the minimum number $\lambda(f)$ of literals appearing in any pure Horn CNF representation of $f$. One calls $\lambda$ the {\it number of literals} measure. Clearly $\lambda (f) \geq s(f)$. For instance, if $H_\Sigma$ from $(18')$ induces $f$, then $\lambda (f) \leq 14$. Similarly $s(f) \leq 11$ in view of (18).
Both rhs-optimization and $\lambda$-optimization are NP-hard, and even {\it approximation} remains hard [BG].

\subsection{Acyclic closure operators and generalizations}

To any family $\Sigma$ of implications on a set $E$ we can associate its {\it implication-graph}\footnote{The terminology is from [BCKK], while $G(\Sigma)$ itself was independently introduced in [W3, p.137] and [HK, p.755].} $G(\Sigma)$. It has vertex set $E$ and arcs $a \ra b$ whenever there is an implication $A \ra B$ in $\Sigma$ with $a \in A$ and $b \in B$. What happens when $\Sigma$ merely has singleton-premise implications was dealt with in 3.2.2. Another natural question is: If $G(\Sigma)$ is acyclic, i.e. has no directed cycles, what does this entail for the closure operator  $X \mapsto c(\Sigma, X)$?
The first problem is that for equivalent families $\Sigma$ and $\Sigma'$ it may occur that $G(\Sigma)$ is acyclic but $G(\Sigma')$ isn't. For instance, in the example from [HK, p.755] one checks that $\Sigma = \{1 \ra 2, 2 \ra 3\}$ and $\Sigma'= \{1 \ra 3, 2 \ra 3, 13 \ra 2 \}$ are equivalent. While $G(\Sigma)$ is acyclic, $G(\Sigma')$ is not because it has the cycle $2 \ra 3 \ra 2$. Observe that $13 \ra 2$ is no prime implicate because
it follows from $1 \ra 2$. 

Indeed, the problem evaporates if one restricts attention to the prime implicates. More precisely, call\footnote{In [HK] the authors talk about the acyclicity of pure Horn formulas (or functions). Recall from  3.4.2 the equivalence between closure operators and pure Horn functions.} a closure operator $c$ {\it acyclic} if there is a base $\Sigma$ of $c$ which has an acyclic implication-graph $G(\Sigma)$. As shown in [HK, Cor.V.3] a  closure operator $c$ is acyclic iff $G(\Sigma)$ is acyclic for each base $\Sigma$ of prime implicates. Hence (consensus method, Expansion 7) for an arbitrary family $\Sigma$ of implications it can be checked in quadratic time whether $c(\Sigma, -)$ is an acyclic closure operator.

{\bf 3.5.1} Let $(E, \leq)$ be any poset and let $c: {\cal P}(E) \ra {\cal P}(E)$ be a closure operator with $c(\emptyset) = \emptyset$ and such that for all $Z \subseteq E$ and $y \in c(Z)$ it follows that $y \in c(\{z \in Z: \ z \geq y \})$. Put another way, $c(Z)$ is always a {\it subset} of the order ideal $Z \downarrow$ generated by $Z$. Following\footnote{This terminology is more telling than ``$G$-geometry'' used in [W3].} [SW] we call such an operator of {\it poset type}.

\begin{tabular}{|l|} \hline \\
{\bf Theorem 3}: A closure operator $c$ is acyclic if and only if it is of poset type.\\ \\ \hline \end{tabular}

{\it Proof.} We shall trim the argument of [W3, Cor.15]. So let $c: {\cal P}(E) \ra {\cal P}(E)$ be acyclic and let $\Sigma$ be any base of $c$ for which $G(\Sigma)$ is acyclic. On $E$ we define a transitive binary relation $>$ by setting $b > a$ iff there is a directed path from $b$ to $a$ in $G(\Sigma)$. By the acyclicity of $G(\Sigma)$ this yields a poset $(E,\leq)$. Consider $Z \subseteq E$ and $y \in E$ such that $y \in c(Z)$. Then $c(Z) = c(\Sigma, Z)$ because $\Sigma$ is a base of $c$. To fix ideas suppose $c(\Sigma, Z) = Z''$ where $Z'$ is as defined in (6), and say that $Z' = Z \cup \{3,4\}$ because $(\{1,2\} \ra \{3,4\}) \in \Sigma$ and $\{1,2\} \subseteq Z$. Further let $Z'' = Z'\cup \{6,y\}$ in view of $(\{3,5\} \ra \{6,y\}) \in \Sigma$ and $3,5 \in Z'$. Then $1, 2, 5 \in Z$ and all of them are $> y$ because $G(\Sigma)$ has directed paths $1 \ra 3 \ra y$ and $2 \ra 3 \ra y$ and $5 \ra y$. Hence $y \in c(\Sigma, \{1,2,5\}) \subseteq c(\{z\in Z: z \geq y\})$. Thus $c$ is of poset type.

Conversely let $c$ be of poset type with underlying poset $(E, \leq)$. Let $\Sigma$ be a base of $c$ whose unit expansion yields a {\it prime} Horn CNF. It suffices to show that $G(\Sigma)$ is acyclic. Suppose to the contrary $G(\Sigma)$ contains a directed cycle, say $1 \ra 2 \ra 3 \ra 4 \ra 1$. By definition of $G(\Sigma)$ there is $(A \ra B) \in \Sigma$ with $4 \in A$ and $1 \in B$, and so $1 \in c(A)$. By assumption $1 \in c(A_0)$ where $A_0 : = \{z \in A:  z \geq 1\}$. If we had $4 \not\in A_0$ then $A_0 \ra \{1\}$ would be an implicate of $\Sigma$, which cannot be since $A \ra \{1\}$ is a prime implicate. It follows that $4 \in A_0$, whence $4 > 1$. By the same token one argues that $3 > 4$, and eventually $1 > 2 > 3 > 4 > 1$, which is the desired contradiction. 
\hfill
\hfill $\square$

According to [HK, p.756] each acyclic closure operator $c$ admits a unique nonredundant base $\Sigma_{acyc}$ of prime implicates. Consequently (why?) $\Sigma_{acyc}$ is rhs-optimal and $\lambda$-optimal. Starting out with any family $\Sigma$ of unit implications for which $G(\Sigma)$ is acylic (and whence $c : = c(\Sigma, -)$ is acyclic), it is easy to calculate $\Sigma_{acyc}$.
To fix ideas, one checks that
$$\Sigma:= \{4 \ra 5, \ \ 6 \ra 1, \ \  23 \ra 4, \ \ 23 \ra 1, \ \ 35 \ra 6, \ \ 34 \ra 6, \ \ 234 \ra 5\}$$
has $G(\Sigma)$ acyclic. Any $A \ra \{b\}$ in $\Sigma$ which is not a prime implicate, can only fail to be one because some $A_0 \varsubsetneqq A$ satisfies $b \in c(\Sigma \backslash \{A \ra \{b\}\}, A_0)$, and so $A \ra \{b\}$ is redundant. Here only $234 \ra 5$ isn't a prime implicate (take $A_0 = \{2,3\}$). But also prime implicates in $\Sigma$ may be redundant. In our case $34 \ra 6$ is a consequence of $4 \ra 5$ and $35 \ra 6$. One checks that $\Sigma \backslash \{234 \ra 5, 34 \ra 6\}$ consists of prime implicates and is nonredundant. Hence it must be $\Sigma_{acyc}$. Obviously $\Sigma_{acyc}$ is not minimum among {\it all} bases of $c$ since $23 \ra 1$ and $23 \ra 4$ can be aggregated to $23 \ra 14$.

{\bf 3.5.2} As to generalizations, two variables $x$ and $y$ of a Boolean formula $F = F(u_1, \cdots, u_n)$ are {\it logically equivalent} if they have the same truth value in every model of (the function induced by) $F$. This amounts to say that both $x\ra y$ and $y \ra x$ are (prime) implicates of $f$. A closure operator $c$ is {\it quasi-acyclic} if there is a base $\Sigma$ of prime implicates such that all elements within a strong component of $G(\Sigma)$ are logically equivalent. Each acyclic closure operator is quasi-acylic because all components of $G(\Sigma)$ are singletons. Also the kind of closure operators $c = c(\Sigma, -)$ considered in 3.2.2 are evidently quasi-acyclic.

A closure operator $c$ is {\it component-wise quadratic} $(CQ)$ if there is a base $\Sigma$ of prime implicates such that $G(\Sigma)$ has the following property. For each prime implicate $A \ra \{y\}$ of $c$ and each strong component $K$ of $G(\Sigma)$ it follows from $y \in K$ that $|A \cap K| \leq 1$. Thus for each component $K$ of $G(\Sigma)$ the ``traces'' of the prime implicates on $K$ are ``quadratic'' in the sense of having cardinality $\leq 2$. Here comes the argument of why quasi-acyclic entails $CQ$. Suppose $A \ra \{y\}$ is a prime implicate of $c$ such that $y \in K$ and $A \cap K \neq \emptyset$. Take $x \in A \cap K$. Because $\{x\} \ra \{y\}$ is an implicate of $c$ by quasi-acyclicity, we must have $A = \{x\}$ (which implies $|A\cap K|=1$).  In a tour de force it is shown in [BCKK] that for each $CQ$ closure operator an rhs-optimum base (i.e. minimizing the number of clauses) can be calculated in polynomial time; many auxiliary graphs beyond $G(\Sigma)$ appear in [BCKK]. The quasi-acyclic case had been dealt with in [HK]. Another way to generalize ``acylcic'' is to forbid so-called $D$-cycles, see Expansion 18.

\subsection{Implications and meet-irreducibles}

First some prerequisites about hypergraphs.
A {\it hypergraph} is an ordered pair $(E, {\cal H})$ consisting of a {\it vertex set} $E$ and  a set of {\it hyperedges} ${\cal H}$. The hypergraph is {\it simple} if $X \not\subseteq Y$ for all distinct $X,Y \in {\cal H}$. (An ordinary simple graph is the special case where $|X| = 2$ for all $X \in {\cal H}$). A {\it transversal} of ${\cal H}$ is a set $Y \subseteq E$ such that $Y \cap X \neq \emptyset$ for all $X \in {\cal H}$. We write ${\cal T}r({\cal H})$ for the set of all transversals. Furthermore, the {\it transversal hypergraph} $mtr({\cal H})$ consists of all {\it minimal} members of  ${\cal T}r({\cal H})$. It is easy to see that ${\cal H} \subseteq mtr(mtr({\cal H}))$. Arguably the single most important fact about general simple hypergraphs is [S, p.1377] that equality takes place:

(20) \quad $mtr(mtr({\cal H})) = {\cal H}$

The {\it transversal hypergraph problem} (or {\it hypergraph dualization}), i.e. the problem of calculating $mtr({\cal H})$ from ${\cal H}$ has many applications and has been investigated thoroughly. See [EMG] for a survey and [MU] for a cutting edge implementation of hypergraph dualization.

Let ${\cal F} \subseteq {\cal P}(E)$ be a closure system and let $M({\cal F}) \subseteq {\cal F}$ be its set of meet-irreducibles (see 2.1). 
Clearly the set max$({\cal F})$ of all maximal members of ${\cal F} \backslash \{E\}$ is a subset of $M({\cal F})$. Adopting matroid terminology (4.1.4) we refer to the members of $\max({\cal F})$ as {\it hyperplanes}. More generally, for any $e \in E$ let 
\begin{center}
${\bf \bmax({\cal F},e)}$ be the set of all $Y \in {\cal F}$ that are maximal with the property that $e \not\in Y$. 
\end{center}
If $\bigcap {\cal F} = \emptyset$ (which we assume to avoid trivial cases) then $\max ({\cal F},e) \neq \emptyset$ for all $ e \in E$. In fact each $Y \in \max ({\cal F}, e)$ is meet-irreducible.
Conversely, every $Y \in M({\cal F})$ belongs to some $\max ({\cal F}, e)$. (See Expansion 12.)  
Therefore:

(21) \quad $M({\cal F}) \quad = \quad \bigcup \{\max ({\cal F}, e): \ e \in E\}$.

It is convenient that the sets $\max ({\cal F}, e)$ can be retrieved from any generating set ${\cal H}$ of ${\cal F}$, i.e. not the whole of ${\cal F}$ is required:

(22) \quad $\max ({\cal F}, e) = \max \{Y \in {\cal H}: e \not\in Y\}$.

The proof is given in Expansion 10. 
The smaller ${\cal H}$, the faster we can calculate the simple hypergraphs

(23) \quad ${\bf cmax ({\cal F}, e)} \ \ : = \ \ \{E \backslash X: \ X \in \max({\cal F}, e)\} \quad (e\in E)$.

\begin{center}
\includegraphics{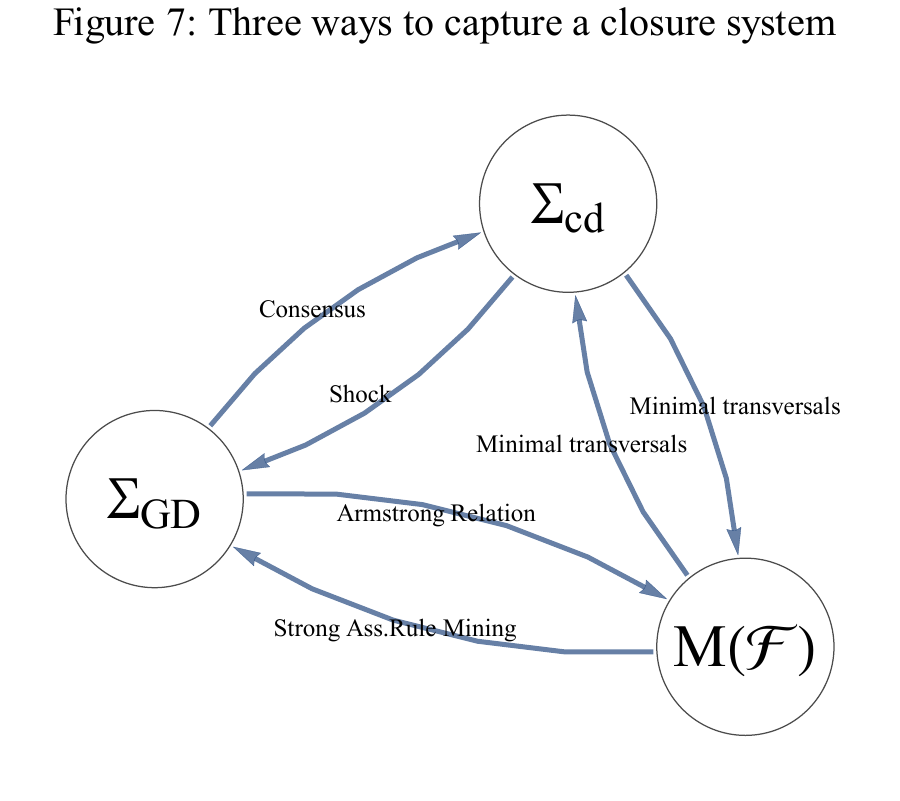}
\end{center}

The next result is crucial for traveling the right hand side of the triangle in Figure 7.

\begin{tabular}{|l|} \hline \\
{\bf Theorem 4:} For any closure system ${\cal F} \subseteq {\cal P}(E)$ with $\bigcap {\cal F} = \emptyset$  one has\\
\\
(a) \ stems$(e) \cup \{e\} = mtr(\mbox{cmax}({\cal F},e)) \quad (e \in E)$\\
\\
(b) \ cmax$({\cal F},e) = mtr(\mbox{stems}(e) \cup \{e\}) \quad (e \in E)$\\
\\
\hline \end{tabular}

{\it Proof.} We draw on [MR2, Lemma 13.3 and Cor.13.1]. We first show that for any fixed $e\in E$ it holds for all $Y \subseteq E$ that:

(24) \quad $e \in c(Y) \ \Leftrightarrow \ Y \in {\cal T}r (\mbox{cmax}({\cal F}, e))$.

{\it Proof of (24)}. Suppose $Y$ is such  that $e \in c(Y) = \cap \{X \in M({\cal F}): X \supseteq Y\}$; see (4). Thus from $X \in M({\cal F})$ and $X \supseteq Y$ follows $e \in X$. For each $X \in \mbox{max}({\cal F}, e) \subseteq M({\cal F})$ (see (21)) we have $e \not\in X$, hence $X \not\supseteq Y$, hence $Y \cap (E \backslash X) \neq \emptyset$, hence $Y \in {\cal T}r(cmax({\cal F}, e))$. Conversely, let $Y$ be such that $e \not\in c (Y)$. Then, because of $c(Y) = \cap \{X \in M(F) : X \supseteq Y\}$, there is $X \in M({\cal F})$ with $e \not\in X \supseteq Y$. We may assume that $X$ is {\it maximal} within $M({\cal F})$ with respect to $e \not\in X$. It then follows from (22) (put ${\cal H} : = M({\cal F})$) that $X \in \max ({\cal F}, e)$. From $X \supseteq Y$ it follows that $Y \cap (E \backslash X) = \emptyset$, and so $Y \not\in {\cal T}r(cmax ({\cal F}, e))$.  This proves (24). 

Let $e \in E$ be fixed. Then the family of minimal $Y$'s satisfying $e \in c(Y)$ is stems$(r) \cup \{e\}$. Likewise the family of minimal $Y$'s satisfying $Y \in {\cal T}r(cmax ({\cal F}, e))$ is $mtr(cmax({\cal F}, e))$. By (24) these two set families coincide, which proves (a).
As to (b), it follows from (a) and (20) that
$mtr(\mbox{stems}(e) \cup \{e\}) = mtr(mtr(\mbox{cmax}({\cal F}, e))) = \mbox{cmax}({\cal F},e)$. \hfill $\square$

As was independently done in [BDVG], let us discuss the six directions in the triangle of Figure 7. Notice that matters don't change much if instead of $M({\cal F})$ we substitute any ``small'' (informal notion) generating set ${\cal H}$ of ${\cal F}$ in Figure 7, and instead of $\Sigma_{GD}$ we sometimes consider any ``small'' (w.r.t. $\Sigma_{GD}$) base $\Sigma$ of ${\cal F}$.
Both practical algorithms illustrated by examples, and theoretic complexity will be discussed. As to going from $\Sigma_{cd}$ to a minimum base $\Sigma$, the most elegant and only slightly sub-optimal method is the one of Shock [Sh]; see Expansion 11. The way from $\Sigma$ to $\Sigma_{cd}$ can be handled by the consensus method (Expansion 7); for another method see [RCEM].  
In Subsections 3.6.1 to 3.6.3 we outline how to travel the remaining four directions, with more details provided in Expansions. 

{\bf 3.6.1} Recall from Theorem 2 that knowing the canonical direct base $\Sigma_{cd}$ means knowing the members of $\bigcup \{\mbox{stems}(e): e \in E\}$. Likewise, by (21) and (23), knowing $M({\cal F})$ amounts to knowing the set collections cmax$({\cal F},e)\, (e \in E)$. Therefore Theorem 4 says that getting $\Sigma_{cd}$ from $M({\cal F})$ or vice versa is as difficult as calculating all minimal transversals of a hypergraph. To fix ideas let us carry out the way from $M({\cal F})$ to $\Sigma_{cd}$ on a toy example. Suppose that $E = [6]$ and ${\cal F}$ is such that

(25) \quad $M({\cal F}) = \{12, 12345, 124, 1245, 13456, 245, 25, 3456, 356\}$.

From (21) and (22) we get

(26) \quad $M({\cal F}) \ \  =  \ \ \max ({\cal F},1) \cup \cdots \cup \max ({\cal F},6)$ 

\hspace*{2.2cm} $= \{245, 3456\} \cup \{13456\} \cup \{1245\} \cup \{25,12, 356\} \cup \{124\} \cup \{12345\}$.

The set union in (26) happens to be disjoint. Generally the union in (21) is disjoint iff $|X^\ast \backslash X| =1$ for all $X \in M({\cal F})$. Here $X^\ast$ is the unique upper cover of $X$ in ${\cal F}$. From say $\max ({\cal F}, 4) = \{25, 12, 356\}$ we get $\mbox{cmax}({\cal F}, 4) = \{1346, 3456, 124\}$, and by Theorem 4(a) we have stems$(4) \cup \{4\} = mtr(\{1346, 3456, 124\})$ which turns out to be $\{4, 13, 16, 23, 26, 15\}$. Dropping $\{4\}$ yields stems$(4)$. Likewise one calculates stems$(1) = \{23, 26\}$, stems$(3) = \{6\}$, stems$(5) = \{3, 6\}$, stems$(2) =$ stems$(6) = \emptyset$. By definition of $\Sigma_{cd}$ in Theorem 2 we conclude that

(27) \quad $\Sigma_{cd} = \{13 \ra 4, \ 16 \ra 4, \ 23 \ra 14, \ 26 \ra 14, \ 15 \ra 4, \  6 \ra 35, \ 3 \ra 5 \}$.

Let us mention a natural enough alternative [W1, Algorithm 3] for $M({\cal F}) \ra \Sigma_{cd}$. 
By processing the members of $M({\cal F})$ one-by-one it updates a corresponding direct base. The worst case complexity being poor, average behaviour still awaits proper evaluation.

{\bf 3.6.2}  How to get $M({\cal F})$ from an {\it arbitrary} (non-direct) implication base $\Sigma$?  One of the first methods was [MR2, Algorithm 13.2], which was improved in [W1, Sec.9]. 
In brief, in view of (21) both methods proceed as follows. For $\Sigma = \{A_1 \ra B_1, A_2 \ra B_2, \cdots, A_n \ra B_n\}$ let $\Sigma_i: = \{A_1 \ra B_1,\cdots, A_i \ra B_i \}$. Then $\max (i, e) : = \max ({\cal F}(\Sigma_i),e)$ can be expressed in terms of the set families $\max (i -1, e)$ and $\max (i-1, a)$ where $a$ ranges over $A_i$. Another idea for $\Sigma \ra M({\cal F})$ in [BMN] features an interesting fixed-parameter-tractability result.
Expansion 8 exhibits a fourth way.

{\bf 3.6.2.1} Unfortunately it is shown in [KKS] that $|M({\cal F})|$ can be exponential with respect to $|\Sigma|$, and vice versa. Furthermore, according to [K] both transitions $\Sigma \ra M({\cal F})$ and $M({\cal F}) \ra \Sigma$ are at least as hard as the transversal hypergraph problem. What's more, whatever the complexity of these transitions, they are equivalent under polynomial reductions. 
Along the way a fifth algorithm [K, p.360-361] to get the {\it characteristic models} (i.e. $M({\cal F})$) from $\Sigma$ is offered. (Some of these results extend to the {\it arbitrary} Horn functions in  4.5.)

{\bf Open Problem 1}: Compare on a common platform and in a careful manner akin to [KuO1], mentioned five methods (and possibly others) for calculating $M({\cal F})$ from $\Sigma$.

{\bf 3.6.2.2} What is the point of calculating $M({\cal F})$ from $\Sigma$? This problem first arose in the vestige of finding an {\it Armstrong Relation} ($=$ short example database) for a given set of functional dependencies. Albeit an Armstrong Relation is not quite the same as $M({\cal F})$, the number of its records is $|M({\cal F})| +1$, see [MR2, Thm.14.4]. Having $M({\cal F})$ enables a ``model-based'' approach to reasoning. For instance, deciding whether $\Sigma \vDash (A \ra B)$ holds, reduces to check whether $A \subseteq X$ entails $B \subseteq X$ for all $X \in M({\cal F})$. This beats the test in (8) when $|M({\cal F})| \ll |\Sigma|$. With the eye on using model-based reasoning in Knowledge Bases article [KR] extends (as good as possible) the concept of characteristic models from Horn functions to arbitrary Boolean functions. Observe that $|M({\cal F})| \ll |\Sigma|$ also occurs in the context of Cayley multiplication tables (4.2.2). Furthermore, many combinatorial problems (e.g. calculating all minimal cutsets of a graph) amount to calculate the subset max$({\cal F}) \subseteq M({\cal F})$ from $\Sigma$.

{\bf 3.6.3} How can one conversely get a small or minimum base $\Sigma$ from $M({\cal F})$ (or from another generating set ${\cal H} \subseteq {\cal F}$)?
 This process is nowadays known as {\it Strong Association Rule Mining} (applications follow in 3.6.3.4). For succinctness, suppose we want $\Sigma = \Sigma_{GD}$. Unfortunately, as shown in [KuO2], not only can $|\Sigma_{GD}|$ be exponential in the input size $|M({\cal F})|\times |E|$, but also calculating the {\it number} $|\Sigma_{GD}| $ is \#$P$-hard.  Despite the exponentiality of $|\Sigma_{GD}|$ one could imagine (in view of (36)) that $\Sigma_{GD}$ can at least be generated in output-polynomial time, given $M({\cal F})$. As shown in [DS], this problem is at least as hard as generating all minimal transversals. Given $M({\cal F})$, the pseudoclosed sets cannot be enumerated in lexicographic order [DS], or reverse lexicographic order [BK], with polynomial delay unless $NP = P$. Several related results are shown in [BK]. For instance, given ${\cal H} \subseteq {\cal P}(E)$ and $A \subseteq E$, it is $coNP$-complete to decide whether any minimum base $\Sigma$ of ${\cal F}({\cal H})$ (see 2.1.1) contains an implication of type $A \ra B$. (Conversely, ${\cal F}$ can also be ``large'' with respect to $\Sigma_{GD}$, see Expansion 4.)
 
{\bf 3.6.3.1}  A different approach to go from ${\cal H}$ to a small base $\Sigma$ of ${\cal F} = {\cal F}({\cal H})$  was hinted at in [W1, p.118] and developed in [RDB]. It essentially amounts to a detour ${\cal H} \ra M({\cal F})$ and then $M({\cal F}) \ra \Sigma_{cd} \ra \Sigma$, but in a clever way that avoids to generate large chunks of $\Sigma_{cd}$. It is argued that even if the resulting base $\Sigma$ is considerably larger than $\Sigma_{GD}$, this is more than offset by the short time to obtain $\Sigma$. A similar approach is taken in [AN2], but instead of $\Sigma_{cd}$ the $D$-basis of 4.3 (a subset of $\Sigma_{cd}$) is targeted. Furthermore the likely superior [MU] subroutine for hypergraph dualization is used.

{\bf 3.6.3.2} In another vein, it was recently observed in [R] that for given ${\cal H} \subseteq {\cal P}(E)$ one can readily exhibit a set $\Sigma'$ of implications based on a superset $E'\supseteq E$ such that ${\cal F}': = {\cal F} (\Sigma')$ satisfies ${\cal F}'[E] = {\cal F}({\cal H})$. Here ${\cal F}'[E] : = \{X \cap E: X \in {\cal F}'\}$ is the {\it projection} of ${\cal F}'$ upon $E$. Furthermore, $|E'| = |E| + |{\cal H}|$ and $\Sigma'$ has a mere $2|E|$ implications. What also is appealing: If ${\cal F}'$ is given by $012n$-rows as in 4.4 then ${\cal F}'[E]$ is smoothly calculated by setting to $0$ all components with indices from $E'\setminus E$, and adapting the other components accordingly.

{\bf 3.6.3.3} A natural variation of the ${\cal H} \ra \Sigma$ theme is as follows. For any ${\cal H} \subseteq {\cal P}[n]$ call a family $\Sigma$ of implications a {\it Horn approximation} of ${\cal H}$ if ${\cal H} \subseteq {\cal F}(\Sigma)$. The intersection of all these ${\cal F}(\Sigma)$ is the smallest closure system ${\cal F}({\cal H})$ that contains ${\cal H}$.  Given ${\cal H} \subseteq {\cal P}[n]$ and any $\varepsilon, \delta \in (0,1]$ there is by [KKS, Thm.15] a randomized polynomial algorithm that calculates a family $\Sigma$ of implications which is a Horn approximation of ${\cal H}$ with probability $1 - \delta$ and moreover satisfies $2^{-n}(|{\cal F}(\Sigma)| - |{\cal F}({\cal H})|) < \varepsilon$.
 
{\bf 3.6.3.4} It should be emphasized that current efforts in data mining do however concern ``approximations'' that involve parameters different from $\varepsilon$ and $\delta$ above. These approximations are called {\it association rules} and they involve a support-parameter $\sigma$ and a confidence-parameter $\gamma$ taking values in the interval $(0,1]$. The association rule $A \ra B$ has {\it confidence} $\gamma = 0.57$ if in 57\% of all situations $A \subseteq X \in {\cal F}$ one has $B \subseteq X$. Our ordinary implications $A \ra B$ coincide with the {\it strong} association rules, i.e. having $\gamma =1$. Even ordinary implications like $\{$butter, bread$\} \ra \{$milk$\}$ in 1.1.2 can have a small {\it support} like $\sigma = 0.15$. Namely, when merely 15\% of all transactions  actually feature {\it both} butter and bread, whereas in the other 85\% the implication ``trivially'' holds. See [B] for an introduction to Association Rule Mining that focuses on the underlying mathematics. See also [PKID2, Section 5.1].

\section{Selected topics}

See the introduction (1.2) for a listing of the five selected topics. More detailed outlooks will be provided at the beginning of each Subsection 4.1 to 4.5.

\subsection{Optimum implicational bases for specific closure operators and lattices}

We first show (4.1.1) that {\it each} lattice ${\cal L}$ is isomorphic to a closure system ${\cal F}_J$ on the set $J({\cal L})$ of its join-irreducibles. It thus makes sense to speak of implicational bases of lattices, and we shall investigate special classes of lattices in this regard. Actually, for some lattices ${\cal L}$ it is more natural to start out with a suitable closure operator $c$ and turn to ${\cal L} \simeq {\cal F}_c$ later. For us these ${\cal F}_c$'s are distributive (4.1.2), geometric (4.1.4) and meet-distributive (4.1.5) lattices respectively.

{\bf 4.1.1} We use a basic familiarity with posets, semilattices and lattices, see e.g. [G]. We denote by $\top$ the largest element of a join semilattice, and by $\bot$ the smallest element of a meet semilattice. Recall that a lattice is a poset $({\cal L}, \leq)$ which is both a join and meet semilattice with respect to the ordering $\leq$. In this case some relevant interplay between the sets $J({\cal L})$ and $M({\cal L})$ of join respectively meet-irreducibles occurs (see Expansion 12).

Each closure system ${\cal F} \subseteq {\cal P}(E)$ yields an example of a meet semilattice: The meet of $A, B \in {\cal F}$ (i.e. the largest common lower bound) obviously is $A \cap B$. The smallest element is $\bot = \bigcap {\cal F}$, and ${\cal F}$ has a largest element $\top = E$ as well. Whenever a meet semilattice  happens to have $\top$ then it automatically becomes a lattice. The most important instance of this phenomenon concerns closure systems:

(28) \quad Each closure system ${\cal F} \subseteq {\cal P}(E)$ is a lattice $({\cal F}, \wedge, \vee)$ with meets and joins given by\\
\hspace*{1cm} $X \wedge Y = X \cap Y$ and $X \vee Y = \bigcap \{Z \in {\cal F}: Z \supseteq X \cup Y\}=c_{\cal F} (X \cup Y)$.

\begin{center}
\includegraphics[scale=0.4]{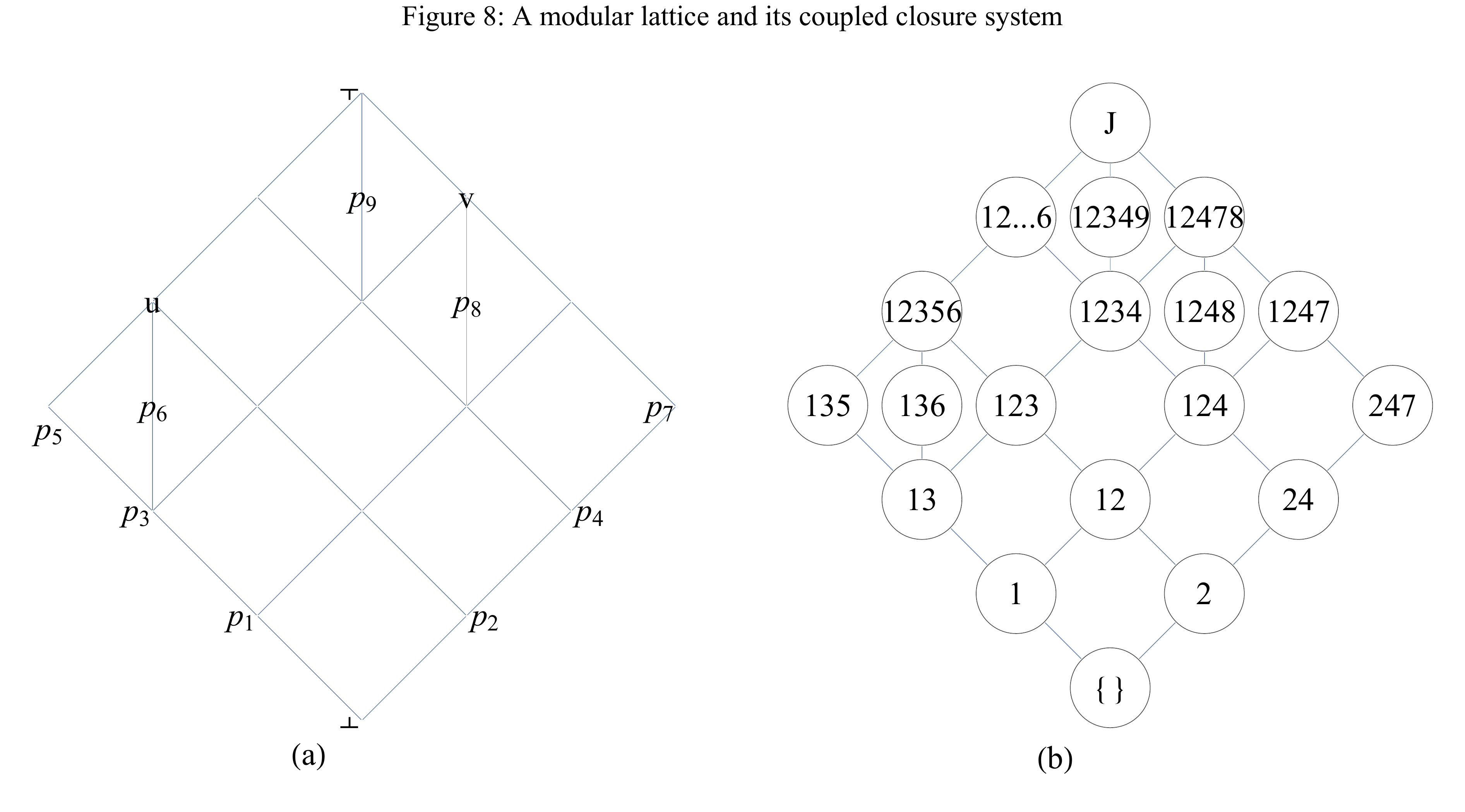}
\end{center}

Let us show that conversely {\it every} lattice ${\cal L}$ arises in this way. What's more, the set $E$ can often be chosen much smaller than ${\cal L}$. Thus for a lattice ${\cal L}$ and any $x \in J := J({\cal L})$ we put
$$J(x) : = \{p\in J : p \leq x\}.$$
We claim that $J(x) \cap J(y) = J(x \wedge y)$. As to $\supseteq$, from $x \wedge y \leq x$ follows $J(x \wedge y) \subseteq J(x)$. Similarly $J(x \wedge y) \subseteq J(y)$, and so $J(x \wedge y) \subseteq J(x) \cap J(y)$. As to $\subseteq$, take $p \in J(x) \cap J(y)$. Then $p \leq x$ and $p \leq y$ which (by the very definition of $\wedge$) implies that $p \leq x \wedge y$, and so $p \in J(x \wedge y)$. If $x \leq y$ then $J(x) \subseteq J(y)$. If $x \not\leq y$ then each $p \in {\cal L}$ minimal with the property that $p \leq x, p \not\leq y$ is easily seen to be join irreducible. Hence $x \not\leq y$ implies $J(x) \not\subseteq J(y)$. Summarizing we see\footnote{Switching from ${\cal F}_J$ to the dually defined ${\cal F}_M$ (see 4.1.6) is sometimes more beneficial.} that:

(29) \quad For each lattice ${\cal L}$ the set system ${\cal F}_J: = \{J(x): x \in {\cal L}\}$ is a closure system and $x \mapsto J(x)$\\
\hspace*{1cm} is a lattice isomorphism from $({\cal L},\wedge, \vee)$ onto $({\cal F}_J, \cap, \vee)$.

Following [AN1] we call ${\cal F}_J$ the {\it standard} closure system coupled to the lattice ${\cal L}$ (recall $J = J({\cal L})$). The standard closure system ${\cal F}_J$ of ${\cal L}$ in Fig.8(a) is shown in Fig.8(b). Now let $c_J: {\cal P}(J) \ra {\cal P}(J)$ be the {\it standard} closure operator coupled to ${\cal F}_J \subseteq {\cal P}(J)$. Explicitely

(30) \quad $c_J(\{ p_1, \cdots, p_n\}) = J(p_1 \vee p_2 \vee \cdots \vee p_n)$

for all subsets $\{p_1, \cdots, p_n\} \subseteq J$.  For instance $c_J(\{p_2, p_5\}) = J(u) = \{p_1, p_2, p_3, p_5, p_6\}$ in Fig.8(a). 
We emphasize that not every closure operator $c$ is ``isomorphic'' to one of type $c_J$, see Expansion 14. Each $c_J$-quasiclosed subset of $J$ clearly is an order ideal of $(J, \leq)$.  This invites to replace each implication $P \ra (c_J (P) \setminus P)$ in $\Sigma'_{GD}$ by $\max (P) \ra \max (c_J(P))$. Along these lines one can associate with each standard closure system ${\cal F}_J$ a (generally not unique) {\it $K$-base} $\Sigma_K$ which stays minimum but satisfies $s(\Sigma_K) \leq s (\Sigma'_{GD})$. See [AN1, Sec.5]. By definition the {\it binary part} of a family $\Sigma$ of implications is $\Sigma^b: = \{(A \ra B) \in \Sigma : |A| = 1\}$. As shown in [AN1, Sec.4], for standard closure spaces the binary parts of implication bases can be ``optimized independently'' to some extent. That relates to Open Problem 3 in Expansion 15.

We now discuss four types of lattices or closure operators for which the structure of the optimum implicational bases is known. These are in turn 
all distributive, all modular, some geometric, and some meet-distributive lattices.

{\bf 4.1.2} A closure operator $c: {\cal P}(E) \ra {\cal P}(E)$ is {\it topological} if $c(X \cup Y) = c(X) \cup c(Y)$ for all $X, Y \in {\cal P}(E)$. For instance, if $\Sigma$ consists of {\it singleton-premise} implications as in 3.2.2 then $c(\Sigma, -)$ is easily seen to be topological. Conversely, if $c$ is topological then by iteration $c(\{x_1, \cdots, x_n\}) = c(\{x_1 \}) \cup \cdots \cup c(\{x_n\})$, and so $\Sigma = \{\{x\} \ra c(\{x\}): x \in E\}$ is a
base for $c$.
Furthermore, for $X = c(X)$ and $Y = c(Y)$ in ${\cal F}_c$ it follows from (28) that
$X \vee Y \  = \  c (X \cup Y) = c(X) \cup c(Y) = X \cup Y.$
By (28) always $X \wedge Y = X \cap Y$, and so ${\cal F}_c$ is a sublattice of the distributive lattice $({\cal P}(E), \cup, \cap)$, which thus must be distributive itself. 
In Expansion 15 we show that conversely {\it every} distributive lattice ${\cal L}$ is isomorphic to a sublattice of ${\cal P}(J)$, and we determine the unique optimum base $\Sigma_J$ of ${\cal L}$.

{\bf 4.1.3} A lattice ${\cal L}$ is {\it modular} if it follows from $x \leq z$ that $(x \vee y) \wedge z = x \vee (y \wedge z)$. For instance the lattice of all submodules of an $R$-module is modular. Furthermore, each distributive lattice is modular. 
The $(n+2)$-element lattice consisting of $n$ atoms and $\bot, \top$ will be denoted by $M_n$.  It is modular but not distributive for $n \geq 3$. In fact every modular but nondistributive lattice has $M_3$ as a sublattice. For any lattice ${\cal L}$ and any $x \in {\cal L}\backslash \{\bot\}$ we define $x_\ast$ as the meet of all lower covers of $x$.  We call $x \in {\cal L}$ an $M_n${\it -element} if the interval $[x_\ast, x]$ is isomorphic to $M_n$ for some $n \geq 3$.  According to [W2] each optimum base $\Sigma$ of a modular lattice is of type $\Sigma = \Sigma_J \cup \Sigma_{HW}$ where $\Sigma_J$ is as in Expansion 15, and the implications constituting  $\Sigma_{HW}$ are as follows. Coupled to each $M_n$-element $x$ choose ${n \choose 2}$ suitable implications of type $\{p,q\} \ra \{v\}$. They are not uniquely determined by $x$ but they all satisfy $p \vee q = x$ among other restrictions. To fix ideas, the lattice ${\cal L}_0$ in Fig. 8(a) is modular and one possible optimum base is $\Sigma = \Sigma_J \cup \Sigma_{HW}$ where $\Sigma_{HW}$ contains the nine implications
$$\begin{array}{lll}
\{p_2, p_5\} \ra \{p_6\}, & \{p_2, p_6\} \ra \{p_5 \}, & \{p_5, p_6\} \ra \{p_2\},\\
\\
\{p_3, p_7\} \ra \{p_8\}, & \{p_3, p_8\} \ra \{p_7 \}, & \{p_7, p_8\} \ra \{p_3\},\\
\\
\{p_6, p_8\} \ra \{p_9\}, & \{p_6, p_9\} \ra \{p_8 \}, & \{p_8, p_9\} \ra \{p_6\},\end{array}$$

It is convenient to think of the $n$ join-irreducibles underlying the ${n \choose 2}$ implications coupled to a fixed $M_n$-element as a {\it line} $\ell$. These lines have properties akin to the lines occuring in projective geometry (see also 4.1.4). Modular lattices which are freely generated by a poset (in a sense akin to 4.2) are economically computed by combining Theorem 5 with the technique of 4.4. A preliminary version of this work in progress is in [arXiv: 1007.1643.v1].

{\bf 4.1.4} A closure operator $c: {\cal P}(E) \ra {\cal P}(E)$ is a {\it matroid} (operator) if it satisfies this {\it exchange axiom} for all $X \subseteq E$ and $x,y \in E$:

(31) \quad $(y \in c(X \cup \{x\})$ and $y \not\in c(X) ) \quad \Ra \quad x \in c(X \cup \{y\})$

As a consequence each minimal generating set of $E$ (or $X = c(X)$) is maximal independent. Thus for matroids the word ``among'' in 3.3.2 can be replaced by ``exactly''. The edge set $E$ of any graph yields a ``graphic'' matroid $c: \ {\cal P}(E) \ra {\cal P}(E)$ whose circuits in the sense of Expansion 5 coincide with the circuits in the usual graph theoretic sense. As another example, let $F$ be any field and let $E \subseteq F^n$ be any (finite) subset which need not be a subspace. If for $X \subseteq E$ we define $c(X) : = \mbox{span}(X) \cap E$, then the restriction $(E, c)$ is an {\it $F$-linear} matroid.  The particular features of  $(E,c)$ depend on the kind of subset $E$ chosen. For instance, if $E$ is a linearly independent set then $c(X) = X$ for all $X \subseteq E$. Another extreme case is $E = F^n$. Then
$\Sigma: = \{ \{x,y\} \ra \mbox{span}(\{x,y\}): \ x, y \in F^n \}$
is a base of $c$ and ${\cal F}(\Sigma)$ is the {\it complemented} modular lattice\footnote{In fact, for {\it any} matroid $c$ the coupled lattice is complemented but usually only {\it semi}-modular. Such lattices are also called {\it geometric}.} of all subspaces of $F^n$, thus a special case of 4.1.3. In fact, the $M_n$-elements of ${\cal F}(\Sigma)$ are the rank two subspaces ($=$ projective lines). The features of  a $F$-linear matroid also depend on the field of scalars $F$. For $F = \Z_2$ one speaks of {\it binary matroids}, in which case the family $\Sigma$ of implications $(K\backslash \{x\}) \ra \{x\}$, where $K$ ranges over all {\it closed} circuits $K$ and $x$ ranges over $K$, is the unique optimum implication base of $(E, c)$, see [W3]. It is well known that each graphic matroid is binary, but not conversely. For the many facets of matroids see [S, Part IV]. We mention in passing that [S] arguably is the most comprehensive, and likely the most readable book on combinatorial optimization around.

{\bf 4.1.5} A closure operator $c: {\cal P}(E) \ra {\cal P}(E)$ is a {\it convex geometry} (operator) if it satisfies this {\it anti-exchange axiom}:

(32) \quad If $x \neq y$ \ and \ $x,y \not\in c(X)$ \ and \ $y \in c(X \cup \{x\}) \ \mbox{then} \ x \not\in c(X \cup \{y\})$.

The kind of operator $c$ in 2.2.5 is the name-giving example of a convex geometry. 
As to another example, it was observed by Bernhard Ganter (around 1990, unpublished) and also follows from [SW, Lemma 7.7] that each closure operator $c$ of poset type (see 3.5.1) is a convex geometry. 

One deduces from (32) that each $X \subseteq E$ contains the {\it unique} minimal generating set $ex(X)$ of $c(X)$. In particular $|{\cal F}_c| = |\mbox{Indep}(c)|$ in 3.3.1.  The elements of $ex(X)$ are the {\it extreme} points of $X$. If $X$ is closed then so is $X \backslash \{x\}$ for all $x \in ex (X)$.
Each circuit $K$ of $c$ (Expansion 5) has a {\it unique} root $e$. If one needs to emphasize $e$, one speaks of the {\it rooted circuit} $(K, e)$. Other than for arbitrary closure operators, if $U$ is a stem of $e$ in a convex geometry then $(U \cup \{e\}, e)$ is a rooted circuit. It follows [W3, Cor.13] that the family of all rooted circuits matches the family $\Sigma_{cd}^u$ of all prime implicates. A rooted circuit $(K, e)$ is {\it critical} if $c(K) \backslash \{e, x\}$ is closed for all $x \in K \backslash \{e\}$. Recall the definition of closure-minimal in 3.3. As we show in Expansion 16, for each rooted circuit $(K, e)$ it holds that:

(33) \quad $(K, e)$ is critical $\Leftrightarrow \ c(K)\backslash \{e\}$ is quasiclosed $\Leftrightarrow$ the stem $K \backslash \{e\}$ of $e$ is closure-minimal

As opposed to the antimatroid side of the coin (Expansion 16), note that the subfamily
$$\Sigma_{crci} : = \{(K \backslash \{e\}) \ra \{e\} : (K, e) \ \mbox{is critical rooted circuit of} \ c\}$$
of $\Sigma_{cd}^u$ usually is {\it no} implicational base of $c$. For instance, the set $\Sigma_{cd}^u$ of prime implicates of the convex geometry $c$ in 2.2.5 is the union of all sets $\{T \ra \{e\} : e \in c(T) \backslash T\}$ where $T$ ranges over ${\cal T}$. If such a rooted circuit $(T,e)$ has $c(T) = T \cup \{e\}$ then $c(T) \setminus \{e\}$ is quasiclosed. Conversely, assume $c(T)$ contains a point $f \neq e$. By considering the triangulation of $ch(T)$ induced by $f$ (as in 2.2.5) one sees that $e \in (c(T) \setminus \{e\})^\bullet$, and so $c(T) \setminus \{e\}$ is {\it not} quasiclosed. It follows from (33) that $\Sigma_{crci} = \{T\ra \{e\} : T \in {\cal T}, c(T) = T \cup \{e\} \}$. Hence $\Sigma_{crci}$ is contained in every base of prime implicates but is not itself a base (unless the point configuration in $\R^2$ is rather trivial). 
We mention that closure-minimality of (order-minimal) stems also features in the so-called $E$-basis of [A] and [AN1]. The convex geometries of type 2.2.5 and 3.5.1 can be generalized (Expansion 18) but the results and proofs become quite technical. This is one reason for dualizing (29) in 4.1.6.

{\bf 4.1.6} For any lattice ${\cal L}$ and $x \in {\cal L}$ put $M(x) : = \{m \in M({\cal L}) : m \geq x\}$. Dually to (29), ${\cal F}_M : = \{M(x) : x \in {\cal L}\}$ is a closure system which is bijective to ${\cal L}$ under the map $x \mapsto M(x)$. In particular, if ${\cal L}$ is meet-distributive (thus ${\cal L}$ ``is'' a convex geometry according to Expansion 16) then a crisp implication base $\Sigma = \Sigma_M \cup \Sigma_{JNW}$ of ${\cal F}_M$ is obtained as follows\footnote{Mutatis mutandis, this is Theorem 2 in [W4]. The acronym JNW means Janssen-Nourine-Wild.}. First $\Sigma_M$ is the dual of $\Sigma_J$ from Expansion 15. Second, each doubleton $\{m, m_0\} \subseteq M({\cal L})$ which admits a (unique if existing) $p\in J({\cal L})$ with $p \updownarrow m$ and $p \updownarrow m_0$, induces two implications. One is $\{m\} \cup \mbox{ucov}(m_0) \ra \{m_0\}$, the other $\{m_0\} \cup \mbox{ucov}(m) \ra \{m\}$. Here $\updownarrow$ is as in Expansion 12, and say $\mbox{ucov}(m)$ is the set of upper covers of $m$ in the poset $(M({\cal L}), \leq)$. All these implications make up $\Sigma_{JNW}$. In view of $|M({\cal L})| \geq |J({\cal L})|$ the philosophy in 4.1.6 is similar to 3.6.3.2 which also trades a larger universe for a smaller implication base.

\subsection{Excursion to universal algebra: Finitely presented semilattices and subalgebra lattices}
 
First we show (4.2.1) that finding an implicational base for a lattice ${\cal L}$ in the sense of 4.1  means finding a presentation for ${\cal L}$, viewed as  $\vee$-semilattice, in the sense of universal algebra. Afterwards we show (4.2.2) how subalgebra lattices and homomorphisms between algebras can be calculated by setting up appropriate implications. 

{\bf 4.2.1} For starters imagine a $\vee$-semilattice that has a set $J$ of (not necessarily distinct) generators $p_1, \ldots, p_6$ that satisfies this set ${\cal R}$ of (inequality) relations:

(34) \quad $p_3 \geq p_5, \quad p_1 \vee p_5 \geq p_4, \quad  p_6 \geq p_3, \quad p_2 \vee p_3 \geq p_1$

 An example of such a semilattice $S_1$ (with say $p_2$ replaced by $2'$) is given in Figure 9 on the left. Notice that all relations hold; e.g. $p_2 \vee p_3 \geq p_1$ holds because $p_2 \vee p_3 = p_2 > p_1$.
It isn't a priori clear whether there is a {\it largest} such semilattice, but universal algebra tells us it must exist.
 It is the so-called {\it relatively free} $\vee$-semilattice $F = FS(J, {\cal R})$ with set of generators $J$ and subject to the relations in ${\cal R}$, shown on the right in Figure 9 (discard $\emptyset$). Every other $\vee$-semilattice satisfying ${\cal R}$ must be an epimorphic image of $F$; in our case the definition of the epimorphism $f: \ F \ra S_1$ is that $\circ$ on the right maps to $\circ$ on the left, $\bullet$ maps to $\bullet$, and so forth.
 
 Each ($\vee$-semilattice) inequality, like $p_1 \vee p_5 \geq p_4$, can be recast as an identity $p_1 \vee p_5 = p_1 \vee p_5 \vee p_4$.  Conversely each identity can be replaced by two inequalities. If in turn inequalities $a_1 \vee \cdots \vee a_s \geq b_1 \vee \cdots \vee b_t$ are viewed as implications $\{a_1, \cdots, a_s \} \ra \{b_1, \cdots, b_t\}$ then we can state the following.

\begin{center}
\includegraphics[scale=0.6]{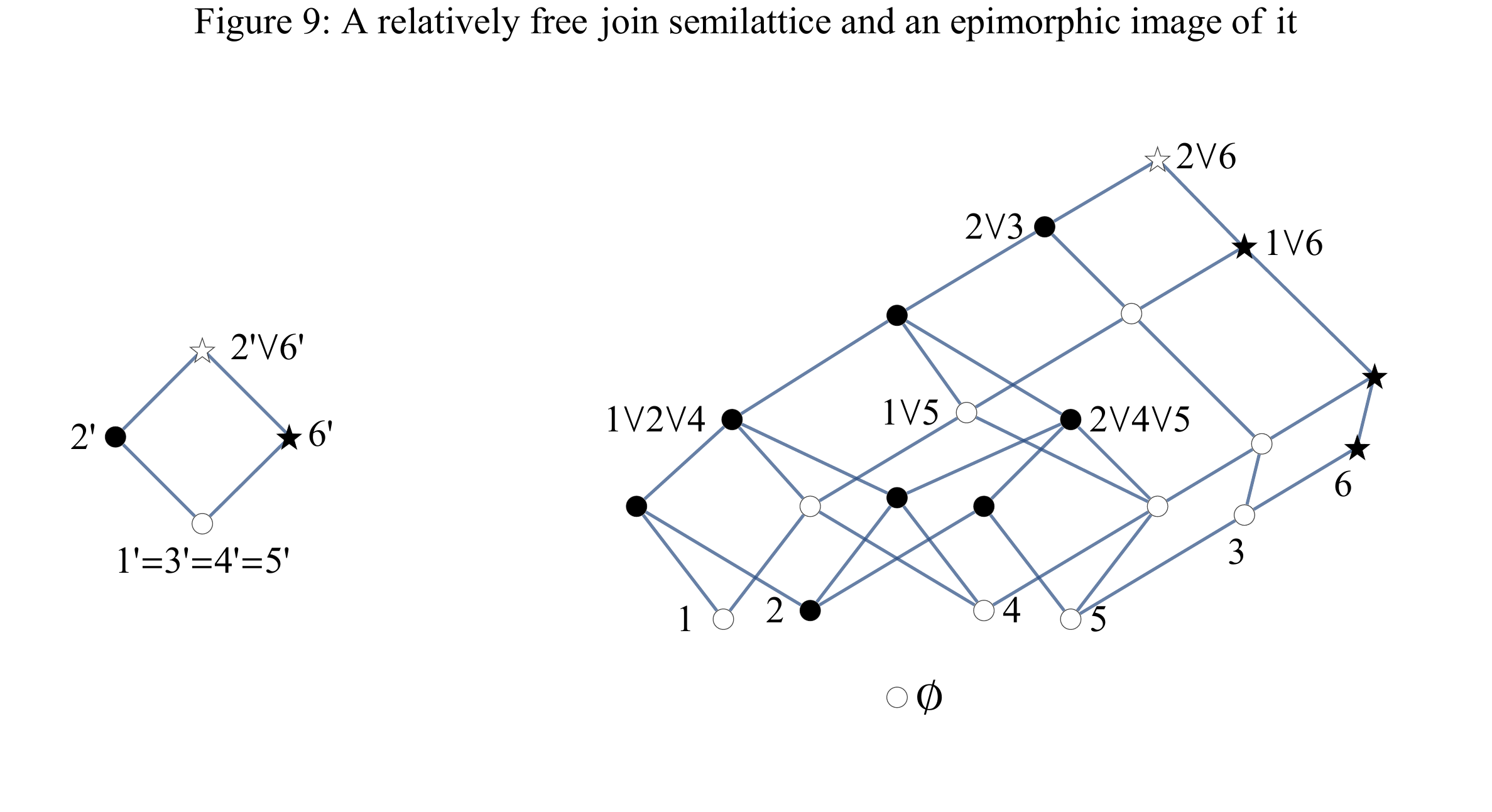}
\end{center}

\begin{tabular}{|l|} \hline \\
{\bf Theorem 5 :} The relatively free $\vee$-semilattice $FS(J, {\cal R})$ is isomorphic to the\\
$\vee$-semilattice ${\cal F}(\Sigma) \backslash \{\emptyset\}$. Here the family $\Sigma$ is obtained from ${\cal R}$ by replacing
each \\ 
inequality in ${\cal R}$ by the matching implication, and each identity in ${\cal R}$ by two\\
implications $A \ra B$ and $B \ra A$.\\
 \\ \\ \hline \end{tabular}

The proof of Theorem 5 is given in [W5, Thm.5]. The closure system ${\cal F}(\Sigma)$ can be calculated from $\Sigma$ in compressed form as explained in 4.4. Specifically for the $\Sigma$ matching the inequalities in (34), thus $\Sigma = \{3 \ra 5, 15 \ra 4, 6 \ra 3, 23 \ra 1\}$, one gets ${\cal F}(\Sigma)$ as $r_9 \cup r_{10} \cup r_{11} \cup r_{12}$ for certain set systems $r_9$ to $r_{12}$ in Table 1 of 4.4.  We mention that $FS(J, {\cal R})$ is also isomorphic to the semilattice $({\cal P}(E) \backslash \{\emptyset\}, \cup )$ modulo a congruence relation $\theta$. Here $E = [6]$ and $\theta$ is as in (11) where $c$ is $c(\Sigma, -)$ with $\Sigma$ from Theorem 5.
See also Expansion 17.

{\bf 4.2.2} As to subalgebra lattices, we only peak at semigroups but the ideas carry over to general algebraic structures (and what concerns homomorphisms, also to graphs). Suppose we know the multiplication table (Cayley table) of a semigroup $(S, \ast)$ where $S = \{a_1, a_2, \cdots, a_n\}$. Obviously the subsets of $S$ closed with respect to the $n^2$ implications $\{a_i, a_j\} \ra \{a_i \ast a_j\}$ are exactly the subsemigroups of $S$. 
The algorithm from 4.4 can thus be invoked to give a compressed representation of all subsemigroups. 

In another vein, sticking again to semigroups $(S, \ast)$ and $(S', \bullet)$ for simplicity, the same algorithm also achieves the enumeration of all homomorphisms $f: S \ra S'$. Namely, these $f$'s are exactly the functions\footnote{More precisely, imposing these $n^4$ implications yields the closure system ${\cal F}$ of all homomorphic {\it relations} $f \subseteq S \times S'$ in output-polynomial time.  True, one needs to sieve the functions among them, but this is often feasible. As to the large cardinality $n^4$ of our family $\Sigma$ of implications, instead of calculating ${\cal F}$ as ${\cal F}(\Sigma)$ one may directly target $M({\cal F})$, see 3.6.2. All of this is work in progress.}  $f \subseteq S \times S'$ which are closed with respect to all $n^4$ implications of type $\{(a,x), (b,y)\} \ra \{(a \ast b, x \bullet y)\}$. How these ideas compete with other computational tools in algebra (e.g. consult the Magma Handbook) remains to be seen. They will fare the better the fewer structural properties of the algebras at hand can be exploited. Put another way, there are greener pastures for our approach than e.g. the beautiful theory of subgroup lattices of Abelian groups [Bu].

\subsection{Ordered direct implicational bases}

We start by introducing order-minimal prime implicates, thus a third kind besides the closure-minimal ones in 3.3 and the strong ones in Expansion 6. To minimize technicalities we focus on the case of standard closure operators $c_J$. Then the prime implicates of $c_J$ are the nonredundant join covers in the lattice ${\cal L}$ that underlies $c_J$. Specifically, $\{2, 5\}$ in Figure 10 (taken from [ANR]) is a {\it join cover} of 6 since $2 \vee 5 \geq 6$. It is nonredundant since $2 \not\geq 6$ and $5 \not\geq 6$. (Generally,  {\it nonredundant} means that no proper subset is  a join cover.) Correspondingly $\{2,5\} \ra \{6\}$ is a prime implicate of $c_J$. However $\{2, 5\} \ra \{6\}$ is not {\it order-minimal} since $4 < 5$ and still $\{2, 4\} \ra \{6\}$ is a prime implicate. The general definition of ``order minimal'' is the obvious one. The relevance this concept was first observed in [N, p.525]. Notice that $\{2, 4\} \ra \{6\}$ is not closure-minimal since $\{2,3\} \ra \{6\}$ is a prime implicate with $2 \vee 3 < 2 \vee 4$. Conversely a closure-minimal prime implicate need not be order-minimal.

We are now in a position to address the topic in the title.
Recall from 3.3 that the {\it direct} basis $\Sigma_{cd}$ of a closure operator $c$ has the advantage that $c(\Sigma_{cd}, X) = X'$ as opposed to $c(\Sigma, X) = X^{'' \cdots '}$ (as to $X'$, see (6)). However the drawback of $\Sigma_{cd}$ is its usually large cardinality. As a kind of compromise we present {\it ordered direct} implicational bases $\Sigma$. The key is a specific {\it ordering} in which the implications of $\Sigma$ must be applied exactly once: For given $X \subseteq E$ applying the first implication $A_1 \ra B_1$ of $\Sigma$ to $X$ yields $X_1 \supseteq X$. Applying $A_2 \ra B_2$ to $X_1$ yields $X_2 \supseteq X_1$. And so forth until applying the last implication $A_n \ra B_n$ to $X_{n-1}$ yields $X_n \supseteq X_{n-1}$ which is the correct closure of $X$. Of course such a $\Sigma$ is also an implication base in the ordinary sense.

Listing (in any order) all\footnote{In certain circumstances,  one or both ``all'' in this sentence can be weakened (by restricting ``any order'').} {\it binary} prime implicates $x \ra y$ (thus $x > y$), and then listing (in any order) all order-minimal prime implicates, yields a particular ordered direct implicational base $\Sigma_D$ which is called a $D${\it -basis}. The ``$D$'' derives from the so-called $D$-relation discussed in Expansion 18.

\begin{center}
\includegraphics[scale=0.6]{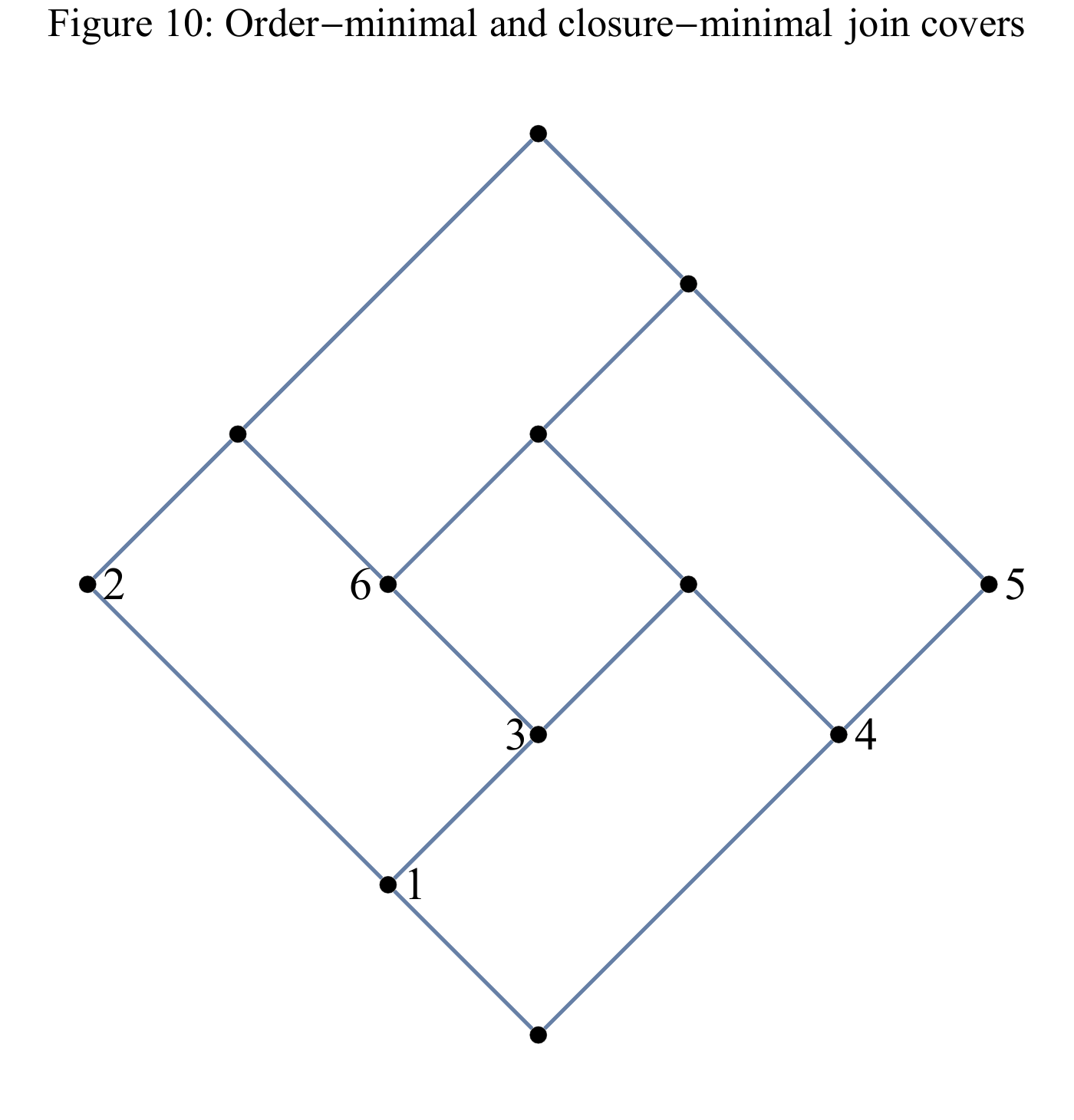}
\end{center}

 In our example one possibility is

(35) \quad $\Sigma_D = (2 \ra 1, \  6 \ra 3, \ 6 \ra 1, \ 5 \ra 4, \ 3 \ra 1, \ 14 \ra 3, \ 24 \ra 5, \ 15 \ra 6, \ 24 \ra 6, \ 23 \ra 6)$.

Applying $\Sigma_D = (A_1 \ra B_1, \cdots, A_{10} \ra B_{10})$ in this order to say $X = \{2,5\}$ yields
$$X_1=X_2 = X_3 = 251, \quad X_4=X_5 = 2514, \quad X_6=X_7 = 25143, \quad X_8 = X_9 = X_{10} = 251436$$
In contrast, ordinary forward chaining (2.2) needs three runs to find the closure:
$$X'= 2514, \quad X'' = 251436, \quad X''' = 251436 = X'' = c(X)$$
Notice that the underlying unordered set of any $D$-basis  coincides with $\Sigma_{cd}^u$ if $J({\cal L})$ is an antichain: Then there is no binary part, and so each member of $\Sigma_{cd}^u$ is trivially order-minimal. There is actually no need to stick to bases of prime implicates. Given any basis $\Sigma$ of $c_J$ one can aim for an ordered direct base by suitably ordering $\Sigma$, and perhaps repeat some implications. Unfortunately the canonical base $\Sigma_{GD}$ needs not be orderable in this sense [ANR, p.719].

\subsection{Generating ${\cal F}(\Sigma)$ in compact form}

Calculating ${\cal F}(\Sigma)$ amounts to generating the model set Mod$(f)$ of a pure Horn function $f$ given in CNF (see 3.4). As glimpsed this has applications in Formal Concept Analysis, Learning Theory, and Universal Algebra. One could be tempted to calculate ${\cal F}(\Sigma)$ from $\Sigma$ with NextClosure (Expansion 4). But this yields the closed sets {\it one-by-one} which is infeasible when ${\cal F}(\Sigma)$ is large.
 In 4.4.1 we thus outline an algorithm for {\it compactly} generating ${\cal F}(\Sigma)$ from $\Sigma$. In 4.4.2 we discuss how to get a compact representation of ${\cal F}$ not from $\Sigma$, but from a generating set ${\cal H} \subseteq {\cal F}$.

{\bf 4.4.1} A $012$-{\it row} like $(0,2,1,1,2,2)$ is a succinct representation for the interval $\{U \subseteq {\cal P}[6]: \{3,4\} \subseteq U \subseteq \{3,4,2,5,6\}\}$, which thus has cardinality $2^3$. Each ``2'' in $(0,2,1,1,2,2)$ is used as a {\it don't care} symbol (other texts use ``$\ast$'') which indicates that both 0 and 1 can be chosen. For instance, if the clause $\ol{x}_1 \vee x_4 \vee \ol{x}_5$ (thus $15 \ra 4$) is viewed as a Boolean function of $x_1, \cdots, x_6$, then Mod$(\ol{x}_1 \vee x_4 \vee \ol{x}_5)$ clearly is the disjoint union of these four $012$-rows:

\begin{tabular}{|c|c|c|c|c|c|}
1 & 2 & 3& 4 & 5 & 6\\ \hline \hline
0 & 2 & 2 & 2& 0 & 2 \\ \hline
0 & 2 & 2 & 2 & 1 & 2 \\ \hline
1 & 2 &  2& 2& 0 & 2\\ \hline
1 & 2 & 2 & 1 & 1& 2 \\ \hline \end{tabular}

If we let the  $n$-{\it bubble} $(n, n, \cdots, n)$ mean ``{\it at least one 0 here}'' then the first three rows can be compressed to the $012n${\it -row} $r_1$ in Table 1. It thus follows that Mod$(\ol{x}_1 \vee x_4 \vee \ol{x}_5)$ is the disjoint union of $r_1$ and $r_2$ in Table 1. Consider the pure Horn function $f: \{0,1\}^6 \ra \{0,1\}$ given by
$$f(x) : = (\ol{x}_1 \vee x_4 \vee \ol{x}_5) \wedge (x_1 \vee \ol{x}_2 \vee \ol{x}_3) \wedge (\ol{x}_3 \vee x_5) \wedge (x_3 \vee \ol{x}_6).$$
In order to calculate Mod$(f)$ we
 need to ``sieve'' from $r_1$, and then from $r_2$, those bitstrings which also satisfy $x_1 \vee \ol{x}_2 \vee \ol{x}_3$. It is evident that this shrinks $r_1$ to $r_3 \cup r_4$ and does nothing to $r_2 =: r_5$.  In $r_3$ the two $n$-bubbles are independent of each other and distinguished by subscripts.

\begin{tabular}{c|c|c|c|c|c|c|}
& 1 & 2 & 3 & 4 & 5 & 6 \\ \hline
& & & & &     &\\ \hline
$r_1=$ & $n$ & 2 & 2 & 2& $n$ & 2\\ \hline
$r_2=$ & 1 & 2 & 2 & 1 & 1& 2\\ \hline
& & & & & & \\ \hline
$r_3=$ & $n_1$ & ${\bf n_2}$ & ${\bf n_2}$ & 2 & $n_1$ & 2\\ \hline
$r_4=$ & 1 & ${\bf 1}$ & ${\bf 1}$ & 2 & 0 & 2 \\ \hline
$r_5=$ & 1 & 2& 2 & 1 & 1& 2 \\ \hline
 & & & & & & \\ \hline
 $r_6=$ & $n$ & 2 & ${\bf 0}$ & 2 & $n$ & 2 \\ \hline
 $r_7=$ & 0 & 0 & ${\bf 1}$ & 2 & 1 & 2\\ \hline
 $r_8=$ & 1 & 2& 2& 1 & 1& 2 \\ \hline
  & & & & & & \\ \hline
$r_9=$ & $n$ & 2 & 0 & 2 & $n$ & 0\\ \hline
$r_{10}=$ & 0 & 0 & 1 & 2& 1 & 2\\ \hline
$r_{11}=$ & 1& 2 & 2 & 1& 1& ${\bf 0}$ \\ \hline
$r_{12}=$ & 1 & 2& 1 & 1 & 1& ${\bf 1}$ \\ \hline   \end{tabular}

Table 1: Using $012n$-rows to compress a closure system

Note that forcing the first component of $n_1 n_1$ to 1 in $r_4$ (due to $23 \ra 1$) forces the second to 0. Imposing the constraint $\ol{x}_3 \vee x_5$ (i.e. $3 \ra 5$) upon $r_3 \cup r_4 \cup r_5$ replaces $r_3$ by $r_6 \cup r_7$, deletes $r_4$, and leaves $r_5= r_8$ unscathed. Imposing the implication $6 \ra 3$ upon $r_6 \cup r_7 \cup r_8$ yields $r_9 \cup r_{10} \cup r_{11} \cup r_{12} = \, \mbox{Mod}(f)$. We were lucky that $n_2 n_2$ didn't clash with $n_1n_1$, otherwise things  would get uglier. Concerning the deletion of $r_4$, with some precautions the deletion of rows can be avoided, which is the main reason making the implication $n$-algorithm output-polynomial [W6]. The implication $n$-algorithm easily extends to a {\it Horn $n$-algorithm} which can handle impure Horn functions in the sense of 4.5.  Concerning a speed-up for singleton-premise implications, see Expansion 19. As to connections to $M({\cal F})$ and CNF $\ra$ DNF conversion, see Expansion 8 and 9 respectively.

{\bf 4.4.2} As to calculating ${\cal F}$ from a {\it generating} set ${\cal H} \subseteq {\cal F}$, the first idea that springs to mind is to use NextClosure or some other algorithm discussed in [KuO1].  However, this as before yields the closed sets one-by-one which is infeasible when ${\cal F}$ is large. Alternatively, one may calculate a base $\Sigma$ of ${\cal F}$ by either proceeding as in 3.6.3.1 or 3.6.3.2. Feeding $\Sigma$ to the implication $n$-algorithm yields a compact representation of ${\cal F}(\Sigma) = {\cal F}$. An analysis of the pro's and con's of these ways to enumerate ${\cal F}$ is  pending.

\subsection{General Horn functions}

We discuss negative functions in 4.5.1 and then use them to define general Horn functions in 4.5.2. Theorem 6 says, in essence, that good old implications suffice to economically capture any impure Horn function; only {\it one} additional impure Horn clause is necessary.

{\bf 4.5.1} For any nonempty ${\cal H} \subseteq {\cal P}(E)$ the set ideal {\it generated} by ${\cal H}$ is ${\cal H}\downarrow \ : = \{U \subseteq E: (\exists U'\in {\cal H}) \ U \subseteq U'\}$. By 3.4.1 a Boolean function $g$ is negative if and only if Mod$(g)$ is a set ideal. Dually one defines {\it set filters}. Consider an arbitrary family $\Gamma$ of sets $A \subseteq E$ which we refer to as {\it complications}\footnote{This is handy ad hoc terminology which conveys a link to ``implications''.}.
Call $X \subseteq E$ a {\it noncover} (of $\Gamma$) if it doesn't cover any complication, i.e. $X \not\supseteq A$ for all $A \in \Gamma$. It is evident that the set ${\cal N}{\cal C}(\Gamma)$ of all noncovers is a set ideal ${\cal G}$. 
Among all families $\Gamma'$ with ${\cal N}{\cal C}(\Gamma') = {\cal G}$ there is smallest one; it obviously is the family $\Gamma_0$ of all minimal members of the set filter ${\cal P}(E)\backslash {\cal G}$. In particular $\Gamma_0$ is an antichain (no two distinct members of $\Gamma_0$ are comparable). Conversely, {\it each} set ideal ${\cal G}$ admits a unique antichain $\Gamma_0 \subseteq {\cal P}(E)$ of complications $A$ that yields ${\cal G}= {\cal N}{\cal C}(\Gamma_0)$. Put another way, each negative Boolean function $g: \{0,1\}^n \ra \{0,1\}$ admits a {\it unique} irredundant CNF of {\it negative clauses}. For instance if $E = [7]$ and by definition the model set of $g: {\cal P}(E) \ra \{0,1\}$ is the set ideal, ${\cal N} {\cal C}(\{\{2,3,5\}, \{2, 4\}\})$, then $g = g(x_1, \ldots, x_7)$ has the unique irredundant CNF  $(\ol{x}_2 \vee \ol{x}_3 \vee \ol{x}_5) \wedge (\ol{x}_2 \vee \ol{x}_4)$. We see that the ``representation theory'' of negative Boolean functions $g$ via complications ($=$ negative clauses) is much simpler than the representation theory of pure Horn functions $f$ via implications ($=$ pure Horn clauses).

{\bf 4.5.2} This leads us to the definition of a {\it Horn function} $h: \{0,1\}^n \ra \{0,1\}$ as one that can be represented as a conjunction $h = f \wedge g$ of a pure Horn function $f$ with a negative function $g$. One checks that pure Horn functions and negative functions are special cases of Horn functions. It is evident that Mod$(h) = {\cal N}{\cal C}(\Gamma) \cap {\cal F}(\Sigma)$ where $\Sigma$ and $\Gamma$ are such that ${\cal F}(\Sigma)= \mbox{Mod}(f)$ and ${\cal N}{\cal C}(\Gamma) = \, \mbox{Mod}(g)$.  
We call $\Sigma \cup \Gamma$ a {\it base} of $h$. Thus our previous bases $\Sigma$ become the special case where $\Gamma = \emptyset$. With Mod$(f)$ and Mod$(g)$ also Mod$(h)$ is a subsemilattice\footnote{The only difference between subsemilattices ${\cal S} \subseteq {\cal P}(E)$ and closure systems ${\cal F} \subseteq {\cal P}(E)$ is that subsemilattices need not contain $E$. The usefulness of meet-irreducible sets, also in the impure case, remains.} of $({\cal P}[n], \cap )$. But Mod$(h)$ can be empty, and so different from 3.4 a general Horn function $h$ need not be satisfiable. The good news is, because ${\cal F}(\Sigma)$ has a {\it smallest} member $\bigcap {\cal F}(\Sigma)$, it follows that Mod$(h) = \emptyset$ iff $\bigcap {\cal F}(\Sigma)$ contains some $A \in \Gamma$. Since $\bigcap {\cal F}(\Sigma)$ can be calculated from $\Sigma$ as $c(\emptyset, \Sigma)$, satisfiability can be tested in linear time. (In plenty texts this simple state of affairs is veiled by clumsy notation.)

Observe that the above representation $h= f\wedge g$ is not unique since the subsemilattice ${\cal S} = \,\mbox{Mod}(h)$ can be written as an intersection ${\cal F} \cap {\cal G}$ of a closure system ${\cal F}$ with a set ideal ${\cal G}$ in many ways. The most obvious way is ${\cal S} = \bot \cap ({\cal S} \downarrow )$ where $\bot$ is the closure system ${\cal S} \cup \{E\}$. (The notation $\bot$ foreshadows the framework (39) in Expansion 20.) 
The parameters defined for pure Horn functions $f$ in 3.4.4.1 carry over to general Horn functions $h$. Here we are only interested in 
$$ca(h): = \min \{|\Sigma  \cup \Gamma |: \ \Sigma \cup \Gamma \ \mbox{is a base of} \ h\}.$$
Note that $ca(h) = \sigma (h)$ in [CH, p.297], i.e. the minimum number of ``source sides'' possible.

\begin{tabular}{|l|} \hline \\
{\bf Theorem 6:} Let $h : {\cal P}(E) \ra \{0,1\}$ be any Horn function, and let $f_\bot$ be the {\it pure} Horn function\\
defined by Mod$(f_\bot) := \, \mbox{Mod}(h) \cup \{E\}$. Then $ca(f_\bot) \leq ca(h) \leq ca(f_\bot) +1$.\\ \\ \hline \end{tabular}

{\it Proof.} Since Mod$(h) \subseteq {\cal P}(E)$ is a subsemilattice, Mod$(h) \cup \{E\}$ is indeed a closure system. Let $f_\bot$ be the induced pure Horn function, and let $\Sigma_0$ be a base of implications for Mod$(h) \cup \{E\}$ of minimum cardinality $ca(f_\bot)$. We claim that $\Sigma_0 \cup \{E\}$ is a base of $h$: Indeed, if say $E=[n]$ then spelling out the complication $E$ gives $\ol{x}_1 \vee \cdots \vee \ol{x}_n$. It kills exactly one $\Sigma_0$-closed set, namely $E$. Therefore $ca(h) \leq ca(f_\bot) +1$. 

Conversely, let $\Sigma \cup \Gamma$ be a base of $h$ of cardinality $ca(h)$. Putting $\Sigma' : = \{A \ra E: A \in \Gamma \}$, it suffices to show that $\Sigma \cup \Sigma'$ is a base of $f_\bot$;  then $ca(f_\bot) \leq |\Sigma \cup \Sigma'| = ca(h)$ as claimed. First, each model $X \subseteq E$ of $\Sigma \cup \Gamma$ remains a model of $\Sigma \cup \Sigma'$ because $A \not\subseteq X$ for all $(A \ra E) \in \Sigma'$. Second, let $X \subseteq E$ be a model of $\Sigma \cup \Sigma'$ which is not a model of $\Sigma \cup \Gamma$. Then $A \subseteq X$ for some $A \in \Gamma$, and so $X = E$ in view of $(A \ra E) \in \Sigma'$. \quad $\square$

Theorem 6 suggests a simple procedure to ``almost minimize'' a given base $\Sigma \cup \Gamma$ of $h$: Take the base $\Sigma \cup \Sigma'$ of $f_\bot$ and replace it by a minimum base $\Sigma_0$ e.g. by using Shock's algorithm (Expansion 11). Then $\Sigma_0 \cup \{E\}$ is a base of $h$ of cardinality at most $ca(h) +1$. 
In Expansion 20 we indicate that calculating the precise value of $ca(h)$ is comparatively tedious.

{\bf 4.5.3} An analogue of the Guigues-Duquenne base (3.2) is introduced in [AB] for general Horn functions $h$. It is shown that a well known query leraning algorithm of Angluin et al. in fact always produces this base, independently of the counterexamples it receives.

\section{Omitted proofs and various expansions}

{\bf Expansion 1}. We note that ${\cal F}_c$ as defined in (3) is a closure system even when $c$ is not idempotent. See [W7, Expansion 1] for details.

{\bf Expansion 2}. As to the algorithmic complexity of calculating $c(\Sigma, S)$, let us merely look at the partial problem of calculating $S'$ from $S$. If $|E| = m$ then it costs time $O(m)$ to check whether or not $A_i \subseteq E$ for some fixed index $i$. Thus for $\Sigma$ as in (5) it costs $O(nm)$ to get $S'$ from $S$ in the ``naive way'' suggested by definition (6). If we think of the premises $A_i$ as the rows of a $n \times m$ matrix $M$ with entries $0$ and $1$, then the naive way amounts to process $M$ row-wise. It isn't hard to see [W1, p.114] how a {\it column-wise} processing of $M$ also yields $S'$. The theoretic cost is the same, i.e. $O(mn) = O(nm)$, but in practise the column-wise way is the better the larger $n/m$. For instance, it takes more time to process a million sets of cardinality 100 (since they need to be ``fetched'' individually) than to process only 100 sets albeit each of cardinality a million. This trick, known as {\it vertical layout} in the Frequent Set Mining community (also observed in [W1]), often works when many but small sets need to be manipulated. In the Relational Database community the algorithm {\it LinClosure} [MR2] to calculate $c(\Sigma, S)$ has become the standard. Whether LinClosure or vertical layout or something else is best, depends on the shape of $\Sigma$ and a smart implementation of vertical layout.

{\bf Expansion 3.} Recall from Boolean logic (or other logic frameworks) that a formula $\psi$ is a ``consequence'' of a formula $\phi$ (written $\phi \vDash \psi$) if every ``structure'' that satisfies $\phi$ also satisfies $\psi$. This is the {\it semantic} level. It contrasts with the {\it syntactic} level where a formula $\psi$ is ``derivable'' from a formula $\phi$ (written $\phi \vdash \psi$) if $\psi$ can be obtained from $\phi$ with certain ``inference rules'' in a step-by-step manner. Two pages of details can be found in [W7, Expansion 3].

{\bf Expansion 4}. One algorithm for enumerating all closed sets, called NextClosure, was devised by B. Ganter in 1984 and became a cornerstone of FCA. Its key idea is to generate the closed sets in lexicographic order. See [GW, Thm.5], from which one also readily deduces the following:

(36) \quad Suppose the closure operator $c: {\cal P}[n] \ra {\cal P}[n]$ is such that calculating $c(X)$ takes time at\\
\hspace*{.9cm} most $T$ for any $X \subseteq [n]$. Then NextClosure enumerates all $N = |{\cal F}_c|$ many closed sets in\\
\hspace*{.9cm} output polynomial time $O(NTn)$.

One benefit of NextClosure is that it doesn't matter in which way the closure operator $c$ is provided. Thus $c$ could be given as $c(U) = \bigcap \{S \in {\cal H}: \ S \supseteq U\}$ where ${\cal H}$ is a $\cap$-generating set of ${\cal F}$ (first way), or $c(U) = c(\Sigma, U)$ where $\Sigma$ is an implication base (second way), or any other way. In fact $c$ itself can be a certain selfmap of ${\cal P}(E)$ more general than a closure operator, see [GR]. 
As to the first way, apart from NextClosure and Dowling's algorithm (2.1.1), many other methods to construct ${\cal F}({\cal H})$ from ${\cal H}$ are evaluated in [KuO1]. As to the second way, it usually cannot compete with the compressed calculation of ${\cal F}(\Sigma)$ in Section 4.4. However, the issue (3.6.3) is often how to find an implication base $\Sigma$ of ${\cal F}$ in the first place. Another popular application of NextClosure is {\it attribute exploration} [GW, p.85]. This particular kind of Query Learning strives to compute the canonical base $\Sigma_{GD}$ of some hidden closure system ${\cal F}$. Unfortunately, as a not always welcome side product, the whole of ${\cal F}$ gets calculated one by one along the way. Impressive strides to avoid this succeed for the kind of ``modern'' attribute exploration proposed in [RDB] and [AN2].

{\bf Expansion 5} A non-independent set is {\it dependent}, and minimal dependent sets are {\it circuits}. This terminology [W3] is motivated by the established use of ``circuit'' for matroids (4.1.4) and convex geometries (4.1.5). Let now $K$ be a circuit of $c$. Since $K$ is dependent there is at least one $e \in K$ with $e \in c(K\backslash \{e\})$. The minimality of $K$ implies that $U : = K\backslash \{e\}$ is a {\it stem} with root $e$. Thus if
$$\mbox{roots}(K) : = \{e \in K: \ e \in c(K \backslash \{e\})\, \},$$
then $|\mbox{roots}(K) | \geq 1$ and each $e \in \, \mbox{roots}(K)$ induces a {\it root-stem-partition} $K = \{e\} \cup U$. Observe that an arbitrary root $e$ with stem $U$ need {\it not} yield a circuit $K = U \cup \{e\}$. For instance, let $c$ be the closure operator induced by the implications $\{1,2\} \ra \{3\}$ and $\{3\} \ra \{2\}$. Then $\{1,2\}$ is a stem for the root $3$ but $\{1,2,3\}$ is no circuit because it contains the proper dependent subset $\{2,3\}$.

{\bf Open Problem 2}: Develop a theory for those closure operators (e.g. their optimum bases), for which each root-stem-partition $U \cup \{e\}$ is a circuit.

Most prominently, matroids and convex geometries belong to this class of closure operators. In the first case each circuit $K$ has roots$(K) = K$, in the second case $|\mbox{roots}(K)| =1$.

{\bf Expansion 6} It is easy to see that neither a properly quasiclosed set $Q$ needs to contain a $\theta$-equivalent stem $X$, nor is a stem $X$ necessarily contained in a $\theta$-equivalent proper quasiclosed set. Nevertheless, those stems $X$ that {\it coincide} with a properly quasiclosed set can be characterized neatly. For starters, since each stem $X$ is independent and a proper subset of an independent set has a strictly smaller closure, we see that:

(37) \quad Each stem which is properly quasiclosed is in fact pseudoclosed.

This raises the problem to grasp the ``pcst-sets'' which by definition are pseudoclosed and a stem (i.e. belong to $\Sigma_{GD}$ {\it and} $\Sigma_{cd}$). If $P$ is pseudoclosed then one can decide whether $P$ is pcst as follows: For all $e \in c(P)\backslash P$ check whether $P$ is {\it minimal} with the property that $e \in c(P)$. No better description of the pcst-sets {\it within the family of all pseudoclosed sets} seems to be known. In contrast, the pcst-sets look neat {\it within the family of all stems}:

\begin{tabular}{|l|} \hline \\
{\bf Theorem 7:} For each stem $X$ of a closure operator $c: \ {\cal P}(E) \ra {\cal P}(E)$ the \\
following properties are equivalent:\\
\\
(i) \ $X$ is pseudoclosed.\\
\\
(ii) $X$ is inclusion-minimal among all stems of $c$.\\
\\
(iii) $X$ is a {\it strong} stem in the sense that roots$(X) = c(X) \backslash X$.\\ \\ \hline \end{tabular}

{\it Proof of Theorem 7.}
As to (i) \ $\Leftrightarrow$ \ (ii), we show that $\neg$(i) \ $\Leftrightarrow \ \neg$(ii), i.e. that
$$X \varsubsetneqq X^\circ \Leftrightarrow Y \varsubsetneqq X \ \mbox{for some stem} \ Y.$$
As to ``$\Ra$'', take $e \in X^\circ \backslash X$. By the definition of $X^\circ$ there is a $Y_o \varsubsetneqq X$ with $e \in c(Y_o) \varsubsetneqq c(X)$. We can shrink $Y_o$ to a stem $Y$ of $e$. As to ``$\Leftarrow$'', because $Y \varsubsetneqq X$ is a stem we can be sure that $c(Y) \backslash Y \neq \emptyset$. If $e \in c(Y) \backslash Y$ then $e \in c(Y) \varsubsetneqq c(X)$, where $\varsubsetneqq$ is due to the independence of $X$. Thus $e \in X^\circ \backslash X$.

As to (i) $\Ra$ (iii), if $Y \varsubsetneqq X$ then again $c(Y) \varsubsetneqq c(X)$ since $X$ (being a stem) is independent. Hence $c(Y) \subseteq X^\bullet = X$. So for {\it each} $e \in c(X) \backslash X$ the set $X$ is minimal w.r.t. the property that its closure captures $e$. As to (iii) $\Ra$ (ii), suppose $Y \varsubsetneqq X$ was a stem, say $Y \in \, \mbox{stems}(e)$. Necessarily $e \in c(X) \backslash X$ since $X$ is independent. But then $e \in \, \mbox{roots}(X)$ by assumption, and so $e \in c(Y)$ is impossible. This contradiction shows that $X$ is inclusion-minimal. \quad $\square$

Theorem 7 draws on [KN]. We changed ``prime stem'' in [KN] to ``strong stem'' in order to avoid confusion with the prime implicates in 3.4.3.

{\bf Expansion 7.} If $f$ is given as a CNF then the well-known {\it consensus method} [CH, 2.7] is applicable to generate all prime implicates of $f$. For instance let $f: \{0,1\}^6 \ra \{0,1\}$ be the conjunction of the four clauses at level $L1$ in Table 2 below (where e.g. $\ol{3} \vee 5$ abbreviates $\ol{x}_3 \vee x_5$). The clauses $C_1= \ol{3} \vee 5$ and $C_2 = \ol{1} \vee 4 \vee \ol{5}$ are such that there is exactly {\it one} literal $x_i$ which appears in one clause and its negation in the other; namely $x_i = x_5$. In this situation we add (while keeping $C_1, C_2$) the {\it consensus} clause $\ol{1} \vee \ol{3} \vee 4$ which is thus obtained by dropping $5$ and $\ol{5}$ from the disjunction $C_1 \vee C_2$. All consensi obtained from level $L1$ are listed in level $L2$. One continues by building consensi between $L1$ and $L2$, and then between $L2$ and $L2$. All of these are listed in $L3$. The list $L1 \cup L2 \cup L3$ is long enough that some of its members get unveiled as redundant; such as $\ol{2} \vee \ol{3} \vee 4 \vee \ol{5}$ which is implied by $\ol{2} \vee \ol{3} \vee 4$. Level $L4$ contains the pruned list. Building consensi within $L4$ (more precisely between the first and second line of $L4$) yields $L5$. Pruning $L4 \cup L5$ yields $L6$.

$\begin{array}{ll}
L1, \ \mbox{start} : & \ol{3} \vee 5, \quad \ol{1} \vee 4 \vee \ol{5}, \quad 3 \vee \ol{6}, \quad 1 \vee \ol{2} \vee \ol{3}\\
\\
L2, \ \mbox{consensus}: & \ol{1} \vee \ol{3} \vee 4, \quad 5 \vee \ol{6}, \quad \ol{2} \vee \ol{3} \vee 4 \vee \ol{5}, \quad 1 \vee \ol{2} \vee \ol{6}\\
\\
L3, \ \mbox{consensus}: &  \ol{2} \vee \ol{3} \vee 4, \quad \ol{1} \vee 4 \vee \ol{6}, \quad \ol{2} \vee 4 \vee \ol{5} \vee \ol{6}, \quad 1 \vee 4 \vee \ol{6}, \quad 2 \vee 4 \vee \ol{5} \vee \ol{6},  \\
& \ol{2} \vee \ol{3} \vee 4; \quad \ol{2} \vee \ol{3} \vee 4 \vee \ol{6}, \quad \ol{2} \vee \ol{3} \vee 4 \vee \ol{6}\\ 
\\
L4, \ \mbox{pruning}: & \ol{3} \vee 5, \quad \ol{1} \vee 4 \vee \ol{5}, \quad 3 \vee \ol{6}, \quad  1 \vee \ol{2} \vee \ol{3}, \quad \ol{1} \vee \ol{3} \vee 4, \quad 5 \vee \ol{6},  \\
& 1 \vee \ol{2} \vee \ol{6}, \quad \ol{2} \vee \ol{3} \vee 4, \quad \ol{1} \vee 4 \vee \ol{6}, \quad \ol{2} \vee 4 \vee \ol{5} \vee \ol{6}\\
\\
L5, \mbox{consensus} : & \ol{2} \vee \ol{3} \vee 4 \vee \ol{6}, \quad \ol{2} \vee \ol{4} \vee \ol{5} \vee \ol{6}, \quad \ol{2} \vee 4 \vee \ol{6}, \quad \ol{2} \vee \ol{3} \vee 4 \vee \ol{6}, \quad \ol{2} \vee 4 \vee \ol{6}\\
\\
L6, \mbox{pruning}: & \ol{3} \vee 5, \quad \ol{1} \vee 4 \vee \ol{5}, \quad 3 \vee \ol{6}, \quad 1 \vee \ol{2} \vee \ol{3}, \quad \ol{1} \vee \ol{3} \vee 4, \quad  5 \vee \ol{6}, \\
 & 1 \vee \ol{2} \vee \ol{6}, \quad \ol{2} \vee \ol{3} \vee 4,\quad \ol{1} \vee 4 \vee \ol{6}, \quad \ol{2} \vee 4 \vee \ol{6}
\end{array}$

Table 2: The consensus  algorithm (simple version)

Now $L6$ yields no new consensi. According to a famous 1959 theorem of Quine [Q] the members in $L6$ hence constitute {\it all} prime implicates of $f(x_1, \cdots, x_6)$. We mention that $L6$ matches $\Sigma_{cd}$ in (27). See [CH, chapter 6.5] for a consensus method working for all Boolean functions $f$, and running in polynomial incremental time in the case of Horn functions $f$. The consensus method can be viewed as a special case of an algorithm [AACFHS] that generates all maximal bicliques ($=$ complete bipartite subgraphs) of a graph $G$. If $G$ itself is bipartite, say with shores $E_1, E_2$ this problem amounts to generate all closed sets of a Galois connection (2.1.2).

{\bf Expansion 8}. We present a novel way for the direction $\Sigma \ra M({\cal F})$. 
Suppose that

(38) \quad $\Sigma := \{\{3\} \ra \{5\}, \quad \{1, 5\} \ra \{4\}, \quad \{6\} \ra \{3\}, \quad \{2, 3\} \ra \{1\}\}.$

Observe that $\Sigma$ is equivalent to $L1$ in Expansion 7 and whence to the family of implications in (27). Hence, if our method is correct, we will wind up with $M({\cal F})$ as in (26). As shown in
Section 4.4 by running the implication $n$-algorithm one can represent ${\cal F} := {\cal F}(\Sigma)$ as a disjoint union of eight $012$-rows, i.e. subcubes of ${\cal P}[6]$, as shown in Table 3. Let us argue that such a representation readily yields $M({\cal F})$ as a side product. 

\begin{tabular}{c|c|c|c|c|c|c|} 
& 1 & 2& 3& 4& 5& 6 \\ \hline
$r'_1=$ & 0 & 2& 0 &2 & 2& 0 \\ \hline
$r'_2 =$ & 1 &2 & 0 & 2 & 0 & 0 \\ \hline
$r'_3=$ & 1 & 2& 0 & 1& 1& 0 \\ \hline
$r'_4=$ & 0 & 0 & 1 & 2 & 1 & 1 \\ \hline
$r'_5 =$ & 1 & 0 & 1 & 1 & 1 &1 \\ \hline
$r'_6=$ & 1 & 1 & 1& 1 & 1& 2 \\ \hline
$r'_7=$ & 0 & 0 & 1 & 2& 1& 0 \\ \hline
$r'_8=$ & 1 & 0 & 1 & 1 & 1 &0 \\ \hline \end{tabular}

Table 3: Getting $M({\cal F})$ by column-wise processing a compressed representation of ${\cal F}$

By (21) it suffices to show how to get $\max ({\cal F}, e)$ for any particular $e \in E = [6]$. Say $e=4$.  If $r'_i$ has its fourth component equal to 1 then $r'_i$ cannot contain a member of $\max({\cal F},4)$. This e.g. happens for $r'_3$. If the fourth component of $r'_i$ is 0 or 2 then at most the unique row-maximal set $\max (r'_i, 4) \in r'_i$ {\it may} belong to $\max ({\cal F}, 4)$. Hence the collection of all maximal row-maximal sets is $\max ({\cal F}, 4)$. Thus
$$\begin{array}{lll}
\max({\cal F}, 4) & =& \max \{\max (r'_1, 4), \max (r'_2, 4), \max (r'_4, 4), \max (r'_7, 4) \} \\
\\
& =& \max \{\{2, 5\}, \{1, 2\}, \{3, 5, 6\}, \{3, 5\}\}\\
\\
& = & \{\{2, 5\}, \{1, 2\}, \{3, 5, 6\} \}. \end{array}$$

Likewise the other collections $\max ({\cal F}, e)$ are obtained, and so we get (matching (26)) that 
$$\begin{array}{lll}
M({\cal F}) & =& \max ({\cal F},1) \cup \cdots \cup \max ({\cal F}, 6)\\
\\
& =& \{245, 3456\} \cup \{13456\} \cup \{1245\} \cup \{25, 12, 356\} \cup \{124\} \cup \{12345\}. \end{array}$$

Let $\max ({\cal F}) = \{H_1, \cdots, H_s\}$ be the set of hyperplanes of $c$. Obviously the minimal keys of $E$ are exactly the minimal transverals of ${\cal H} = \{E \backslash H_1, \cdots, E \backslash H_s\}$, and so any good algorithm for mtr$({\cal H})$ yields them, provided the hyperplanes are known. In particular, the $H_i$'s can be gleaned from a table like Table 3 since $\max ({\cal F}) \subseteq M({\cal F})$.

{\bf Expansion 9.} Here we present another view of Table 3 in Expansion 8. But first we need to dualize some concepts from 3.4.1. 
Thus a conjunction of literals is called a {\it term}. The model set of a term $T$, viewed as a Boolean function $T: \{0,1\}^n \ra \{0,1\}^n$, is an interval in the lattice $\{0,1\}^n = {\cal P}[n]$. (It is also common, although less precise, to speak of ``subcubes'' instead of intervals.) For instance if $T$ is $x_1 \wedge \ol{x}_3 \wedge \ol{x}_5 \wedge \ol{x}_6$ then Mod$(T) = (1,2, 0, 2, 0,0)$. This $012$-{\it row} is a succinct notation for the interval  $\{U\subseteq {\cal P}[6] : \{1\} \subseteq U \subseteq \{1,2,4\}\}$.  A {\it disjunctive normal form} (DNF) is any disjunction of terms.

Now back to Table 3. The pure Horn function matching $\Sigma$ in (38) is
$$f(x_1, \cdots, x_6) = (\ol{x}_3 \vee x_5) \wedge (\ol{x}_1 \vee \ol{x}_5 \vee x_4) \wedge (\ol{x}_6 \vee x_3) \wedge (\ol{x}_2 \vee \ol{x}_3 \vee x_1).$$
We aim to convert this CNF into a DNF. Because Mod$(f) = {\cal F}(\Sigma)$ is represented as the union of the $012$-rows $r'_i$ in Table 3, and because $r'_i = \, \mbox{Mod}(T_i)$ for obvious terms $T_i$, one DNF for $f(x_1, \cdots, x_6)$ is
$$T_1 \vee \cdots \vee T_8 : = (\ol{x}_1 \wedge \ol{x}_3 \wedge \ol{x}_6) \vee (x_1 \wedge \ol{x}_3 \wedge \ol{x}_5 \wedge \ol{x}_6) \vee \cdots \vee (x_1 \wedge \ol{x}_2 \wedge x_3 \wedge x_4 \wedge x_5 \wedge \ol{x}_6).$$
The above DNF is {\it orthogonal} [CH, chapter 7] in the sense that Mod$(T_i) \cap \, \mbox{Mod}(T_j)=\emptyset$ for $i \neq j$. It would be interesting to know how to exploit the orthogonality of a DNF in a (dual) consensus method.

{\bf Expansion 10}. {\it Proof of (22)}. As to $\subseteq$, from $X \in \max ({\cal F}, e)$ follows that $X$ is maximal within ${\cal F}$ w.r.to $e \not\in X$. A fortiori $X$ is maximal within ${\cal H} \subseteq {\cal F}$ w.r.to $e \not\in X$, {\it provided} $X$ belongs to ${\cal H}$ at all. But this follows from (21) and $M({\cal F}) \subseteq {\cal H}$. As to $\supseteq$, let $X \in {\cal H}$ be maximal w.r.to $e \not\in X$. Then there is $Y \in {\cal F}$ which is maximal w.r.to $Y \supseteq X$ and $e \not\in Y$. Hence $Y \in \max ({\cal F}, e)$ by definition of the latter, and so $Y \in M({\cal F}) \subseteq {\cal H}$ by (21). By the maximality property of $X$, we have $X = Y \in \max ({\cal F}, e)$.

{\bf Expansion 11}. As to going from $\Sigma_{cd}$ (or in fact from any base) to a minimum base $\Sigma_0$, we illustrate the method of Shock [Sh], which first demands to replace, for each $A \ra B$ in $\Sigma_{cd}$, the conclusion $B$ by $c(B)$ where $c$ is the closure operator induced by $\Sigma_{cd}$. For $\Sigma_{cd}$ in (27) we get an equivalent family of full implications
$$\Sigma^\ast_{cd} = \{13 \ra 1345, \ 16 \ra 16435, \ 23 \ra 23145, \ 26 \ra 261435, \ 15 \ra 154, \ 6 \ra 635,\ 3 \ra 35 \}.$$
Recall from (8) that $(A \ra B) \in \Sigma^\ast_{cd}$ is redundant iff $B$ is contained in the $(\Sigma^\ast_{cd} \backslash \{A \ra B\})$-closure of $A$. Incidentally $\Sigma (A)$, as defined before (12), is $\{A \ra B\}$ for all$(A \ra B) \in \Sigma^\ast_{cd}$, and so the $(\Sigma^\ast_{cd} \backslash \{A \ra B\})$-closure of $A$ is $A^\bullet$ by (12). Because of $1345 \subseteq 13^\bullet = 1354$ we can thus drop $13 \ra 1345$ from $\Sigma^\ast_{cd}$. Further $16 \ra 16435$ can be dropped because of $16435 \subseteq 16^\bullet = 16354$, and $26 \ra 261435$ can be dropped because of $261435 \subseteq 26^\bullet = 263514$. The resulting base
$$\Sigma_0 = \{23 \ra 23145, \ 15 \ra 154, \ 6 \ra 635, \ 3 \ra 35\}$$
is nonredundant and whence minimum by Theorem 1(d). The kind of minimum base $\Sigma_0$ obtained by Shock can by Theorem 1 easily be ``blown up'' to $\Sigma_{GD}$.

{\bf Expansion 12}.  In [W7, Expansion 13] it is shown how $\max ({\cal F}, e)$ relates to lattice theory, in particular to the relations $\uparrow, \downarrow, \updownarrow$ which originated in [D1] and are akin to the ones in [GW, p.31]. Coupled to each lattice ${\cal L}$ there is an importatn bipartite graph with shores $J({\cal L})$ and $M({\cal L})$.

{\bf Expansion 13}.  In [W7, Expansion 14] we show the well known fact [CM] that the collection ${\cal C}$ of all closure systems ${\cal F} \subseteq {\cal P}(E)$ is itself a closure system, in fact (viewed as a lattice) it is meet-distributive. Furthermore the technical proof of property (39) in Expansion 20 features there.

{\bf Expansion 14.} For any closure operator $c : {\cal P}(E) \ra {\cal P}(E)$ consider these properties:

\begin{tabbing}
123456\=\kill
$(T 0)$ \> $(\forall p, q \in E) \ \ (p \neq q \ \Ra \ c(\{p\}) \neq c(\{q\}))$\\
\\
$(T \frac{1}{2})$ \> $(\forall p \in E) \ \ \ c(\{p\}) \backslash \{p\}$ \ is closed\\
\\
$(T 1)$ \> $(\forall p \in E) \ \ \ c(\{p\}) = \{p\}$\\ 
\end{tabbing}

The properties $(T 0)$ and $(T 1)$ are well known ``separation axioms'' from topology. For instance ${\cal F}$ in Figure 4(a) violates $(T0)$.
The notation $(T \frac{1}{2})$ stems from [W5] but the property was previously considered. All three axioms make sense for non-topological operators $c$. It is an exercise to verify $(T 1) \Ra (T \frac{1}{2}) \Ra (T 0)$; furthermore $c(\emptyset) = \emptyset$ when $(T \frac{1}{2})$ holds. In fact, as shown in [W5, Thm.8], $c$ is isomorphic to a standard operator $c_J$ as in (30) iff $c$ satisfies $(T \frac{1}{2})$. It is easy to ``boil down'' any closure operator $c$ on a set $E$ to an operator $\ol{c}$ of type $(T 0)$ on a smaller set $\ol{E}$, and $\ol{c}$ to $c_J$ of type $(T \frac{1}{2})$ on a still smaller set $J$, in such a way that the lattices ${\cal F}_c$ and ${\cal F}_J$ are isomorphic.
See [W5, p.165] or [GW, ch.1.1, 1.2] for details.  
Albeit the lattices ${\cal F}_c$ and ${\cal F}_J$ are isomorphic, this may be of little help to get a good base of $c$ from one of $c_J$. For instance it takes some effort to find an optimum base for the closure system ${\cal F} ={\cal F}_c$ in Figure 4(a). In contrast ${\cal F}_J$ is a Boolean lattice and whence has the empty set as an optimum base! (See also Open Problem 4 in Expansion 15.)

{\bf Expansion 15}. Recall from (29) that $x \mapsto {\cal J}(x)$ is a lattice isomorphism from ${\cal L}$ onto ${\cal F}_J  = \{{\cal J}(x) : x \in {\cal L}\}$ and that ${\cal J}(x\wedge y) = {\cal J}(x) \cap {\cal J}(y)$ but usually ${\cal J}(x \vee y) \varsupsetneqq {\cal J}(x) \cup {\cal J}(y)$. To see that ``$=$'' takes place in the distributive case, fix $p \in {\cal J}(x \vee y)$. Then $p \leq x \vee y$ and distributivity imply that $p = p\wedge (x \vee y) = (p \wedge x) \vee (p \wedge y)$. Since $p$ is join irreducible this forces $p = p \wedge x$ or $p =p \wedge y$, whence $p \leq x$ or $p \leq y$, whence $p \in {\cal J}(x) \cup {\cal J}(y)$. Hence ${\cal F}_J$ is a sublattice of $({\cal P}(J), \cap , \cup)$. Consequently the closure operator $c_J$ from (30) is topological, in fact $c_J(\{p_1, \cdots, p_t\}) = J(p_1) \cup \cdots \cup J(p_t)$. 
Therefore ${\cal F}_J$ is the lattice ${\cal L}({\cal J}, \leq)$ of all order ideals of the poset $({\cal J}, \leq)$.
In particular, since ${\cal L} \simeq {\cal F}_J$ by (29), we have ${\cal L} \simeq {\cal L}(J, \leq)$. This is Birkhoff's Theorem, see [Bi, p.59]. 

As to implicational bases, for any lattices ${\cal L} \simeq {\cal F}_J$ it is natural to consider the set  
of implications
$$\Sigma_J : = \{\{p\} \ra \ell cov(p): \ p \in J^\ast \},$$
where $\ell cov(p)$ is the set of lower covers of $p$ within $(J, \leq)$ and $J^\ast$ is the set of all non-minimal members of $(J, \leq)$. It is clear that ${\cal F}(\Sigma_J)$ is the collection of all order ideals of $(J, \leq)$. Hence $\Sigma_J$ is a base of ${\cal L}$ iff ${\cal L}$ is distributive.  Actually $\Sigma_J$ is the unique {\it optimum} base for each distributive lattice ${\cal L}$. That follows immediately from 3.2.2 (all circle formations are points here). Note that $\Sigma_J = \emptyset$ when ${\cal L} \simeq {\cal P}(J)$ is Boolean. For nondistributive lattices $\Sigma_J$ may constitute a relevant part of larger bases. Most prominently, according to 4.1.3 each optimum base of a modular lattice includes $\Sigma_J$.  On the downside, $\Sigma_J$ needs {\it not} be part of every optimum base of a lattice. For instance the lattice ${\cal L}_0$ in Figure 11 has $\Sigma_J = \{\top \ra 23, 2 \ra 4\}$ whereas one optimum base of ${\cal L}_0$ is $\{\top \ra 34, 2 \ra 4, 34 \ra 2\}$.

{\bf Open Problem 3}: Determine the class ${\cal K}$ lattices ${\cal L}$ (among which all modular ones) for which $\Sigma_J$ in Expansion 15 is part of every optimum base of ${\cal L}$.

\begin{center}
\includegraphics[scale=0.7]{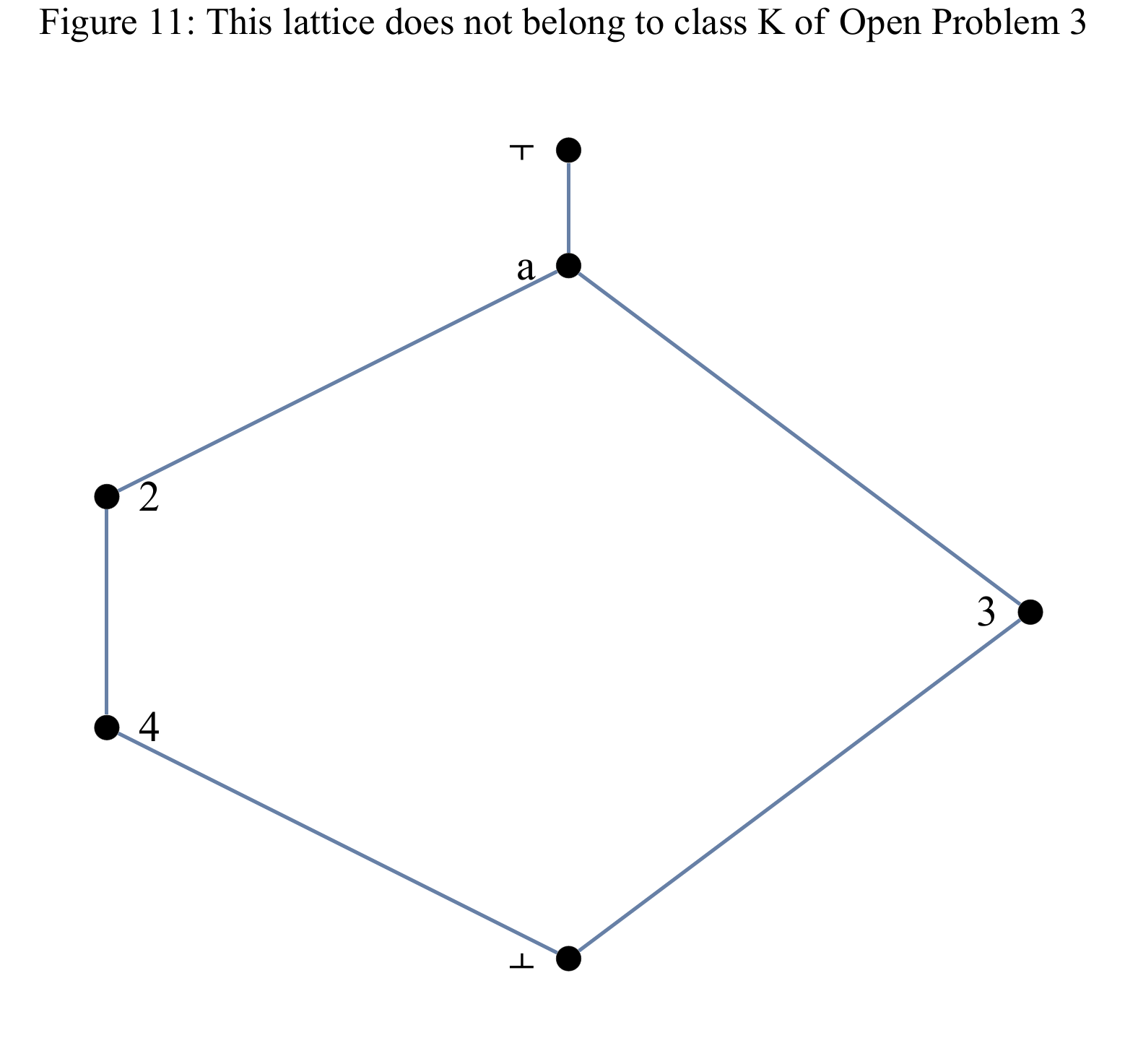}
\end{center}

As seen above, for topological operators $c$ the lattice ${\cal F}_c$ is a sublattice of ${\cal P}(E)$, and whence distributive. However as seen in 3.1, ${\cal F}_c$ can be distributive without being a sublattice of ${\cal P}(E)$.

{\bf Open Problem 4:} Let $c : {\cal P}(E) \ra {\cal P}(E)$ have a distributive lattice ${\cal F}_c$ which is {\it not} a sublattice of ${\cal P}(E)$. Can one find an optimum base of $c$ in polynomial time?

{\bf Expansion 16.} We start by proving (33) in 4.1.5. So let $(K, e)$ be critical, i.e. $c(K)\backslash \{e, x\}$ is closed for all $x \in K \backslash \{e\}$. In order to show that $S: = c(K) \backslash \{e\}$ is quasiclosed\footnote{Notice that when $S$ is quasiclosed then it is properly quasiclosed since $c(S) =c(K) \neq S$.} we take (in view of (9)) $U \subseteq S$ with $c(U) \neq c(S)$ and aim to show that $c(U) \subseteq S$. There must be an $x \in K\backslash \{e\}$ with $x \not\in U$ (otherwise $K \backslash \{e\} \subseteq U$ yields the contradiction $c(U) = c(K)$). But then $U \subseteq c(K) \backslash \{e, x\}$, and so by assumption $c(U) \subseteq c(K) \backslash \{e, x\} \subseteq S$.

Next, assuming $S = c(K) \backslash \{e\}$ is quasiclosed, we show that $K \backslash \{e\}$ is a closure-minimal stem of $e$ in the sense of Expansion 6. Suppose to the contrary there was a stem $U$ of $e$ with $c(U) \varsubsetneqq c(K \backslash \{e\}) = c(K)$. From $U \subseteq S$ and $c(U) \neq c(S)$ follows (since $S^\bullet = S$) that $c(U) \subseteq S$. This is impossible since $e \in c(U)$ (by the definition of stem).

 Finally, letting $K \backslash \{e\}$ be closure-minimal, assume by way of contradiction that $Y : = c(K) \backslash \{e, x\}$ is not closed for some $x \in K \backslash \{e\}$. First, $c(K) \backslash \{x\} = Y\cup \{e\}$ is closed because $x \in ex(c(K)) = ex (K)$. Hence $c(Y) = Y \cup \{e\}$, and so there is a stem $U \subseteq Y$ of $e$. It satisfies $c(U) \subseteq Y \cup \{e\} \varsubsetneqq c(K) = c(K \backslash \{e\})$, and thus $K \backslash \{e\}$ is not closure-minimal.  This proves (33). \quad $\square$
 
 Yet another (equivalent) definition of ``critical'' is given in [W3, p.136]. Furthermore $(K, e)$ is called {\it extra-critical} in [W3] if the quasiclosed set $c(K) \backslash \{e\}$ in 
 (33) coincides with $(K \backslash \{e\})^\bullet$.
  
  If $c: {\cal P}(E) \ra {\cal P}(E)$ is a convex geometry, then the set system ${\cal A}_c : = \{E \setminus X: X \in {\cal F}_c\}$ is a so called {\it antimatroid}. One can define antimatroids independent of $c$ as union-closed set systems ${\cal A} \subseteq {\cal P}(E)$ which are hereditary in the sense that for each nonempty $A \in {\cal A}$ there is some $x \in A$ with $A \setminus \{x\} \in {\cal A}$. What we defined as a rooted circuit $(K, e)$ in 4.1.5 relates as follows to ${\cal A}_c$: Whenever $e \in A \in {\cal A}$ then $(K \setminus \{e\}) \cap A \neq \emptyset$; and $K$ is minimal with this property. In fact this is the {\it original} definition of a rooted circuit [KLS, p.28]. Apart from rooted circuits our definition of a critical circuit $(K, e)$ in 4.1.5 similarly matches the definition given in [KLS, p.31]. Each antimatroid ${\cal A}$ can (apart from the set system view) equivalently be rendered as a certain {\it hereditary language}. From this perspective the critical circuits provide an optimal representation of ${\cal A}$, see [KLS, Thm.3.11]. This contrasts with the fact that $\Sigma_{crci}$ usually is {\it no} implicational base (see 4.1.5). Antimatroids and convex geometries arise in many contexts, often related to combinatorial optimization, see [KLS, III.2].

A lattice ${\cal L}$ is {\it meet-distributive} if the interval $[x_\ast, x] \subseteq {\cal L}$ is Boolean for all $x \in {\cal L} \backslash \{\bot\}$. (Many equivalent characterizations exist.) Every convex geometry $c$ has a meet-distributive lattice ${\cal F}_c$. Conversely, if ${\cal L}$ is meet-distributive then $c_J$ is a convex geometry. 
 The dual concept of meet-distributivity is {\it join-distributivity}, i.e. when 
$[x,x^\ast]$ is a Boolean interval for all $x \in {\cal L} \backslash \{\top\}$.  A lattice which is both meet and join-distributive must be distributive, and conversely.

A lattice ${\cal L}$ is {\it join-semidistributive} $(SD_\vee)$ if for all $x, y, z \in {\cal L}$ it follows from $x \vee y = x \vee z$ that $x \vee z = x \vee (y \wedge z)$. 
In such a lattice $|J({\cal L})| \leq |M({\cal L})|$. See also [W7, Expansion 13].  Notice that ``meet-distributive $\Ra$ join-semidistributive''. In fact, the $SD_\vee$ lattices ${\cal L}$ of length $d({\cal L}) = 
|J({\cal L})|$ are exactly the meet-distributive lattices. If ${\cal L}$ is $SD_\vee$ then by [AN1, prop.49] every essential set $X$ of $c_J$ has a {\it unique} quasi-closed generating set $Q$ (which equals $ex(X)$ in the meet distributive case). Conversely such a {\it unique-criticals} lattice need not be $SD_\vee$. See Figure 12.   Further topics in [AN1] include the uniqueness of the $K$-basis (see 4.1.1) for $SD_\vee$ standard closure systems ${\cal L}$, and the fact that such ${\cal L}$ generally don't belong to the class ${\cal K}$ in Open Problem 3 of Expansion 15. Dually to $SD_\vee$ one defines {\it meet-semidistributivity} $(SD_\wedge)$.  It comes as no surprise that ``join-distributive $\Ra$ 
meet-semidistributive''. Results about bases of $SD_\wedge$-lattices are given in [JN], and exploited in [W4].  See also Expansion 18.

{\bf Expansion 17.} As a variation of Theorem 5, $\vee$-semilattices (in particular lattices ${\cal L}$) can also be described as systems of {\it restricted} order ideals of a poset. This generalizes the representation of  distributive lattices, for which {\it all} order ideals are used (Expansion 15). The restriction imposed on the order ideals is governed by core$({\cal L}) : = \, \mbox{core}(c_J)$ where $c_J$ is as in (30) and core$(c)$ as in (14). We mention that in [D] core$({\cal L})$ is determined for many types of lattices ${\cal L}$. Notice that $|\Sigma_{GD}| \geqq |\mbox{core}({\cal L})|$ and that from core$({\cal L})$ alone one cannot obtain $\Sigma_{GD}$. See [W7, Expansion 18] for more details.

{\bf Expansion 18.} The $D$-relation, which is of importance in the study of free lattices, is defined as follows. For $p, q \in J({\cal L})$ put $pDq$ if $q$ appears in some order-minimal join cover $A$ of $p$. A {\it $D$-cycle} is a configuration of type $p_1Dp_2D \cdots p_n Dp_1$. For instance the convex geometry in 2.2.5 has the $D$-cycle $6D 8D6$ because 146 is a minimal join cover of 8 and 238 is a minimal join cover of 6.
Each $D$-cycle induces a cycle in $G(\Sigma)$ for each base $\Sigma$ of $c_J$, but not conversely. Hence closure operators without $D$-cycles are strictly more general than acyclic operators.
Indeed, the former have $SD_\vee$ closure systems by [FJN], the latter meet-distributive ones by Theorem 3 (see also Expansion 16). While the presence of $D$-cycles can be decided from $\Sigma_{GD}$ in polynomial time [AN1, Thm.43], this is unknown for checking $SD_\vee$.

Likewise the {\it affine} convex geometries (as 2.2.5 but in $\R^n$, not just $\R^2$) can be generalized, i.e. to convex geometries satisfying  the so-called $n$-{\it Carousel Property}. This property was crucial in article [AW] that dealt with the realizability (in $\R^2$) of convex geometries. Implication bases of convex geometries with the $n$-Carousel Property can  be optimized in polynomial time [A, Thm.12], but the arguments get 	``uglier'' than the deliberations in 2.2.5. Notice that checking the $n$-Carousel property ($n$ fixed), as opposed to checking realizability, is ``straightforward'' albeit tedious. Furthermore, optimization of implication bases of order-convex\footnote{By definition the closed sets of an {\it order-convex} geometry are all intervals of some poset.} geometries is polynomial-time [A, sec.6].

\begin{center}
\includegraphics{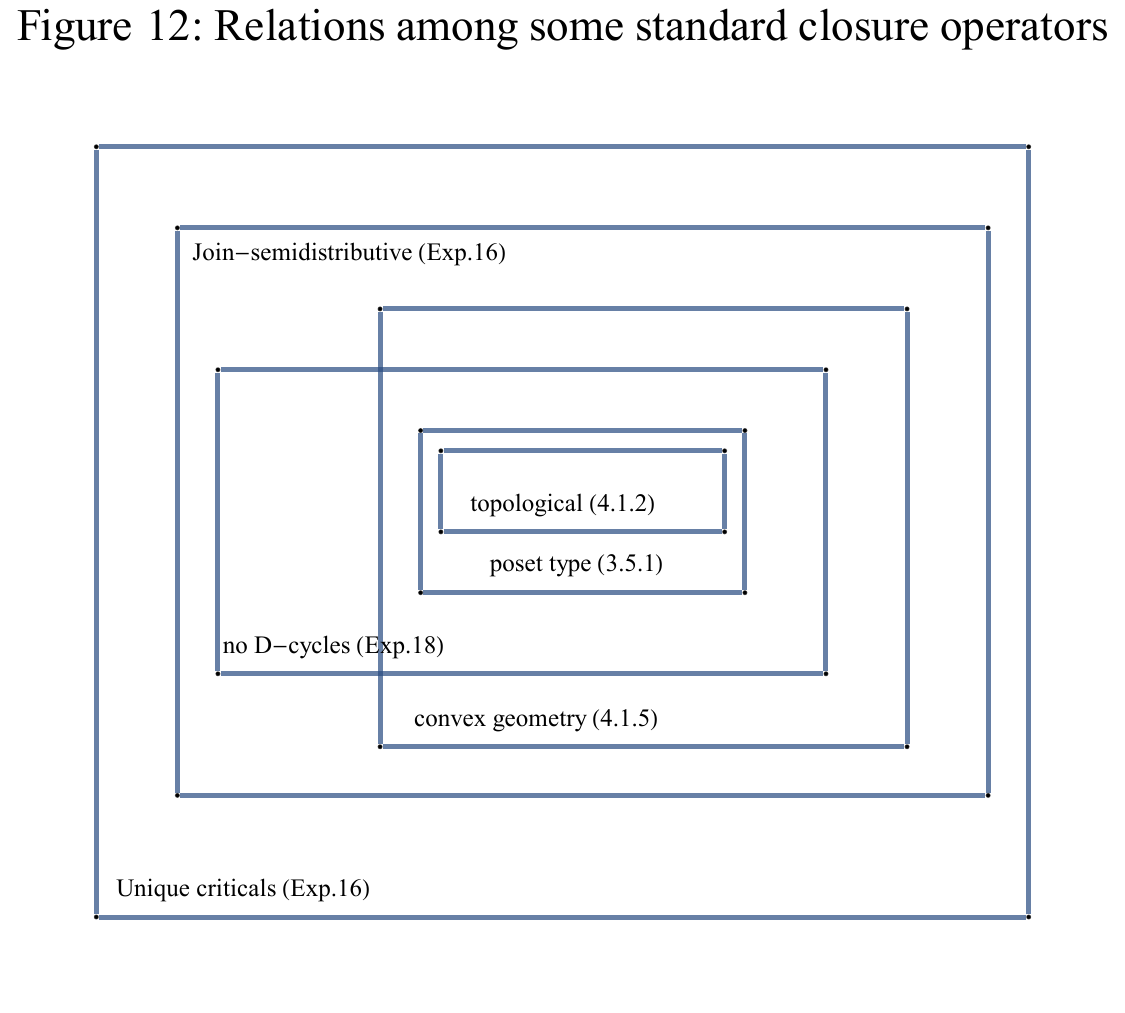}
\end{center}
 
{\bf Expansion 19}. It is easy to replace a $012n$-row by a couple of disjoint $012$-rows. For instance $(n,n,n)$ is the same as $(0,2,2) \cup (1,0,2) \cup (1,1,0)$. 
Sometimes $012$-rows are easier to handle, if only for pedagogical reasons as in Table 3 of Expansion 8. Conversely, a random collection of $012$-rows usually cannot be compressed to fewer $012n$-rows. As seen in 4.4 the $n$-algorithm produces its rows ``from scratch'' without an intermediate state of $012$-rows.
Further fine-tuning is possible. For instance, instead of replacing $r_8$ by $r_{11} \cup r_{12}$ in Table 1 we could have replaced it by the single row $(1, 2, b, 1, 1, a)$ where generally the wildcard $abb\cdots b$ signifies that either ${\bf 0}22\cdots 2$ or ${\bf 1}11\cdots 1$ must take place. 
The author exploited this idea in the special case where all $(A\ra B) \in \Sigma$ are of type $\{a\} \ra B$ in the first place; this essentially amounts to enumerating all order ideals of a poset. 
In a similar manner all anticliques ($=$ independent vertex sets) of a graph can be enumerated in a compact manner (work in progress).

{\bf Expansion 20.} Let us sketch how to (a) get a $ca$-minimum base of a Horn function $h$, and (b) how to merely calculate $ca(h)$.

As to (a), it relates to [W7, Expansion 14] where we showed that for any $\cap$-subsemilattice ${\cal S} \subseteq {\cal P}(E)$ the collection

(39) \quad ${\cal C}({\cal S}): = \{{\cal F} \subseteq {\cal P}(E) \ \mbox{closure system}\ | \ \ (\exists \ \mbox{set ideal} \ {\cal G} \subseteq {\cal P}(E))\ {\cal F} \cap {\cal G} = {\cal S} \}$

is a sublattice of the lattice ${\cal C}$ of all closure systems on $E$. Clearly $\bot = {\cal S} \cup \{E\}$ is the smallest element of ${\cal C}({\cal S})$. Let $f_\top$ be the pure Horn function matching the {\it largest} element $\top$ of ${\cal C}({\cal S})$. Albeit $\top$ as a subset of ${\cal P}(E)$ cannot be described as simply as $\bot$, it is shown in [HK, Lemma 4.2] that $f_\top$ must be the conjunction of all {\it pure} prime implicates of $h$. Once calculated (consensus method), this {\it pure Horn part} $f_\top$ of $h$ can be used as follows to minimize $h$. Compute all {\it negative} prime implicates ($=$ complications) $A_1, A_2, \cdots$ of $h$. Take them as the vertices of a graph $G(h)$ which has an arc from $A_i$ to $A_j$ iff $A_j$ is a consequence of $A_i \wedge f_\top$. Let $P_1, \cdots, P_s$ be the strong components of $G(h)$ that have in-degree $0$ when viewed as elements of the induced factor poset. Now let $\Gamma$ be any transversal of $\{P_1, \cdots, P_s\}$ and let $\Sigma_\top$ be any minimum base of $f_\top$. Then $\Sigma_\top \cup \Gamma$ is a minimum base of $h$ [HK,Theorem 6.2]. 

As to (b), up to duality in [CH, 6.7.3] one associates with an impure Horn function $h$ in $n$ variables a {\it pure} Horn function $h'$ in $n+1$ variables as follows. Take any base $\Sigma \cup \Gamma$ of $h$ and let $h'$ be the function induced by $\Sigma \cup \Sigma^\ast$ where $\Sigma^\ast := \{A \ra \{x_{n+1} \}: A \in \Gamma \}$. According to [CH, Lemma 6.8, Thm.6.15] this is well-defined, i.e. independent of the chosen base $\Sigma \cup \Gamma$ of $h$. Furthermore $ca(h') = ca (h)$. The intricasies of proving $ca(h') =ca(h)$ are not mirrored on the algorithmic side: Switching from $\Sigma \cup \Gamma$ to $\Sigma \cup \Sigma^\ast$ is trivial, and minimizing $\Sigma \cup \Sigma^\ast$ to $\Sigma_0$ works in quadratic time (Expansion 11) and yields $ca(h) = ca(h') = |\Sigma_0|$.

\section*{Acknowledgement:} I am grateful for comments from  Kira Adaricheva, Roni Khardon, Sergei Kuznetsov, Jos\'{e} Balc\'{a}zar, Ron Fagin,  Gert Stumme,  Sergei Obiedkov, Sebastian Rudolph, Hiroshi Hirai, Giorgio Ausiello, Bernard Monjardet.

\section*{References}
\begin{enumerate}
\item[{[A]}] K. Adaricheva, Optimum basis of finite convex geometry, to appear in Disc. Appl. Mathematics.
	\item[{[AN1]}] K. Adaricheva, J.B. Nation, On implicational basis of closure systems with unique critical sets, Appl. Math. 162 (2014) 51-69.
\item[{[AN2]}] K. Adarichva,  J.B. Nation, Discovery of the $D$-basis in binary tables based on hypergraph dualization, arXiv:1504.02875v2.
	\item[{[ANR]}] K. Adaricheva, J.B. Nation, R. Rand, Ordered direct implicational basis of a finite closure system, Discrete Appl. Math. 161 (2013) 707-723.
\item[{[AACFHS]}] G. Alexe, S. Alexe, Y. Crama, S. Foldes, P.L. Hammer, B. Simeone, Consensus algorithms for the generation of all maximal bicliques, Disc. Appl. Math. 145 (2004) 11-21.
	\item [{[AB]}] M. Arias, J.L. Balcazar, Canonical Horn representations and Query Learning, Lecture Notes in Computer Science 5809 (2009) 156-170.
	\item[{[ADS]}] G. Ausiello, A. D'Atri, D. Sacca, Minimal representation of directed hypergraphs, SIAM J. Comput. 15 (1986) 418-431.
\item[{[AW]}] K. Adaricheva, M. Wild, Realization of abstract convex geometries by point configurations, Europ. J. Comb. 31 (2010) 379-400.
\item[{[B]}] J. L. Balc\'{a}zar, Redundancy, deduction schemes and minimum-size bases for association rules, Logical Methods in Computer Science 6 (2010) 1 - 33.
\item[{[Bi]}] G. Birkhoff, Lattice Theory, AMS 1984.
\item[{[BCKK]}] E. Boros, O. Cepek, A. Kogan, P. Kucera, A subclass of Horn CNFs optimally compressible in polynomial time, Annals Math. Artif. Intelligence (2009) 249-291.
\item[{[BDVG]}] K. Bertet, C. Demko, J.F. Viaud, C. Gu\'{e}rin, Lattices, closure systems and implication bases: a survey of structural aspects and algorithms, arXiv.
	\item[{[BG]}] E. Boros, A. Gruber, Hardness results for approximate pure Horn CNF Formulae minimization, Ann. Math. Artif. Intell. 71 (2014) 327-363.
\item[{[BK]}] M.A. Babin, S.O. Kuznetsov, Computing premises of a minimal cover of functional dependencies is intractable, Disc. Applied Math. 161 (2013) 742-749.
	\item[{[BM]}] K. Bertet, B. Monjardet, The multiple facets of the canonical direct unit implicational basis, Theoretical Computer Science 411 (2010) 2155-2166.
\item[{[BMN]}] L. Beaudou, A. Mary, L. Nourine, Algorithms for $k$-meet semidistributive lattices, arXiv.
\item[{[Bu]}] L.M. Butler, Subgroup lattices and symmetric functions, Memoirs AMS 539 (1994).
\item[{[C]}] N. Caspard, A characterization theorem for the canonical basis of a closure operator, Order 16 (1999) 227-230.
\item[{[CH]}] Y. Crama, P.L. Hammer, Boolean Functions, Encyc. of Math. and Appl. 142, Cambridge Univ. Press 2011.
\item[{[CM]}] N. Caspard, B. Monjardet, The lattices of closure systems, closure operators, and implicational systems on a finite set: a survey. Discrete Applied Mathematics 127 (2003) 241-269.
\item[{[D]}] V. Duquenne, The core of finite lattices, Discrete Mathematics 88 (1991) 133-147.
\item[{[D1]}] A. Day, Characterization of finite lattices that are bounded-homomorphic images or sublattices of free lattices, Can. J. Math. 31 (1979) 69-78.
\item[{[D2]}] A. Day, The lattice theory of functional dependencies and normal decompositions, International Journal of Algebra and Computation 2 (1992) 409-431.
\item[{[DHO]}] P.O. Degens, H.J. Hermes, O. Opitz (eds), Die Klassifikation und ihr Umfeld, Indeks Verlag, Frankfurt 1986.
\item[{[DS]}] F. Distel, B. Sertkaya, On the complexity of enumerating pseudo-intents, Disc. Appl. Math. 159 (2011) 450-466.
\item[{[EMG]}] T. Eiter, K. Makino, G. Gottlob, Computational aspects of monotone dualization: A brief survey, Discrete Appl. Math. 156 (2008) 2035-2049.
\item[{[F]}] R. Fagin, Functional dependencies in a relational data-base and propositional logic, IBM. J. Res. Develop. 21 (1977) 534-544. (Cited in [W7].)
\item[{[FV]}] R. Fagin, M.Y. Vardi, The theory of database dependencies - a survey. Mathematics of Information Processing, Proceedings of Symposia in Applied Mathematics 34 (1986) 19-71. (Cited in [W7].)
\item[{[FD]}] J.C. Falmagne, J.P. Doignon, Learning Spaces, Springer-Verlag Berlin Heidelberg 2011.
\item[{[FJN]}] R. Freese, J. Jezek, J.B. Nation, Free lattices, Math. Surveys and Monographs 42, Amer. Math. Soc. 1995.
\item[{[G]}] G. Gr\"{a}tzer, Lattice Theory: Foundation, Birkh\"{a}user 2011.
\item[{[GD]}] J.L. Guigues, V. Duquenne, Familles minimales d'implications informatives r\'{e}sultant d'une table de donn\'{e}es binaires, Math. Sci. Hum. 95 (1986) 5-18.
\item[{[GW]}] B. Ganter, R. Wille, Formal Concept Analysis, Springer 1999.
	\item[{[GR]}] B. Ganter, K. Reuter, Finding all closed sets: A general approach, Order 8 (1991) 283-290.
\item[{[HK]}] P.L. Hammer, A. Kogan, Quasi-acyclic propositional Horn knowledge bases: Optimal compression, IEEE Trans. on knowledge and data engineering 7 (1995) 751-762.
\item[{[JN]}] P. Jansen, L. Nourine, Minimum implicational bases for $\wedge$-semidistributive lattices, Inf. Proc. Letters 99 (2006) 199-202.
\item[{[KKS]}] H. Kautz, M. Kearns, B. Selman, Horn approximations of empirical data, Artificial Intelligence 74 (1995) 129-145.
		\item[{[K]}] R. Khardon, Translating between Horn Representations and their characteristic models, Journal of Artificial Intelligence Research 3 (1995) 349-372.
\item[{[KLS]}] B. Korte, L. Lova\'{a}sz, R. Schrader, Greedoids, Springer-Verlag 1991.
\item[{[KN]}] K. Kashiwabara, M. Nakamura, The prime stems of rooted circuits of closure spaces, The electronic journal of combinatorics 20 (2013), Paper 22, 13 pages.
\item[{[KR]}] R. Khardon, D. Roth, Reasoning with models, Artificial Intelligence 87 (1996) 187-213.
\item[{[KuO1]}] S.O. Kuznetsov, S. Obiedkov, Comparing performance of algorithms for generating concept lattices, J. Expt. Theor. Art. Intelligence 14 (2002) 189-216.
\item[{[KuO2]}] S.O. Kuznetsov, S.A. Obiedkov, Some Decision and Counting Problems of the Duquenne-Guigues Basis of Implications. Discrete Applied Mathematics 156 (2008) 1994-2003.
\item[{[M]}] D. Maier, The Theory of Relational Databases, Computer Science Press 1983.
\item[{[Ma]}] D. Marker, Model Theory: An Introduction, Springer Verlag 2002. (Cited in [W7].)
\item[{[MR1]}] H. Mannila, K-J. R\"{a}ih\"{a}, Design by example: An application of Armstrong Relations, Journal of Computer and System Sciences 33 (1986) 126-141.	
		\item[{[MR2]}] H. Mannila, K.J. R\"{a}ih\"{a}, The design of relational databases, Addison-Wesley 1992.
\item[{[MU]}] K.Murakami, T. Uno, Efficient algorithms for dualizing large scale hypergraphs, Disc. Appl. Math. 170 (2014) 83-94.
\item[{[N]}] J.B. Nation, An approach to lattice varieties of finite height, Algebra Universalis 27 (1990) 521-543.
\item[{[P]}] J. Paredaens, About functional dependencies in a database structure and their coverings, Philips MBLE Lab. Report 342, Brussels 1977.
\item[{[PKID1]}] J. Poelmans, S.O. Kuznetsov, D.I. Ignatov, G. Dedene, Formal Concept Analysis in Knowledge Processing: A survey on models and techniques.
\item[{[PKID2]}] J. Poelmans, S.O. Kuznetsov, D.I. Ignatov, G. Dedene, Formal Concept Analysis in Knowledge Processing: A survey on applications, Expert Systems with Applications 40 (2013) 6538-6560.
\item[{[Q]}] WV. Quine, On cores and prime implicants of truth functions, Amer. Math. Monthly 66 (1959) 755-760.
\item[{[R]}] S. Rudolph, Succinctness and tractability of closure operator representations, arXiv.
\item[{[RCEM]}] E. Rodriguez-Lorenzo, P. Cordero, M. Enciso, A. Mora, A logical approach for direct-optimal basis of implications, Bull. Eur. Assoc. Theor. Comp. Sci. 116 (2015) 204-211.
\item[{[RDB]}] U. Ryssel, F. Distel, D. Borchmann, Fast algorithms for implication bases and attribute exploration using proper premises, Ann Math Artif Intell 70 (2014) 25-53.
\item[{[RN]}] S. Russell, P. Norvig, Artificial Intelligence: A modern approach, Prentice Hall 2003.
	\item[{[S]}] A. Schrijver, Combinatorial Optimization (three volumes), Springer 2003.
	\item[{[Sh]}] R.C. Shock, Computing the minimum cover of functional dependencies, Inf. Proc. Letters 22 (1986) 157-159.
	\item[{[SW]}] L. Santocanale, F. Wehrung, Lattices of regular closed subsets of closure spaces, Internat. J. Algebra Comput. 24 (2014) 969-1030.
	\item[{[W1]}] M. Wild, Computations with finite closure systems and implications, Lecture Notes in Computer Science 959 (1995) 111-120. (An extended version, available as pdf, is the Tech. Hochschule Darmstadt Preprint Nr. 1708 from 1994.)
	\item[{[W2]}] M. Wild, Optimal implicational bases for finite modular lattices, Quaestiones Mathematicae 23 (2000) 153-161.
	\item[{[W3]}] M. Wild, A theory of finite closure spaces based on implications, Advances in Mathematics 108 (1994) 118-139.
\item[{[W4]}] M. Wild, Compressed representation of Learning Spaces. To appear in the Journal of Mathematical Psychology.
\item[{[W5]}] M. Wild, Implicational bases for finite closure systems, {\it Arbeitstagung, Begriffsanalyse und K\"{u}nstliche Intelligenz}, Informatik-Bericht 89/3 (1989), pp.147-169, Institut f\"{u}r Informatik, Clausthal. (The article is downloadable from the ResearchGate.)
\item[{[W6]}] M. Wild, Compactly generating all satisfying truth assignments of a Horn formula, Journal on Satisfiability, Boolean Modeling and Computation 8 (2012) 63-82.
\item[{[W7]}] M. Wild, The joy of implications, aka pure Horn formulas: mainly a survey. This is a preliminary version (arXiv: 1411.6432v2) of the present article. It features the full versions of Expansions 1, 3, 12, 13, 17.
\item[{[Wi]}] R. Wille, Subdirect decomposition of concept lattices, Algebra Universalis 17 (1983) 275-287. (Cited in [W7].)
\end{enumerate}

\end{document}